\documentclass[useAMS,usenatbib]{mn2e}
\usepackage{times}
\usepackage{graphicx}

\def\fH{\hbox{$^{\rm h}$}}

\def\etal{\hbox{et al.}~}
\def\nad{\rm Na\,{\sc i} D}
\def\nai{\rm Na\,{\mbox{\sc i}}}
\def\kms{\,km\,s$^{-1}$}
\def\cm2{\,cm$^{-2}$\,}
\def\logn{\,log\,$N_{\nai}$\,}
\def\vlsr{\,$V_{lsr}$\,}
\defcitealias{corradi97}{Paper~I}

\begin{document}

\title[Kinematics towards the Coalsack and Chamaeleon-Musca]{Local 
Interstellar Medium Kinematics towards the Southern Coalsack and 
Chamaeleon-Musca dark clouds}

\author[Corradi, Franco \& Knude]{W. J. B. Corradi$^1$, G. A. P. Franco$^1$, 
        J. Knude$^2$\\
        $^1$Departamento de F\'{\i}sica - ICEx - UFMG, Caixa Postal 702,
         30.123-970 - Belo Horizonte - MG, Brazil \\
         $^2$Niels Bohr Institute for Astronomy, Physics and Geophysics --
         Juliane Maries Vej 30 -- DK 2100 -- Copenhagen \O, Denmark}

\date{}
\pagerange{}
\pubyear{2003}

\maketitle
\label{firstpage}

\begin{abstract}
The results of a spectroscopic programme aiming to investigate the 
kinematics of the local interstellar medium components towards the Southern 
Coalsack and Chamaeleon-Musca dark clouds are presented. The analysis is based
upon high-resolution ($R \approx 60\,000$) spectra of the insterstellar NaI D 
absorption lines towards 63 B-type stars ($d \leq$ 500 pc) selected to cover 
these clouds and the connecting area defined by the Galactic coordinates: 
$308\degr \geq l \geq 294\degr$ and $-22\degr \leq b \leq 5\degr$. The radial 
velocities, column densities, velocity dispersions, colour excess and 
photometric distances to the stars are used to understand the kinematics and
distribution of the interstellar cloud components. The analysis indicates that
the interstellar gas is distributed in two extended sheet-like structures
permeating the whole area, one at $d \leq$ 60 pc and another around 120-150 pc
from the Sun. The nearby feature is approaching to the local standard of rest
with average radial velocity of $-$7 \kms, has low average column density 
\logn $\approx$ 11.2 \cm2 and velocity dispersion b $\approx$ 5 \kms. The more 
distant feature has column densities between 12.3 $\leq$ \logn\
$\leq$ 13.2, average velocity dispersion b $\approx$ 2.5 \kms\ and seems
associated to the dust sheet observed towards the Coalsack, Musca and
Chamaeleon direction. Its velocity is centered around 0 \kms, but there is a
trend for increasing from $-$3 \kms\ near $b = 1\degr$ to 3 \kms\ near
$b = -18\degr$. The nearby low column density feature indicates a general 
outflow from the Sco-Cen association, in aggreement with several independent 
lines of data in the general searched direction. The dust and gas feature 
around 120 -- 150 pc seem to be part of an extended large scale feature of 
similar kinematic properties, supposedly identified with the interaction zone 
of the Local and Loop I bubbles. Assuming that the interface and the ring-like
volume of dense neutral matter that would have been formed during the 
collision of the two bubbles have similar properties, our results rather 
suggest that the interaction zone between the bubbles is twisted and folded.
\end{abstract}


\section{Introduction}

The present work is part of a comprehensive investigation of the local
interstellar medium (ISM) towards the Southern Coalsack and the
Chamaeleon-Musca dark clouds. A comparison of the colour excess $E(b-y)$ vs.
distance diagrams for the Chamaeleon-Musca complex \citep{franco91} and the
Southern Coalsack \citep{franco89} shows remarkable similarities. The jump of
the colour excess to higher values occurs approximately at the same distance,
and the observed minimum value of this rise is almost the same: $\Delta E(b-y)
\approx 0\fm100$. Although the clouds are apart by more than {15\degr} these
facts suggested that they might be dense condensations embedded in an extended
interstellar structure.

In order to investigate the possible physical association between these two
clouds, a new photometric investigation was carried out. The campaign resulted
on very precise $uvby\beta$ data for 1017 stars covering these clouds and the
connecting area: $307\degr \geq l \geq 294\degr$ and $-20\degr \leq b \leq
5\degr$ \citep*{corradi95}. 

Analysis of the colour excess $vs.$ distance diagrams for these stars indicated
that exists a foreground region with very little reddening ($E(b-y)$ $\approx$
0\fm006), that is bound, at the distance range of 150 $\pm$ 30 pc, by a
transition region where the colour excesses suddenly increase to a
mininum value of $E(b-y)$ $\approx$ 0\fm050 for the most diffuse lines of
sight and $E(b-y)$ $\approx$ 0\fm100 for the densest ones.
Beyond this transition region the diagrams suggest the existence of a second
low column density volume, as the complete range of colour excess from $\approx$
0\fm050 to $0\fm275$, shown by the stars in the very narrow distance slot
centered on 150 pc, remains unchanged for at least another 350 pc 
\citep[hereafter Paper I]{corradi97}.

The onset of this minimum colour excess for the whole surveyed area suggests
that the obscuring material is distributed in an extended interstellar
sheet-like structure. The existence of such an absorbing feature, at a distance
identical to the molecular clouds, also suggests that Coalsack, Chamaeleon and
Musca can be dense condensations embedded on the diffuse medium composing the
sheet-like feature.

It has been also noted in Paper I that the jump in $E(b-y)$, caused by the dust
feature, seems to have a dependence on the galactic latitude that can be
represented by $[E(b-y)_{\rm min}, b] = [0\fm050;0\degr]
\rightarrow [0\fm100;-8\degr] \rightarrow [0\fm150;-15\degr]$. 
The quoted dependence might indicate, if we postulate that the dust sheet is
roughly perpendicular to the galactic plane, that the sheet does not have the
same optical thickness, because $E(b-y)_{\rm min}$ {\it vs.} $(b)$ does not 
follow a simple $E(b-y) \sec (b)$ law. On the other hand, it might also be 
the result of approaching the tangential point of a warped sheet-like 
structure with the same column density, curved away from the Sun. 

As pointed out in Paper I, when viewed in connection to the other data on the
local ISM, the existence of these two low-reddening volumes has led to the idea
that the dust sheet could be part of a large scale structure, probably related
to the interface of the Local and Loop I Bubbles. Such an interface of dense
matter may have resulted from the compression of the ISM by the action of
energetic events of the Sco-Cen OB association from the far side 
(\citealt{weaver,iwan,crawford,degeus}) and either by a supernova explosion 
near the Sun or by whatever created the local low density region from the near 
side (e.g. \citealt{cowie,cox,bochk,frisch86,gehrels,hartquist,bruh96}). If 
the dust sheet is part of a small bubble, it could be associated 
to the LCC-shell proposed by \citet{degeus}. Note however that \citet{genova} 
find that gas is moving out from Sco-Cen in the fourth Galactic quadrant, but
towards the association in the first quadrant. 

Recently, observational evidences of an annular volume of dense neutral matter,
that supposedly would have been formed during the collision of the two bubbles,
was found on the ROSAT all-sky survey data by \citet{egger}.
Remarkably, as noticed by \citet{dame}, the molecular gas from $l \approx
360\degr$ to $290\degr$ and $-25\degr \leq b \leq 25\degr$ seemingly define a
large complex of clouds around 100-150 pc from the Sun, comprising, e.g., $\rho$
Oph, R CrA, G317-4, Musca, Coalsack, Chamaeleon and the Lupus clouds. 

\begin{figure}
\centering
\includegraphics[width=\hsize]{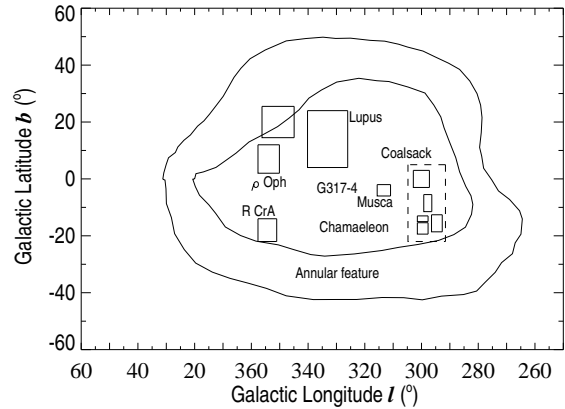}
\caption{Schematic representation of the interaction zone between the Local
and Loop I Bubbles with the dark clouds towards the Sco-Cen association.
The position and sizes of the dark clouds were obtained from the survey by 
\citet{dame}. The ring-like contours, taken from the ROSAT all-sky survey data
by \citet{egger}, represent the annular volume of dense neutral
matter supposedly formed during the collision of the two bubbles.The dashed
square delineates the region surveyed in this work}
\label{ring}
\end{figure}

In this case a schematic drawing of the interaction zone between the two bubbles
and the dark clouds towards the Sco-Cen association would be like 
Fig.~\ref{ring}. The position and sizes of the dark clouds were obtained from 
\citet{dame} and the ring-like contours from \citet{egger}. 
The dashed square delineates the region surveyed in this work.

To investigate the kinematics of the interstellar gas components towards the
Coalsack, Chamaeleon and Musca dark clouds we have obtained high-resolution
($R \approx 60\,000$) spectra of the interstellar \nad\ absorption lines. From
the radial velocities, column densities, velocity dispersions and the known
distance of the stars we aim to study the physical association proposed for
these clouds.

\section{Observations}

\subsection{The Program Stars}

In order to investigate the interstellar gas components towards the Coalsack,
Chamaeleon and Musca dark clouds 60 B-type stars covering the area defined by
the Galactic coordinates: $308\degr \geq l \geq 294\degr$ and $-22\degr \leq b
\leq 5\degr$ have been selected from the photometric $uvby\beta$ sample by
\citet{corradi95}. All targets are located within 500 pc from the Sun and have 
well determined colour excesses $E(b-y)$ and photometric distances 
\citepalias{corradi97}. SAO\,257142, SAO\,258697 and PPM\,377063 were observed 
to complement the data at more negative latitudes. 

\begin{figure}
\centering
\includegraphics[width=\hsize]{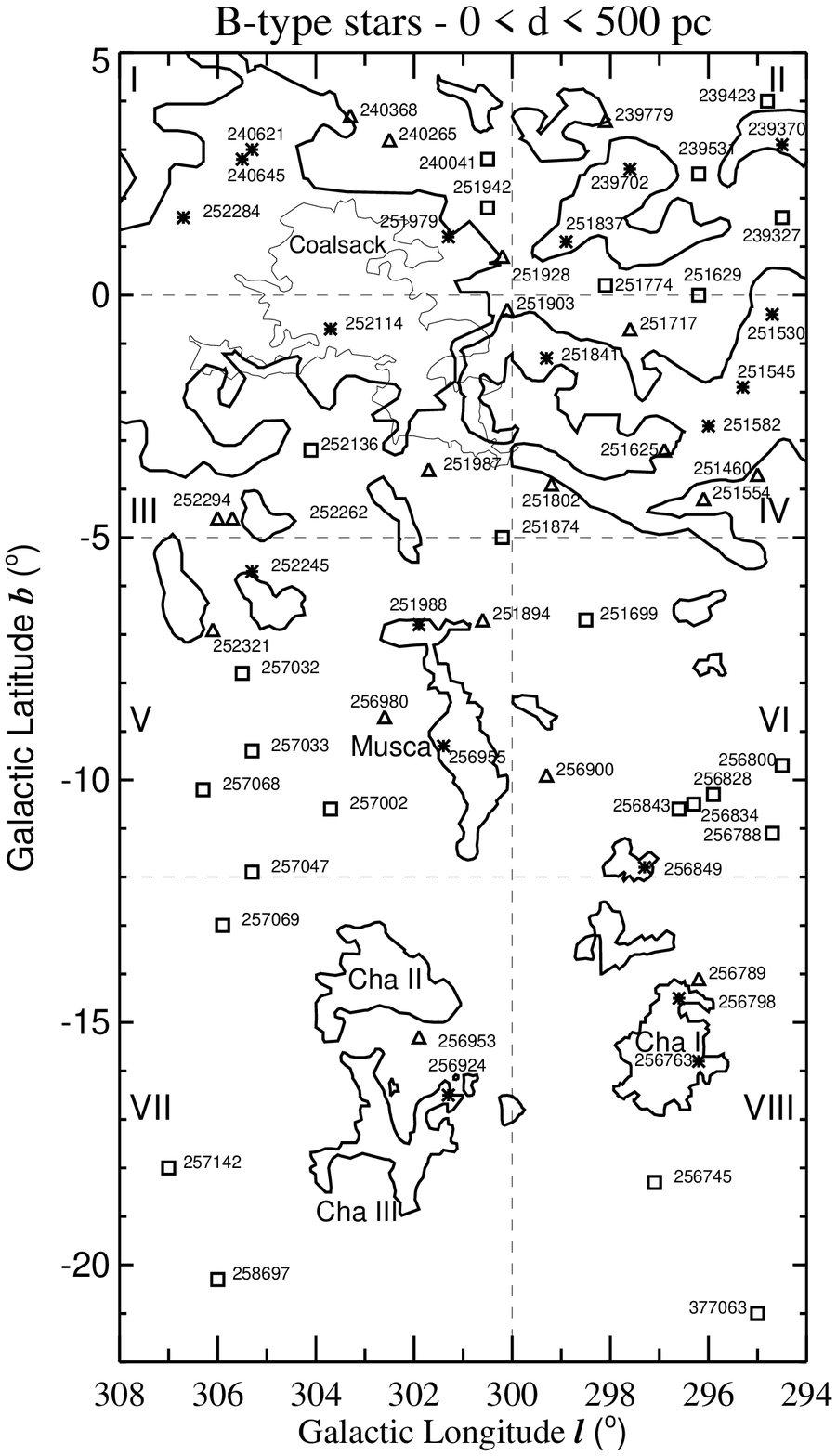}
\caption{Location of the stars within the studied area. The clouds' contours
defined by the thick lines are the lowest opacity level of the photographic
catalogue by \citet{feitzinger} and the thin one is the outer 2 K
km s$^{-1}$ velocity integrated CO emission contour for the Coalsack. The
($*$), ($\triangle$) and ($\sq$) signs indicate the stars with line-of-sight
{\em inside} the clouds' contours, in their {\em outskirts} and towards the
{\em sheet-like} structure, respectively. The horizontal and vertical dashed
lines delineate eight sub-areas, identified by the roman numbers, that are used
to discuss the details of the velocity structure.}
\label{obpos}
\end{figure}

The basic information on the 65 stars is listed in Table 1. For each one the
visual magnitude $V$, the Galactic longitude and latitude, the colour excess
and the distance are given in successive columns. Based on the $uvby\beta$
data the mean errors in the colour excess and distance values are 0\fm018 and
15-30\%, respectively. The positions of the stars within the surveyed area are
given in Fig.~\ref{obpos}. The clouds' contours defined by the thick lines 
are the lowest opacity level of the photographic catalogue by 
\citet{feitzinger} and the thin one is the outer 2 K km s$^{-1}$ velocity 
integrated CO emission contour for the Southern Coalsack \citep{nyman}.

The data sample consists of three general divisions based upon the star's
location within the area: {\em inside} the clouds' contours, in their {\em
outskirts} and towards the {\em sheet-like} structure. They are, respectively,
indicated by the ($*$), ($\triangle$) and ($\sq$) signs. The horizontal and
vertical dashed lines delineate eight sub-areas, identified by the roman
numbers, that will be used to discuss the details of the velocity structure.
Note that 5 stars are within 100 pc, while 7 other are between 100 and 150 pc
and another 7 between 150 and 200 pc from the Sun.

\begin{figure}
\centering
\includegraphics[width=\hsize]{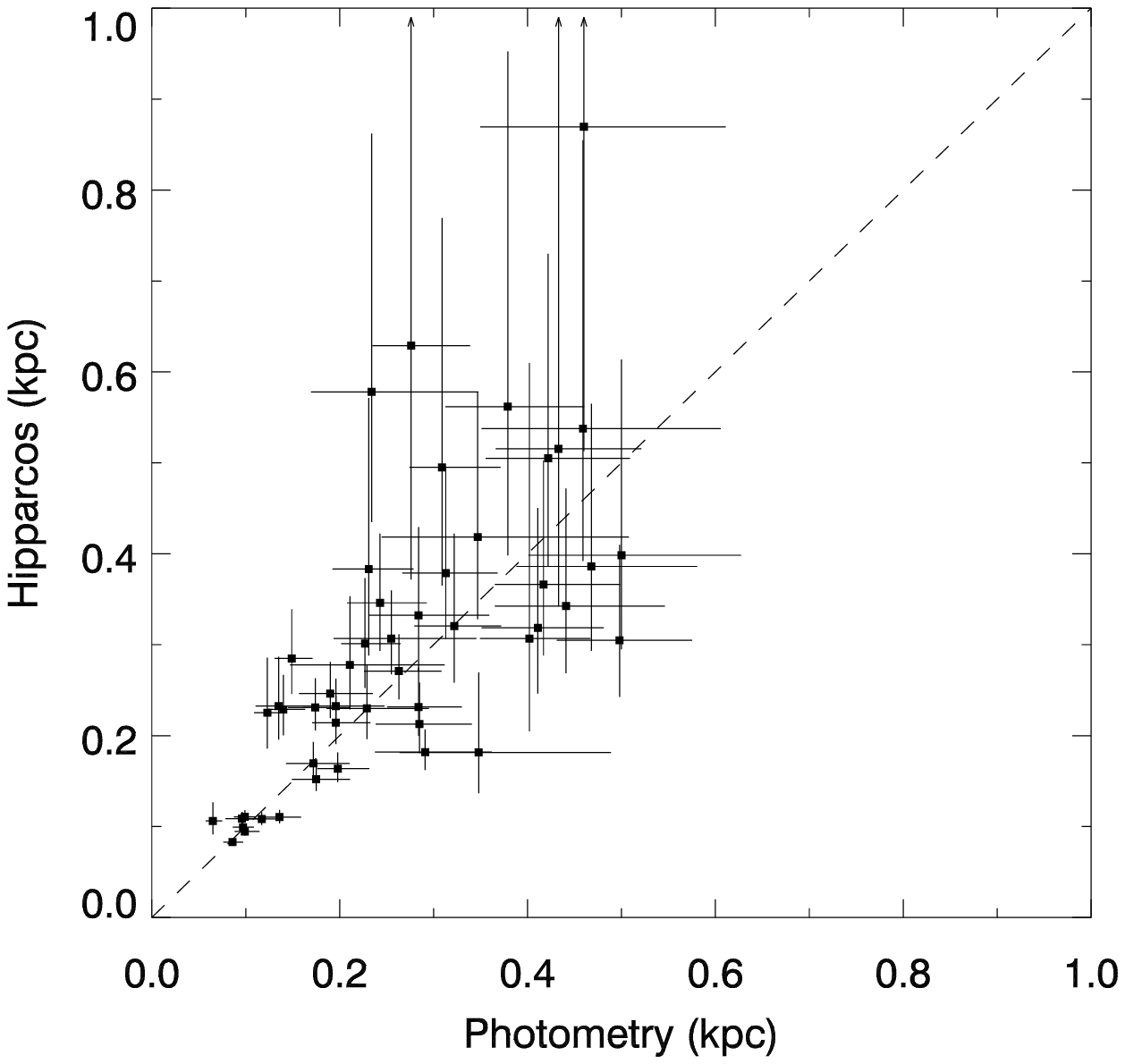}
\caption{Comparison between the trigonometric distances obtained from the
{\it Hipparcos} catalogue \citep{esa} and the photometric distances based
on $uvby\beta$ data.}
\label{hipparcos}
\end{figure}

\subsection{The Instrumental Setup}

\begin{table*}
\caption{Basic information on the program stars}
\scriptsize
\begin{tabular}{rcrrcrclcrrcl} \hline
\multicolumn{1}{c}{Star} & $V$ & $l$\,(\degr) & $b$\,(\degr) & $E(b-y)$ & $d$(pc) & &
\multicolumn{1}{c}{Star} & $V$ & $l$\,(\degr) & $b$\,(\degr) & $E(b-y)$ & $d$(pc) \\ \hline
\multicolumn{6}{c}{Towards the dark clouds contours} & \multicolumn{1}{c}{} &
\multicolumn{6}{c}{Outskirts of the dark clouds} \\ \hline
239370 & 8.754 & 294.5 &   3.1 &  0.132 & 427 & & 239779 & 7.743 & 298.1 &   3.6 &  0.107 & $\mbox{\hskip.8em}$468 \\
239702 & 7.160 & 297.6 &   2.6 &  0.079 & 379 & & 240265 & 7.183 & 302.5 &   3.2 &  0.106 & $\mbox{\hskip.8em}$489 \\
240621 & 8.621 & 305.3 &   3.0 &  0.274 & 454 & & 240368 & 4.617 & 303.3 &   3.7 &  0.014 & $\mbox{\hskip.8em}$136 \\
240645 & 4.624 & 305.5 &   2.8 &  0.022 &  65 & & 251460 & 7.167 & 295.0 &  $-$3.7 &  0.199 & $\mbox{\hskip.8em}$312 \\
251530 & 8.217 & 294.7 &  $-$0.4 &  0.105 & 285 & & 251554 & 8.215 & 296.1 &  $-$4.2 &  0.272 & $\mbox{\hskip.8em}$291 \\
       &       &       &       &        &     & &        &       &       &       &        &     \\
251545 & 8.018 & 295.3 &  $-$1.9 &  0.115 & 417 & & 251625 & 7.332 & 296.9 &  $-$3.2 &  0.149 & $\mbox{\hskip.8em}$347 \\
251582 & 6.880 & 296.0 &  $-$2.7 &  0.164 & 291 & & 251717 & 4.724 & 297.6 &  $-$0.7 &  0.072 & $\mbox{\hskip.8em}$174 \\
251837 & 7.519 & 298.9 &   1.1 &  0.090 & 309 & & 251802 & 7.111 & 299.2 &  $-$3.9 &  0.103 & $\mbox{\hskip.8em}$422 \\
251841 & 4.053 & 299.3 &  $-$1.3 &  0.006 &  99 & & 251894 & 7.897 & 300.6 &  $-$6.7 &  0.083 & $\mbox{\hskip.8em}$348 \\
251979 & 8.378 & 301.3 &   1.2 &  0.139 & 483 & & 251903 & 4.868 & 300.1 &  $-$0.3 &  0.015 & $\mbox{\hskip.8em}$124 \\
       &       &       &       &        &     & &        &       &       &       &        &     \\
251988 & 8.838 & 301.9 &  $-$6.8 &  0.139 & 402 & & 251928 & 6.920 & 300.2 &   0.8 &  0.096 & $\mbox{\hskip.8em}$313 \\
252114 & 8.092 & 303.7 &  $-$0.7 &  0.226 & 498 & & 251987 & 6.266 & 301.7 &  $-$3.6 &  0.021 & $\mbox{\hskip.8em}$117 \\
252245 & 6.674 & 305.3 &  $-$5.7 &  0.114 & 123 & & 252262 & 7.258 & 305.7 &  $-$4.6 &  0.050 & $\mbox{\hskip.8em}$172 \\
252284 & 4.529 & 306.7 &   1.6 &  0.018 &  96 & & 252294 & 7.810 & 306.0 &  $-$4.6 &  0.108 & $\mbox{\hskip.8em}$441 \\
256763 & 7.978 & 296.2 & $-$15.8 &  0.465 & 229 & & 252321 & 6.198 & 306.1 &  $-$6.9 &  0.108 & $\mbox{\hskip.8em}$196 \\
       &       &       &       &        &     & &        &       &       &       &        &     \\
256798 & 7.675 & 296.6 & $-$14.5 &  0.220 & 198 & & 256789 & 7.970 & 296.2 & $-$14.1 &  0.161 & $\mbox{\hskip.8em}$227 \\
256849 & 6.883 & 297.3 & $-$11.8 &  0.286 & 234 & & 256900 & 8.113 & 299.3 &  $-$9.9 &  0.194 & $\mbox{\hskip.8em}$211 \\
256924 & 4.240 & 301.3 & $-$16.5 &  0.007 &  86 & & 256980 & 8.440 & 302.6 &  $-$8.7 &  0.159 & $\mbox{\hskip.8em}$460 \\
256955 & 3.841 & 301.4 &  $-$9.3 &  0.003 &  97 & & 256953 & 7.777 & 301.9 & $-$15.3 &  0.160 & $\mbox{\hskip.8em}$284 \\
\hline \multicolumn{13}{c}{Towards the sheet} \\ \hline
239327 & 8.684 & 294.5 &   1.6 &  0.068 & 500 & & 256828 & 6.399 & 295.9 & $-$10.3 &  0.157 & $\mbox{\hskip.8em}$196 \\
239423 & 8.239 & 294.8 &   4.0 &  0.036 & 411 & & 256834 & 5.562 & 296.3  &$-$10.5 &  0.192 & $\mbox{\hskip.8em}$263 \\
239531 & 7.455 & 296.2 &   2.5 &  0.056 & 459 & & 256843 & 6.079 & 296.6  &$-$10.6 &  0.237 & $\mbox{\hskip.8em}$135 \\
240041 & 7.758 & 300.5 &   2.8 &  0.123 & 231 & & 257002 & 9.163 & 303.7  &$-$10.6 &  0.148 & $\mbox{\hskip.8em}$368 \\
251629 & 8.121 & 296.2 &  $-$0.0 &  0.105 & 433 & & 257032 & 7.628 & 305.5  & $-$7.8 &  0.093 & $\mbox{\hskip.8em}$284 \\
       &       &       &       &        &     & &        &       &        &      &        &     \\
251699 & 5.896 & 298.5 &  $-$6.7 &  0.006 &  99 & & 257033 & 6.045 & 305.3  & $-$9.4 &  0.145 & $\mbox{\hskip.8em}$140 \\
251774 & 9.160 & 298.1 &   0.2 &  0.107 & 384 & & 257047 & 6.618 & 305.3  &$-$11.9 &  0.020 & $\mbox{\hskip.8em}$175 \\
251874 & 8.124 & 300.2 &  $-$5.0 &  0.157 & 322 & & 257068 & 8.747 & 306.3  &$-$10.2 &  0.126 & $\mbox{\hskip.8em}$276 \\
251942 & 8.072 & 300.5 &   1.8 &  0.094 & 454 & & 257069 & 6.308 & 305.9  &$-$13.0 &  0.069 & $\mbox{\hskip.8em}$149 \\
252136 & 9.000 & 304.1 &  $-$3.2 &  0.078 & 353 & & 257142 & 5.000 & 307.0  &$-$18.0 &   --   & $\mbox{\hskip.8em}$169$^a$ \\
       &       &       &       &        &     & &        &       &        &      &        &     \\
256745 & 5.964 & 297.1 & $-$18.3 &  0.076 & 255 & & 258697 & 6.800 & 306.0  &$-$20.3 &   --   & $\mbox{\hskip.8em}$270$^a$ \\
256788 & 6.759 & 294.7 & $-$11.1 &  0.143 & 190 & & 377063$^b$ & 5.100 & 295.0  &$-$21.0 &   --   & $\mbox{\hskip.8em}$165$^a$ \\
256800 & 5.585 & 294.5 &  $-$9.7 &  0.106 & 243 & &        &       &        &      &        &     \\
\hline 
\multicolumn{12}{l}{$^a$ Based on trigonometric parallax from {\it Hipparcos}
\citep{esa}.} \\
\multicolumn{12}{l}{$^b$ Identification from the PPM catalogue.} 
\end{tabular}
\end{table*}

High resolution ($R \approx$ 60\,000) spectra of the \nai\ D$_{1}$ and D$_{2}$
absorption lines superposed on the continua spectra of the selected stars have
been obtained with the 1.4m Coud\'e Auxiliary Telescope (CAT) and the Coud\'e
Echelle Spectrometer (CES) at the European Southern Observatory (ESO), in La
Silla (Chile). The observations were carried out during 5 nights in April 1996
and 6 nights in April 1997, using the remote control facilities at the ESO's
HeadQuarters in Garching bei M\"unchen (Germany).

The CES optics was optimized to give a reciprocal dispersion of 1.8 \AA/mm at
the central wavelength of 5890 \AA, resulting in a spectral coverage of nearly
60 \AA. The entrance slit width of 459 $\mu$m, used for all the observations,
yielded an actual instrumental resolution (FWHM) $\Delta\lambda = 105$ m\AA\ or
a velocity resolution $\Delta v$ = 5.3 km s$^{-1}$ (FWHM). This corresponds to
an actual resolving power of $R \approx$ 56\,000, as determined by measurements
of the intrinsic width of the thorium lines.

The CCD was a LORAL/LESSER 2688 x 512, with 16 pre-scan pixels. It has a pixel
size of 15 $\mu$m $\times$ 15 $\mu$m and is relatively free of blemishes. Since
the optical train selects only a single order, it was always possible to
position the spectrum on a clean portion of the CCD. The dark current and the
readout noise are very low, 1.8 $e^{-}$/pixel/hour at 164 K and
8.3 $e^{-}$, respectively. The combined efficiency of the CAT/CES using the
long camera is about 8\% in the used central wavelength. Additional information
regarding the CAT/CES can be found in \citet{kaper}.

Multiple sets of bias, dark and flat-field exposures were obtained each night
for calibration purposes. Exposures of a thorium lamp were taken before and
after each object exposure to ensure a better wavelength calibration. The CES,
however, showed to be a very stable instrument.

Exposure times ranged from 2 minutes to 2.5 hours, depending on the visual 
magnitude and spectral type. In all cases signal-to-noise ratios (S/N) 
greater than 150 in the raw spectra were achieved. At least two exposures 
were obtained at each setup position and the integration times were limited 
to a maximum of 30 minutes to minimize the effects of cosmic ray events in 
the individual spectra.

\subsection{Stellar Distances}

Although {\it Hipparcos} satellite was launched in 1989 August, the catalogue
only became available in 1997 \citep{esa}. For this reason, the observing
sample was selected based on our photometric distance determination. However,
the paralactic distances may now serve as a basis for comparison with our
photometric distances. A search in the {\it Hipparcos} catalogue provided us 
51 stars in common with our sample -- one of them, SAO\,239370, appears with
negative value for its trigonometric parallax, and was excluded from the
comparison. Fig.~\ref{hipparcos} shows a plot containing the distances based
on the {\it Hipparcos} parallax vs. photometric distances. The denoted errors
for the photometric distances were estimated for each star based on the 
individual uncertainties of their photometric measurements \citep{corradi95}, 
and ranges from 11 to 49 per cent. In general, the agreemment is very good
testifying the high quality of our photometric distance determination. There
are few stars showing rather large discrepancy between their estimated
photometric and parallatic distances, usually those beyond 200\,pc and large 
uncertainties in their trigonometric parallax. However, as will be seen
further, the main interstellar structures discussed in this work are located
nearer than that, and consequently these uncertainties in the distance
estimation do not affect substantially our conclusions. 

\section{Data Reduction}

The initial processing of the CCD frames employed IRAF routines to
subtract the bias, divide by a normalized flat-field, and remove cosmic rays
from the rows occupied by the stellar spectrum and the background regions. The
one-dimensional spectra were then extracted using IRAF routines.

It is worth reiterating that usually the dome flat-fields provided the
best correction for the strong vignetting present in the spectra borders,
besides that the affected parts were trimmed off the images. The wavelength
calibration was established with a set of 20 to 30 lines, identified using the
thorium line wavelengths of \citet{dodorico}. A second-order Chebyshev
function was fitted to the line wavelength as a function of the pixel number,
with rms scatter of the residuals about the fit smaller than 0.002 \AA\ (0.05 km
s$^{-1}$) in all cases. The dispersion solution was applied to all science
exposures with no rebinning of the data, and the final spectra normalized to
unity continuum with cubic splines.

The individual normalized spectra were then co-added on a pixel-by-pixel basis,
weighted by the inverse of the rms deviation of the continuum fit. No noticeable
degradation in spectral resolution resulted from the co-addittion procedure,
since the individual spectra have been obtained within an hour or two of each
other, and the velocity changes in the observer rest frame, over this time
scale, are very small compared to the instrumental resolution.

Numerous absorption lines arising from the Earth's atmosphere occur in this
region of the spectrum, and two lines in particular (at 5889.637 \AA\ and
5890.09 \AA) are very close to the rest wavelength of the D$_2$ line. 
\citet{hobbs} discusses the origin and identification of these atmospheric 
(telluric) lines in the \nad\ wavelength region. To reduce telluric line 
contamination the stellar spectra were divided out by a purely atmospheric 
absorption template spectrum obtained from a fast rotating, unreddened 
early-type star with no interstellar sodium, and taken at similar air mass. 
For both observing runs spectra of $\alpha$\,Leo and $\gamma$\,Lup were taken 
each night. The equivalent width of the \nad\ lines for these stars are 
estimated to be $\leq$ 1 m\AA\ and 1.4 m\AA, respectively (\citealt{welsh91, 
crawford}). The stars SAO\,251050 and SAO\,224833 ($\iota$\,Lup), initially 
chosen as candidates for telluric line correction, showed the presence of 
interstellar \nad\ lines.

Since most of the telluric lines arise from water molecules, their strengths 
change throughout the night as the water content of the Earth's atmosphere
changes. Hence, the template spectra were properly scaled to match the
strengths of the atmospheric lines in the object spectra. The success of the
atmospheric correction procedure is apparent from the fact that all the
telluric lines, which are clear of the interstellar lines, have been
satisfactorily removed (see Fig.~\ref{spectra}).

Finally the observed wavelengths were brought to the velocity scale and
converted to the Local Standard of Rest (LSR) velocity frame, assuming a solar
motion of 20 \kms\ towards $\alpha$ = 18$\fH$ and $\delta$ = 30$\degr$ 
\citep{mihalas}. To access the zero point of the velocities the wavelengths of
some of the numerous telluric lines gathered in the observed spectral range were
measured, providing an external accuracy of the order of 4 m\AA\ and showing
that the lines match quite well, seemingly with no systematic offset.

\begin{figure*}
\hbox{
\includegraphics[width=5.6cm]{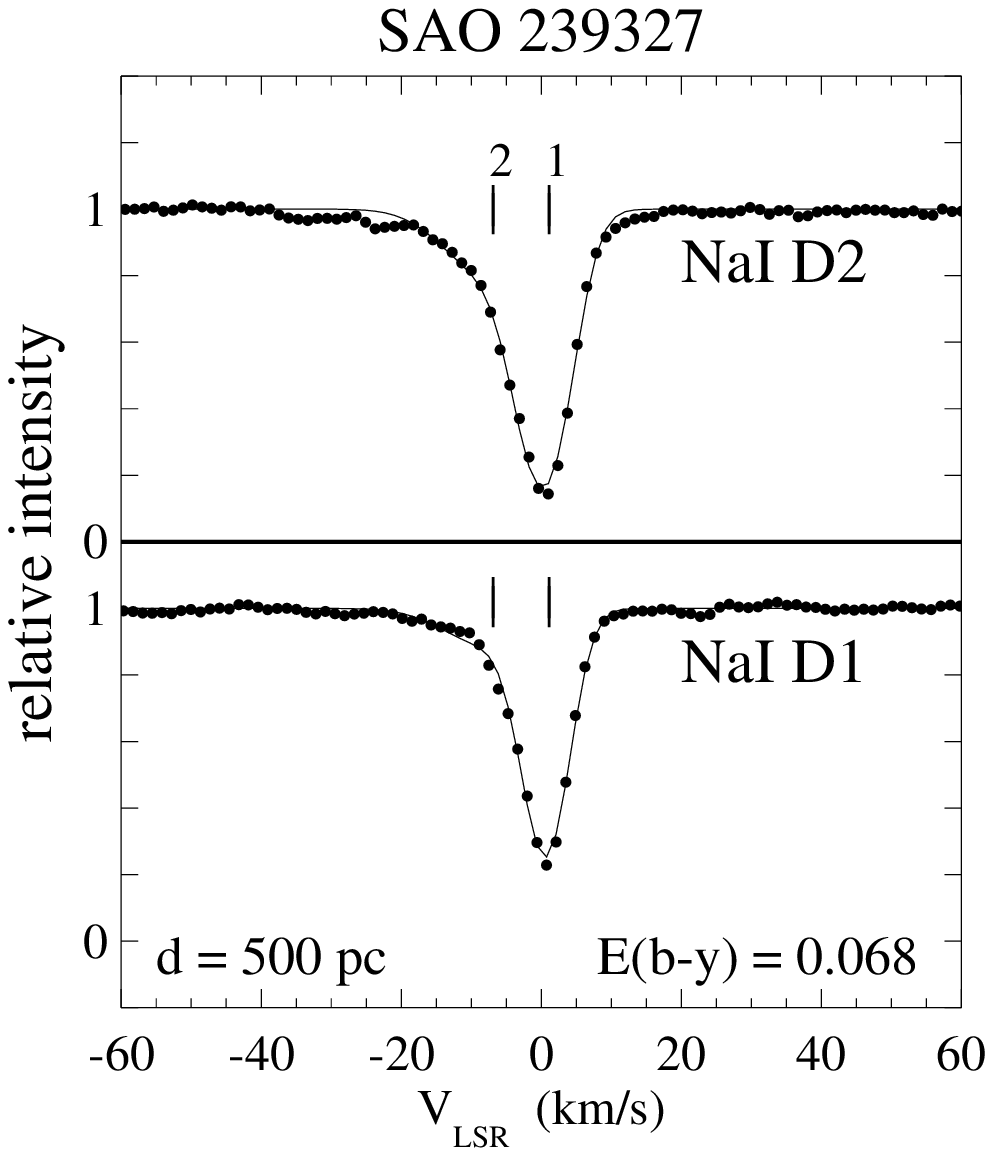} \hfill
\includegraphics[width=5.6cm]{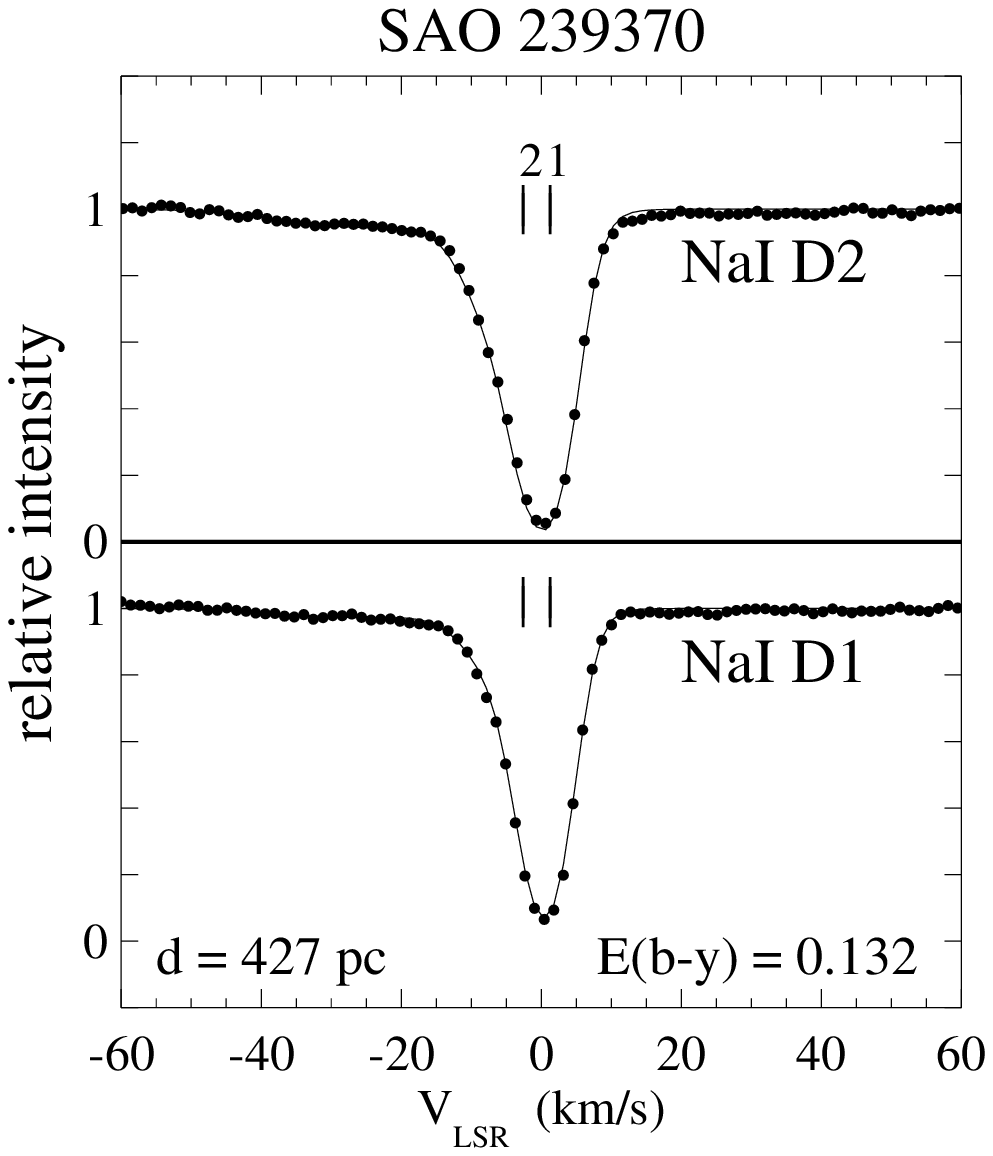} \hfill
\includegraphics[width=5.6cm]{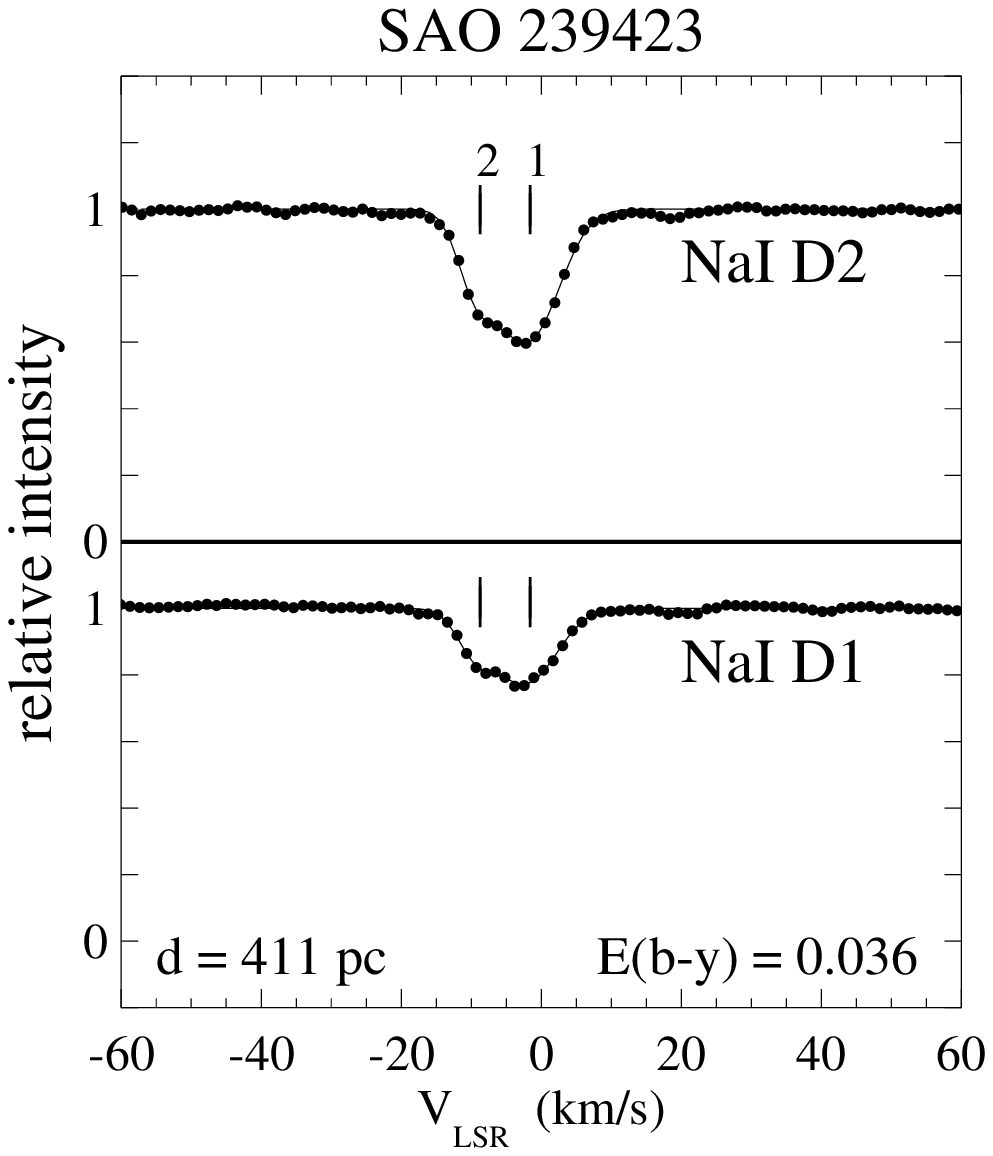} }
\vskip.3cm
\hbox{
\includegraphics[width=5.6cm]{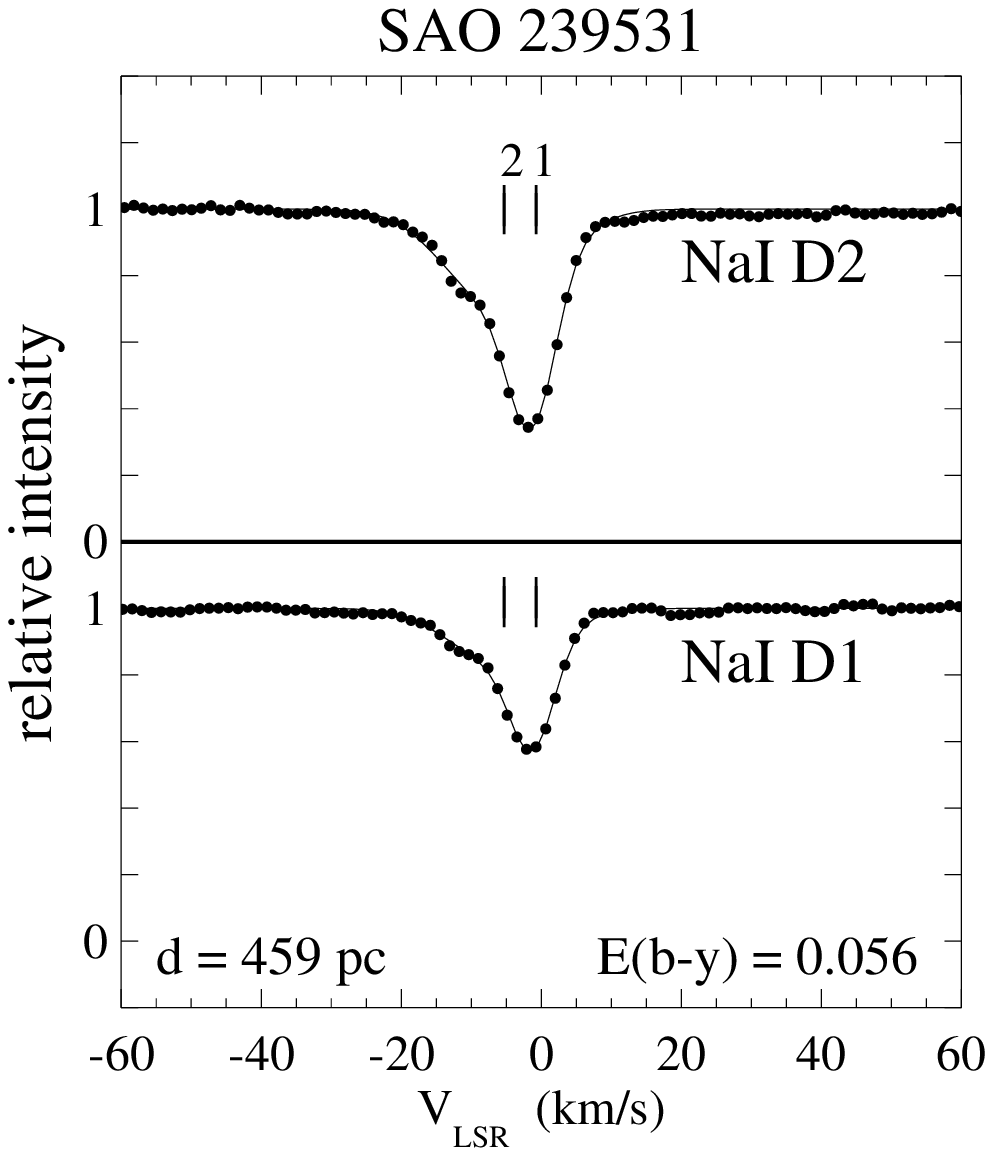} \hfill
\includegraphics[width=5.6cm]{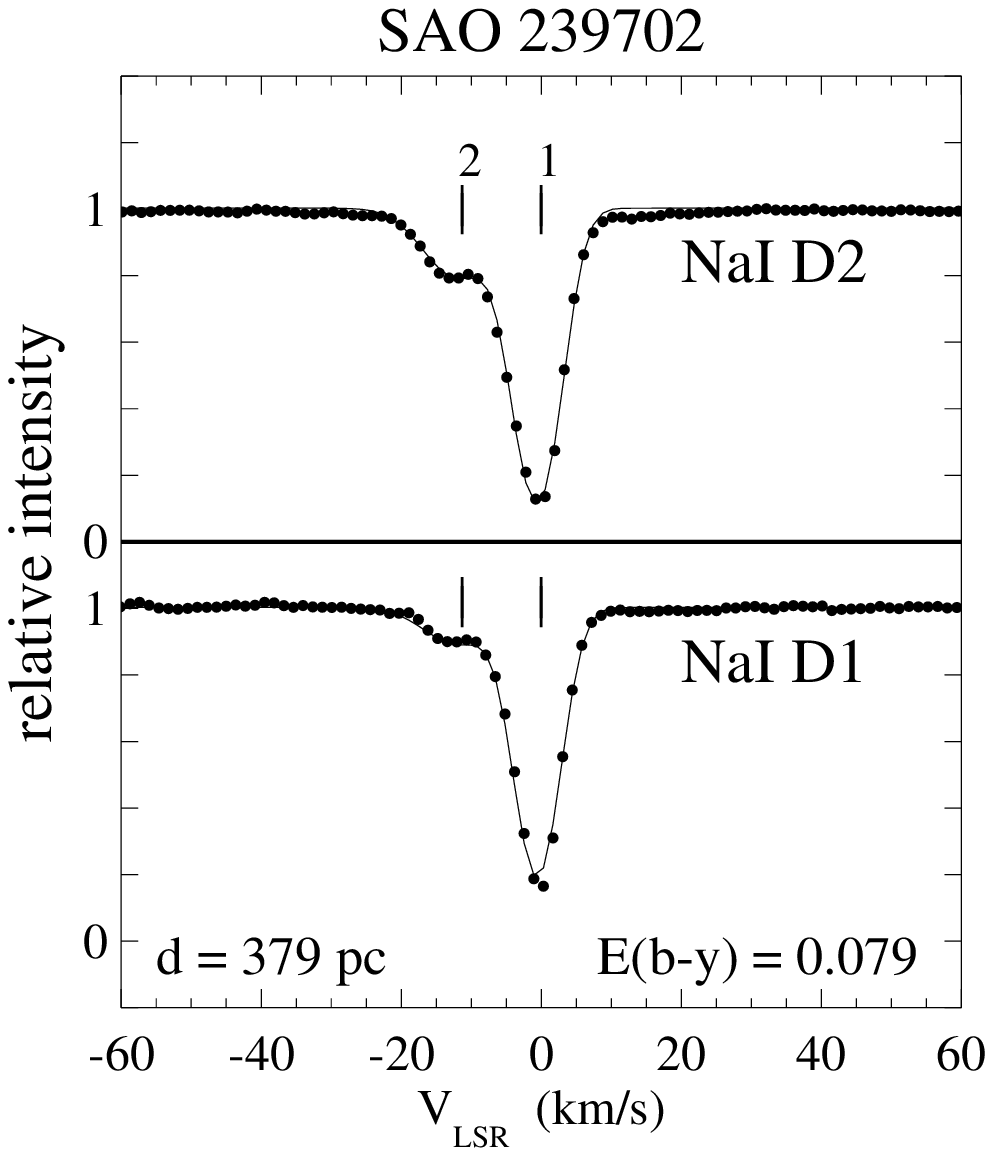} \hfill
\includegraphics[width=5.6cm]{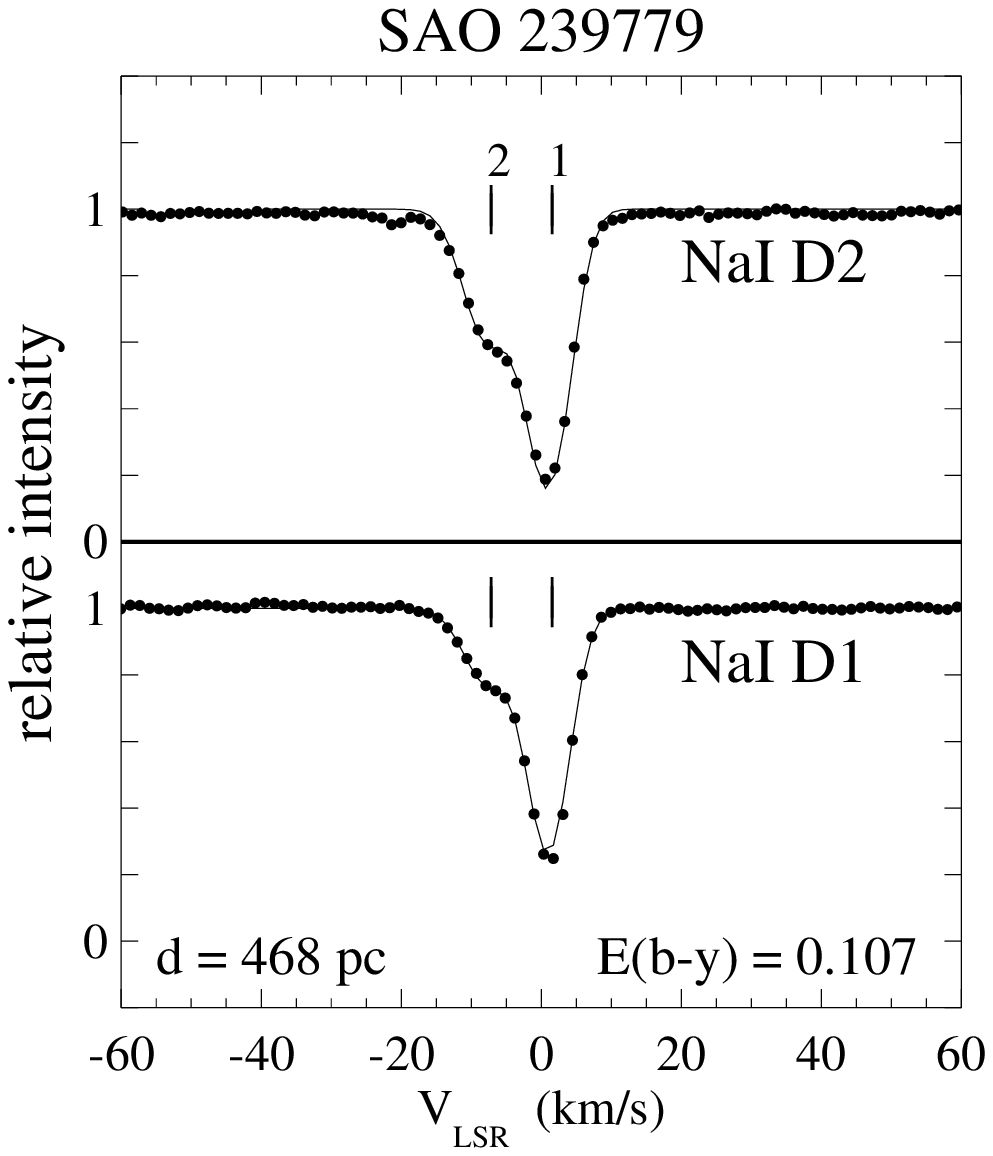} }
\vskip.3cm
\hbox{
\includegraphics[width=5.6cm]{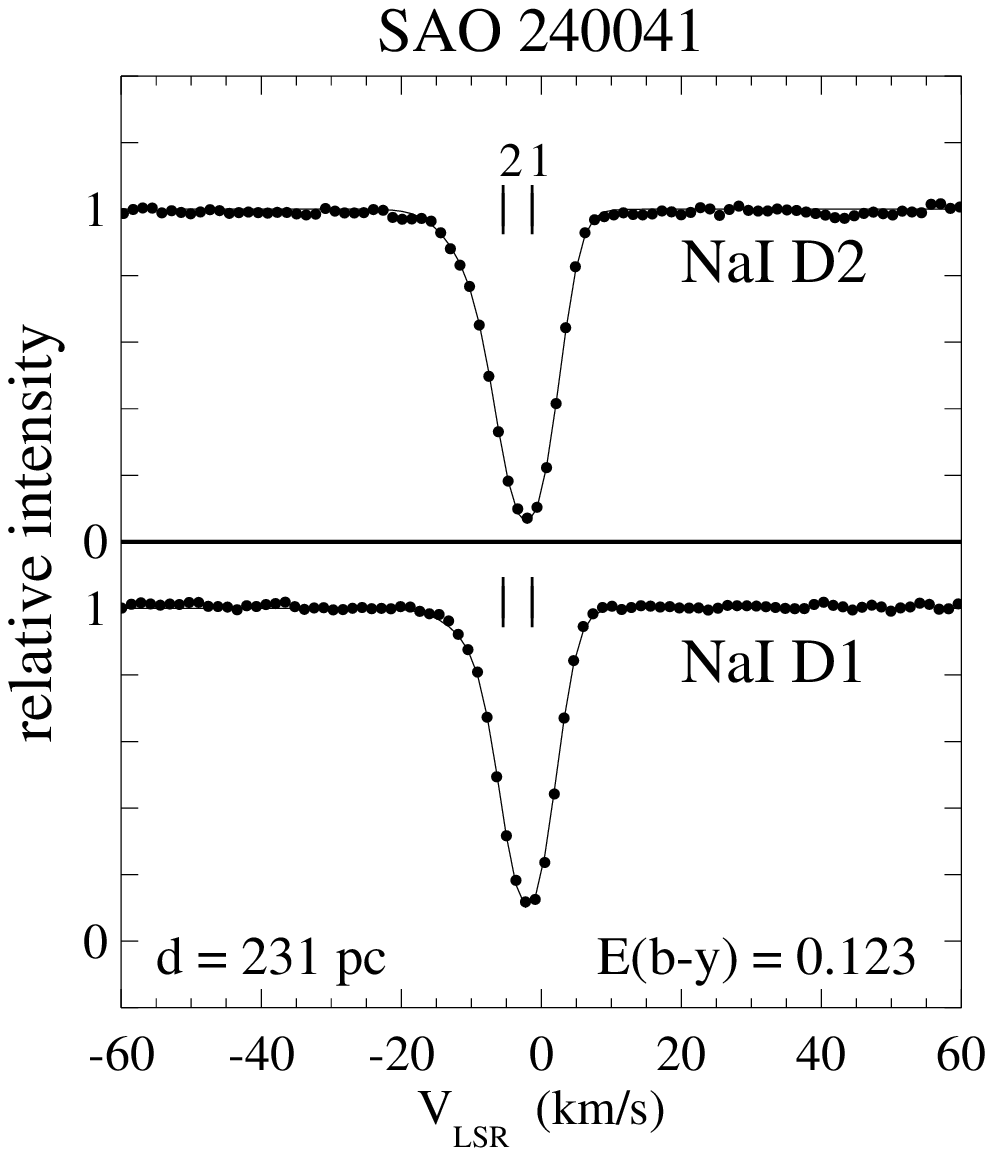} \hfill
\includegraphics[width=5.6cm]{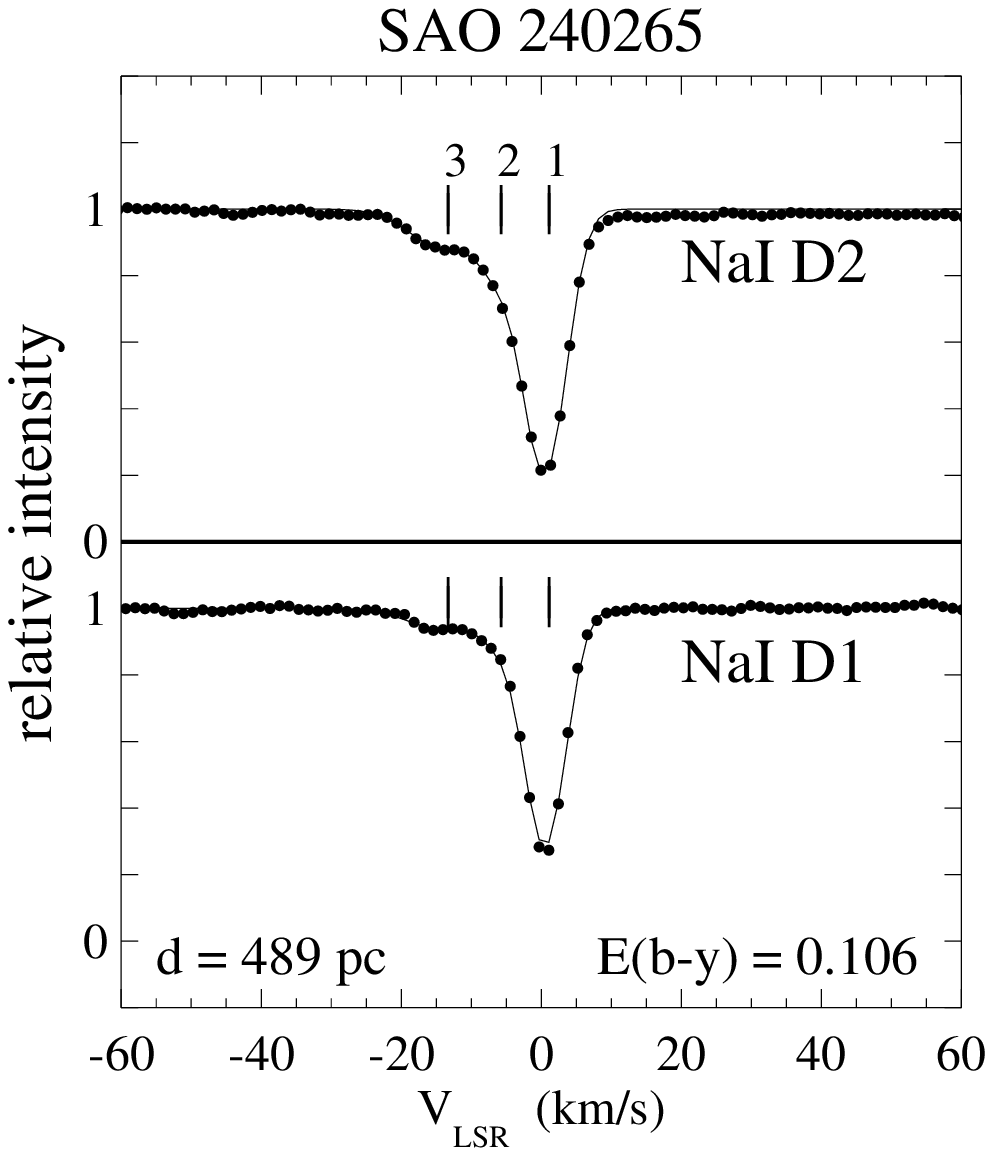} \hfill
\includegraphics[width=5.6cm]{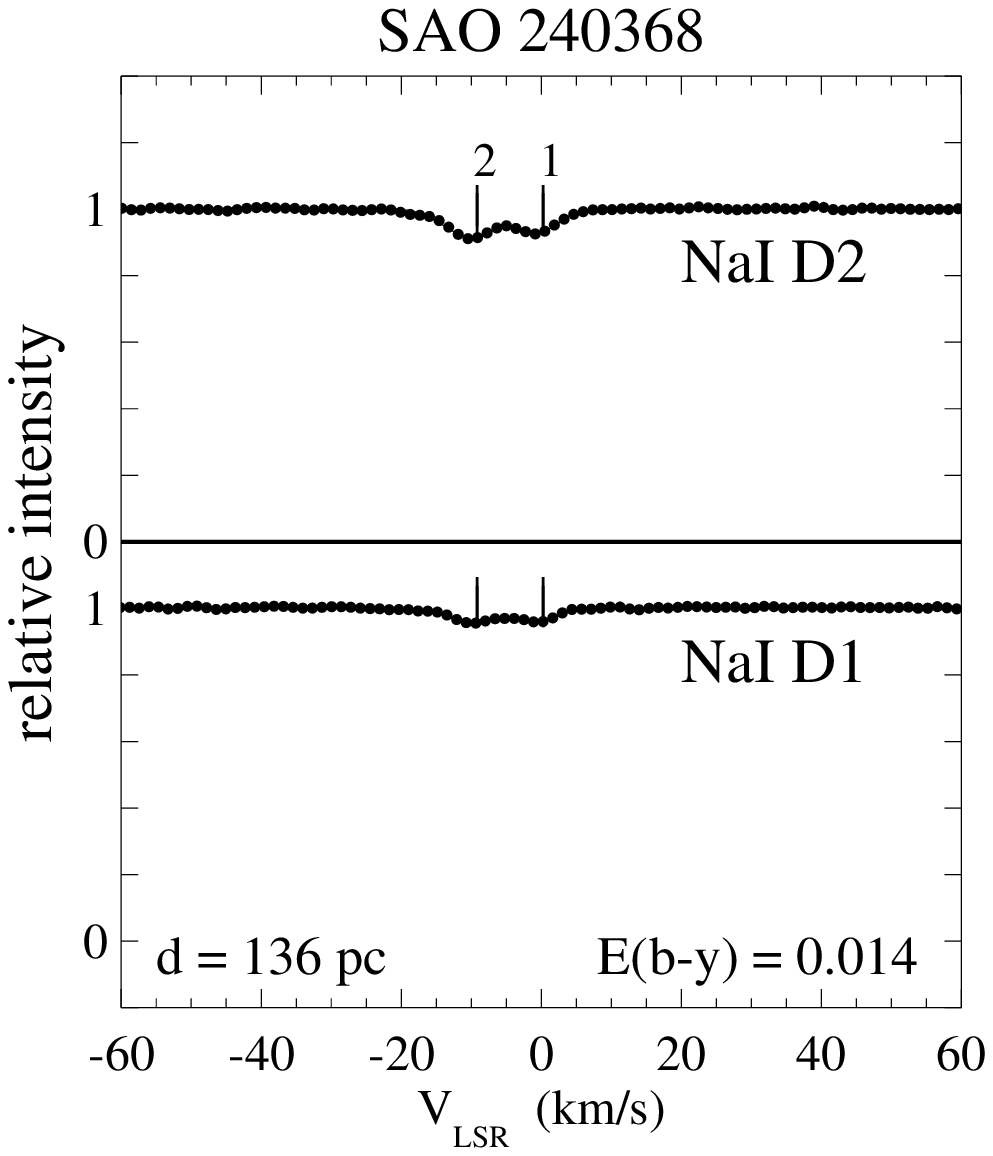} }
\caption{The observed \nad\ absorption lines. The observed intensities are
plotted as dots; the smooth curves are theoretical line profiles with the
parameters given in Table 3. The tick marks and the corresponding number 
identifies the fitted components. The zero point of the velocity scale is 
referred to the rest wavelength of the D$_2$ and D$_1$ lines}
\label{spectra}
\end{figure*}
\addtocounter{figure}{-1}
\begin{figure*}
\hbox{
\includegraphics[width=5.6cm]{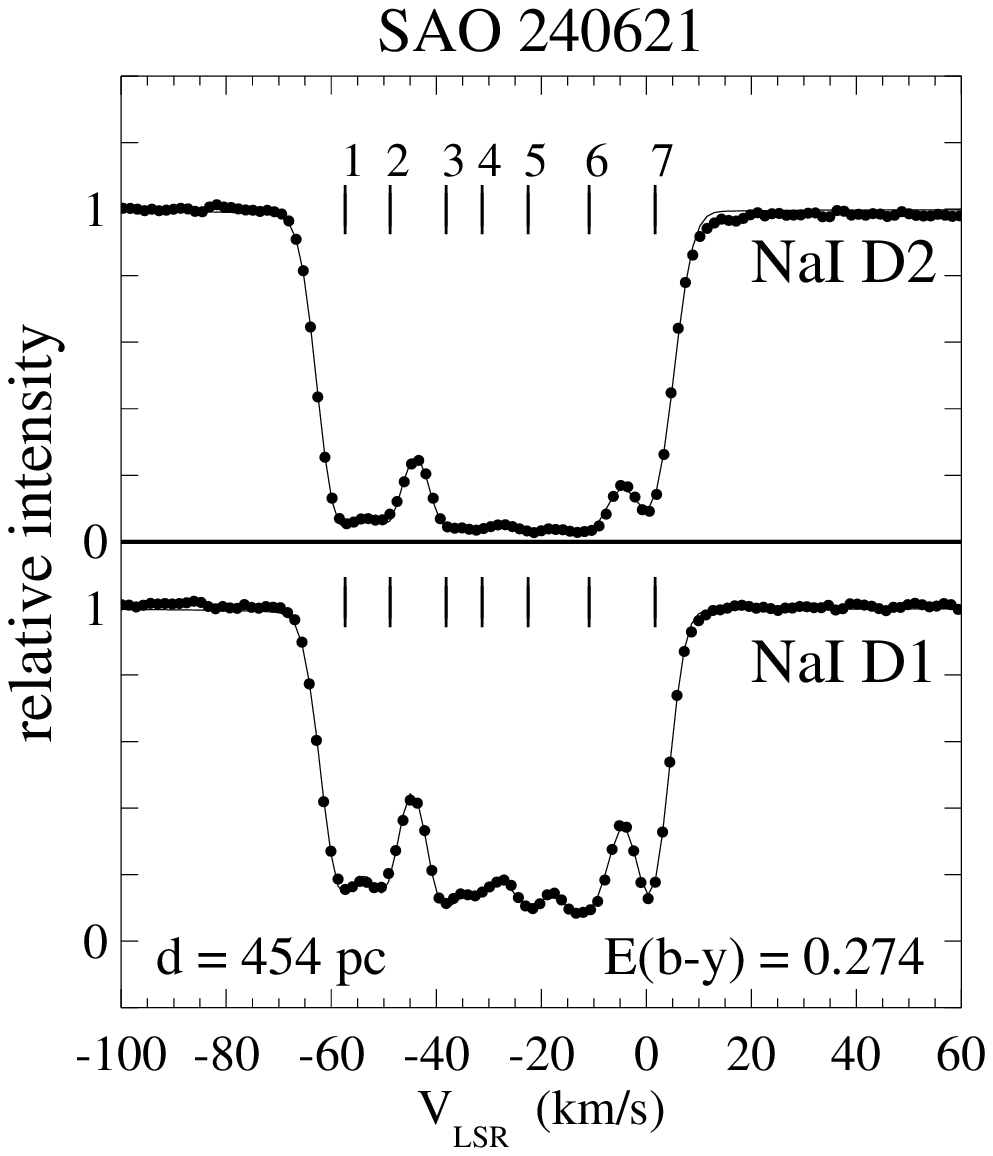} \hfill
\includegraphics[width=5.6cm]{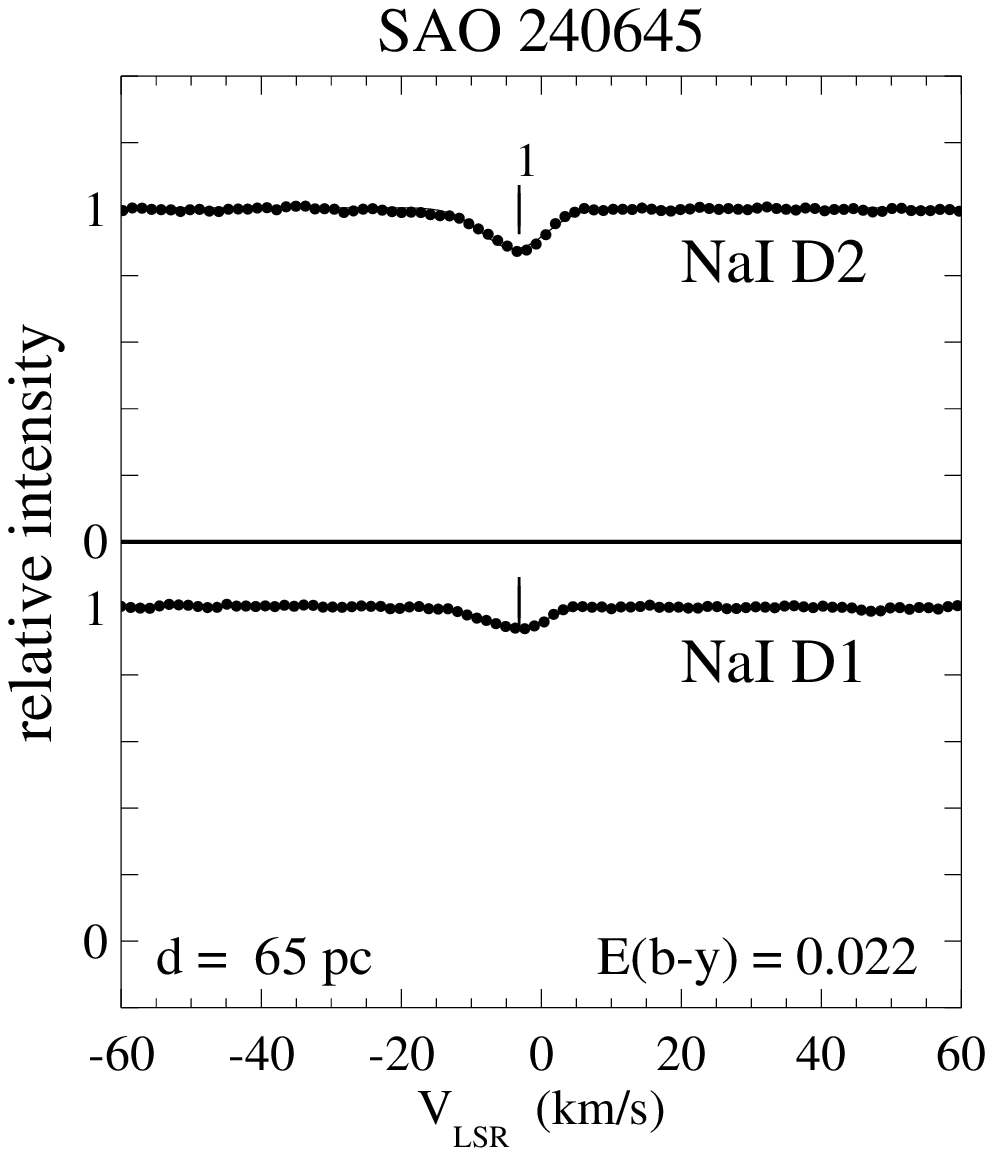} \hfill
\includegraphics[width=5.6cm]{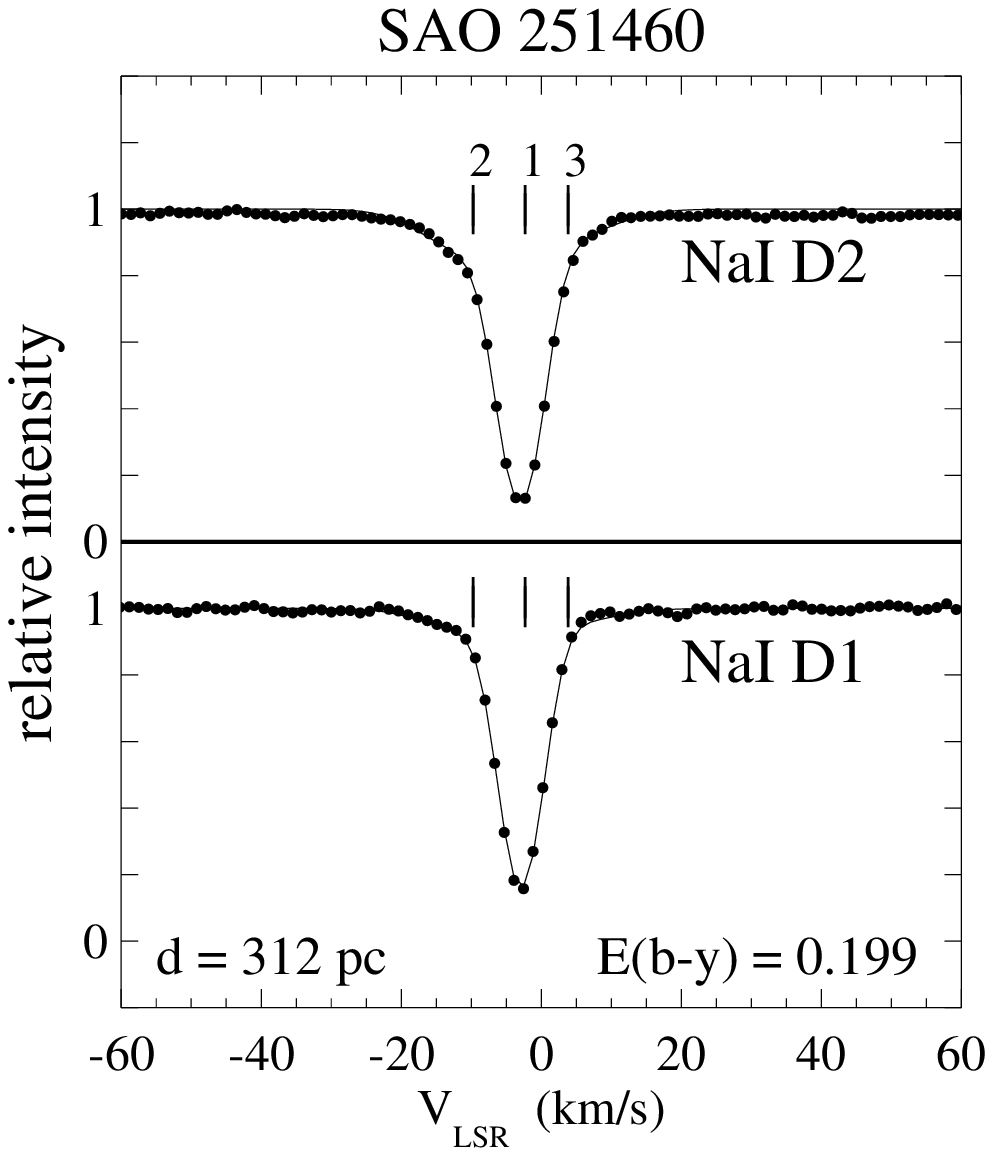} }
\vskip.3cm
\hbox{
\includegraphics[width=5.6cm]{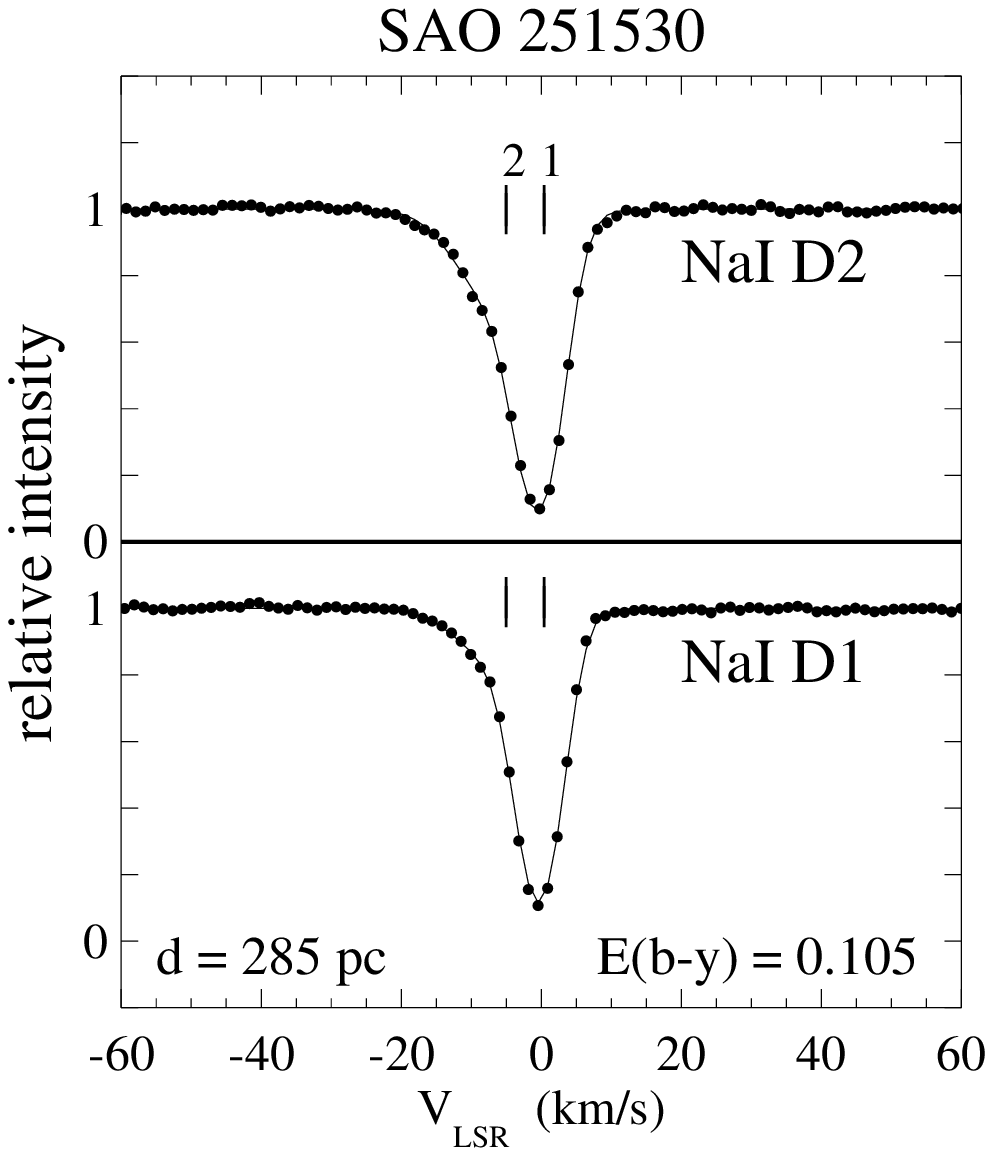} \hfill
\includegraphics[width=5.6cm]{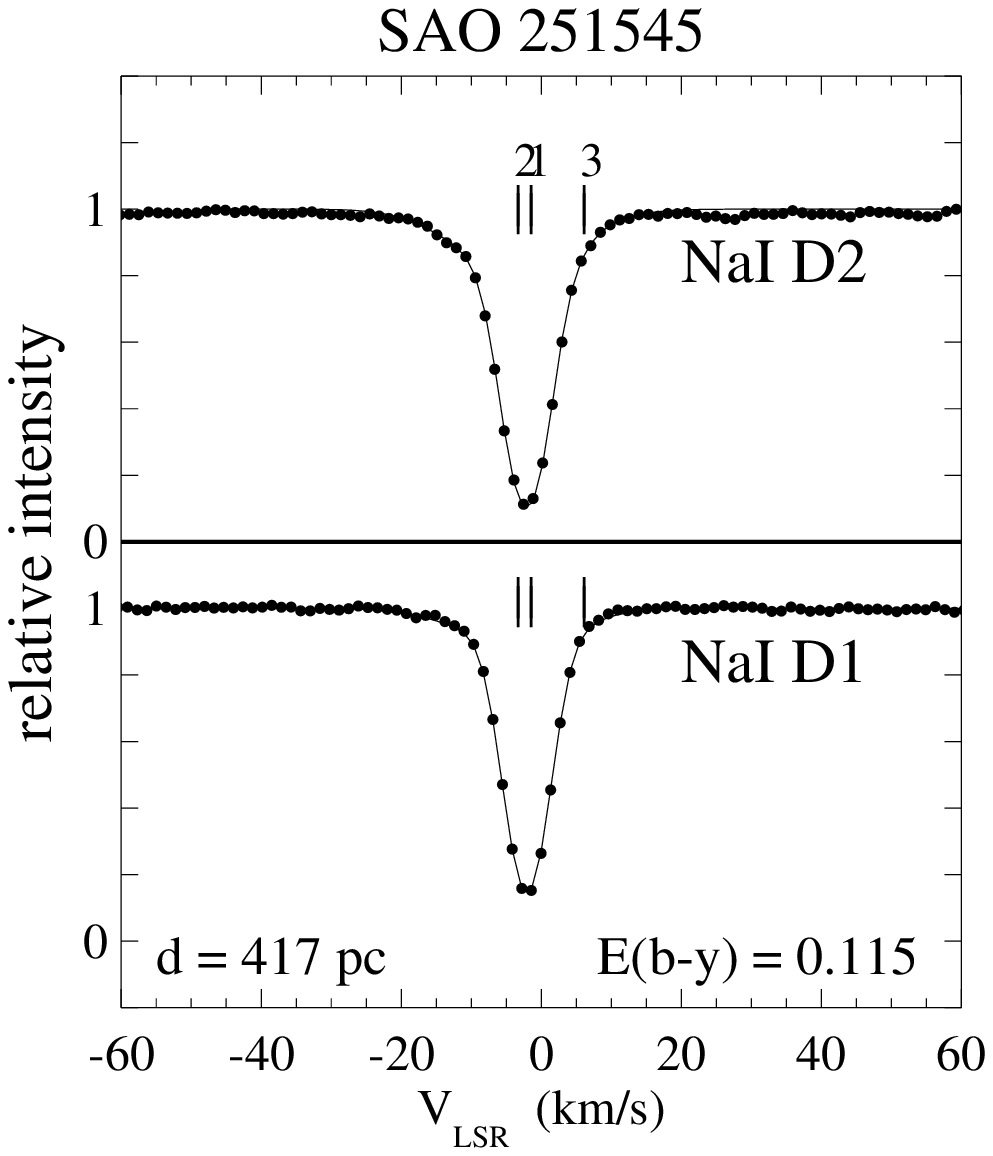}\hfill
\includegraphics[width=5.6cm]{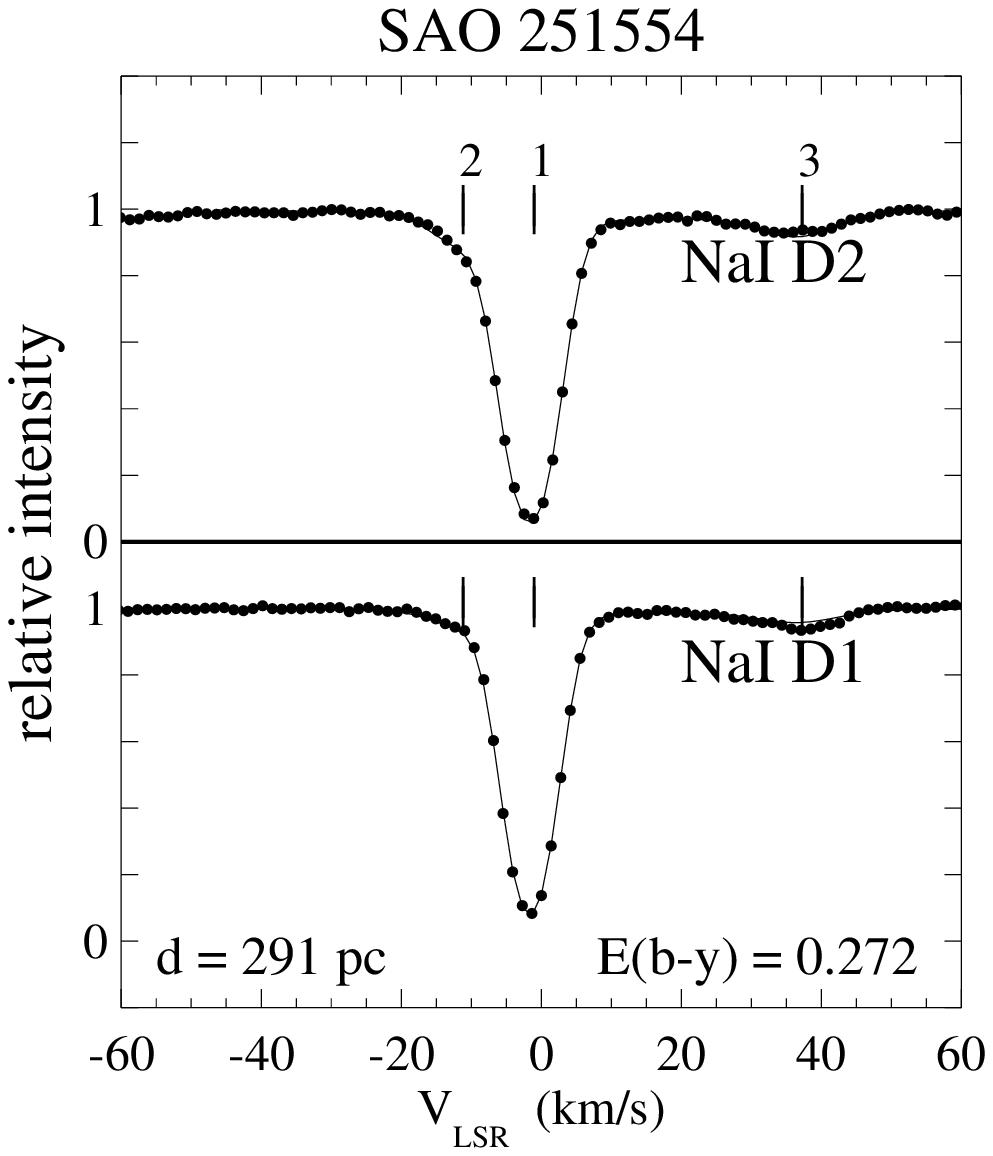} }
\vskip.3cm
\hbox{
\includegraphics[width=5.6cm]{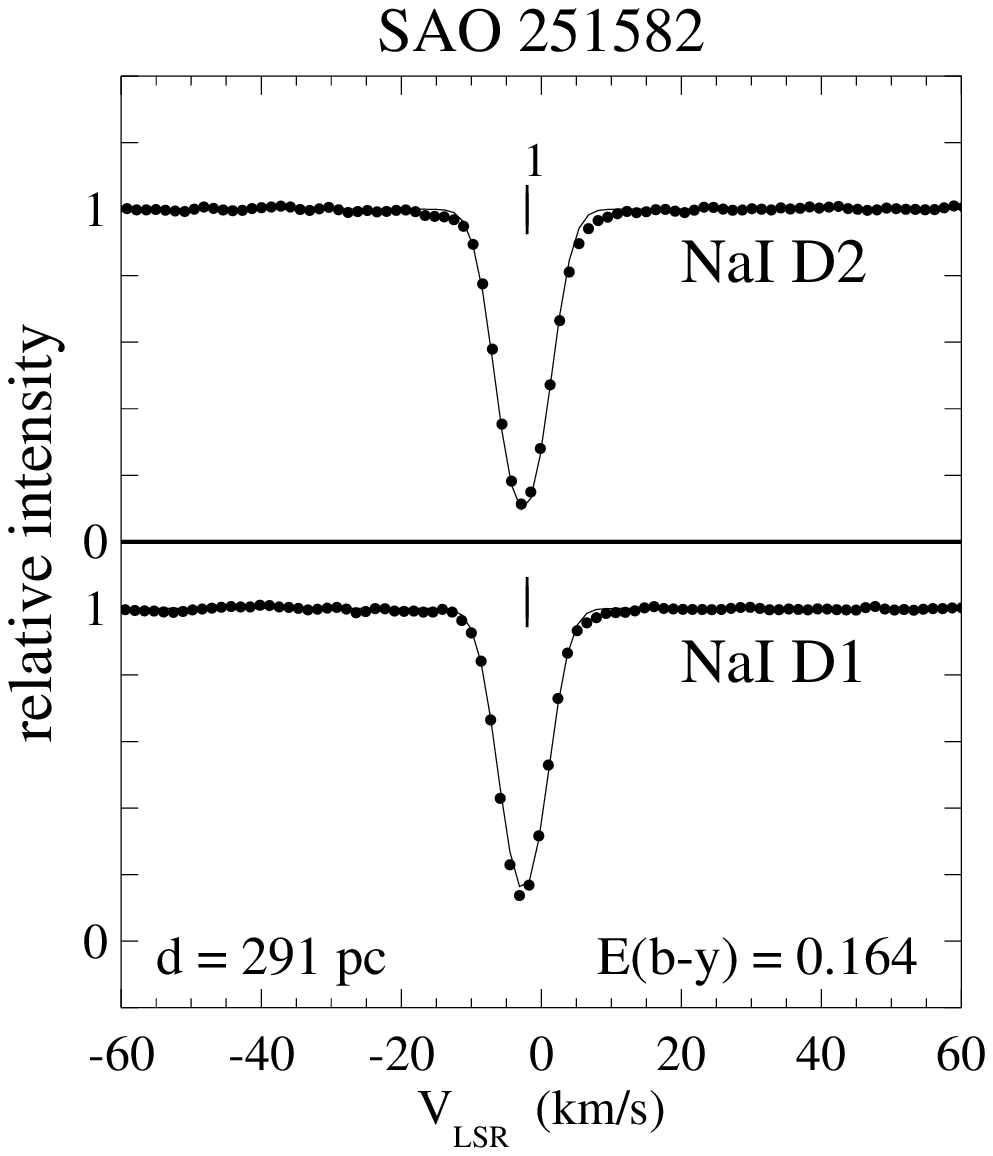} \hfill
\includegraphics[width=5.6cm]{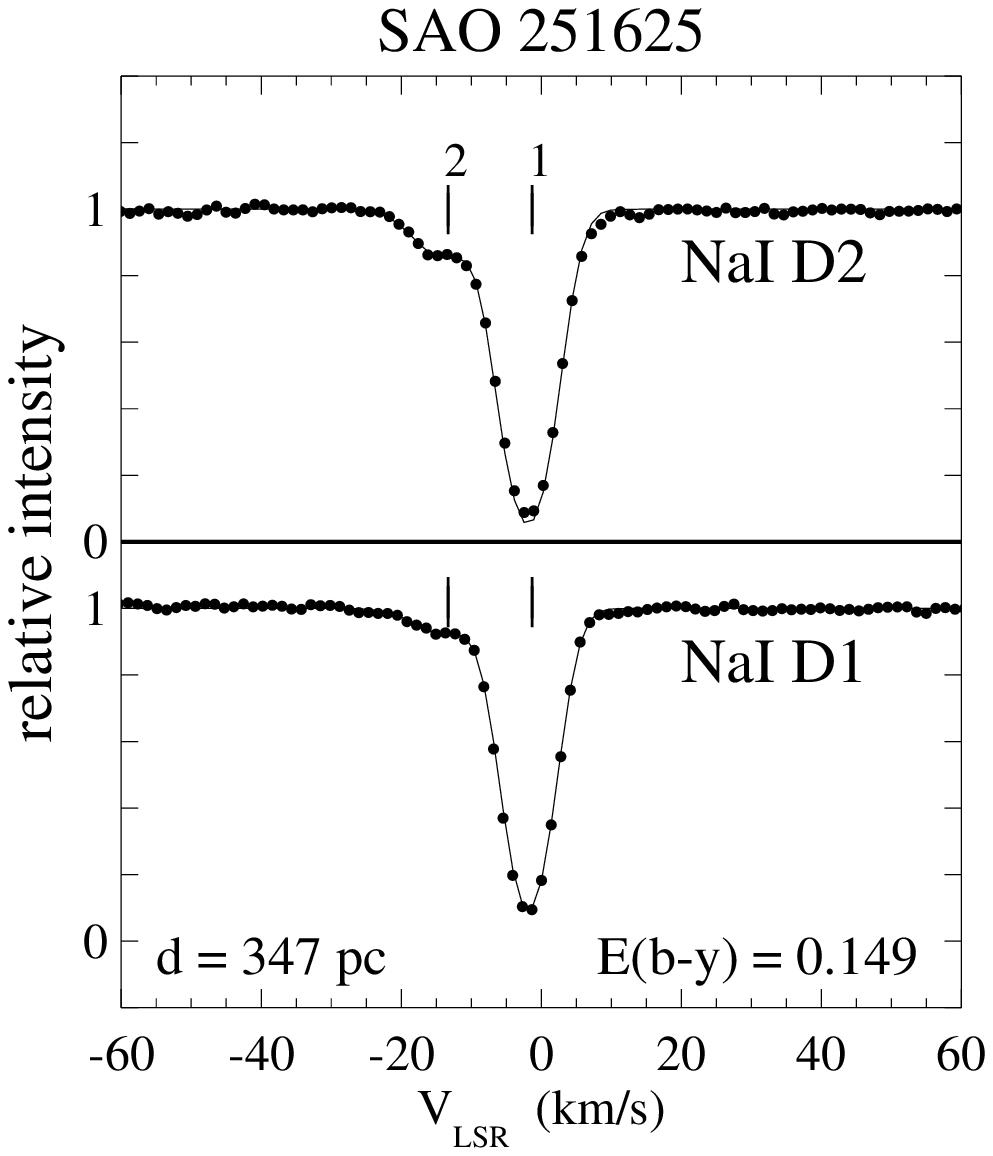} \hfill
\includegraphics[width=5.6cm]{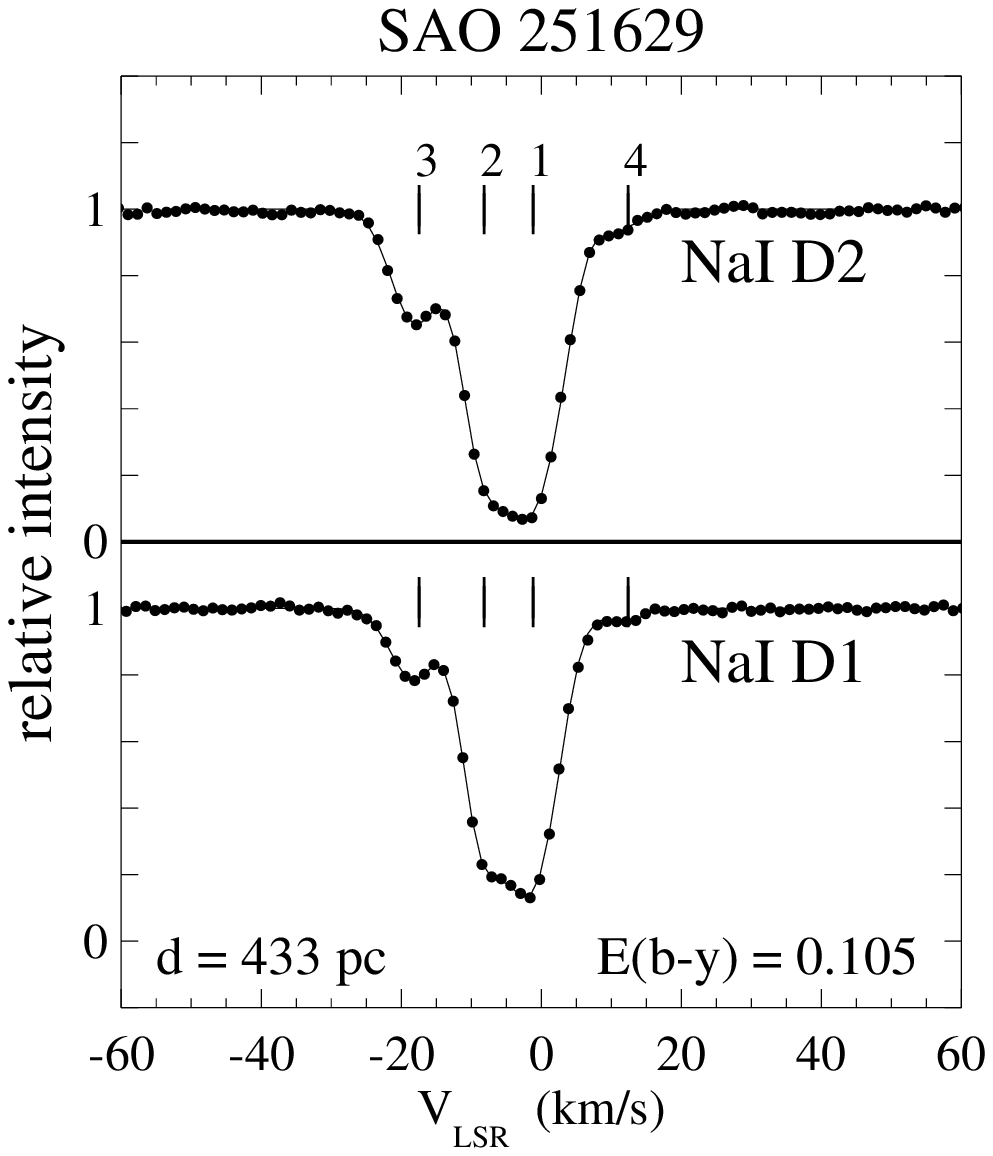} }
\caption{(Continued)}
\end{figure*}
\addtocounter{figure}{-1}
\begin{figure*}
\hbox{
\includegraphics[width=5.6cm]{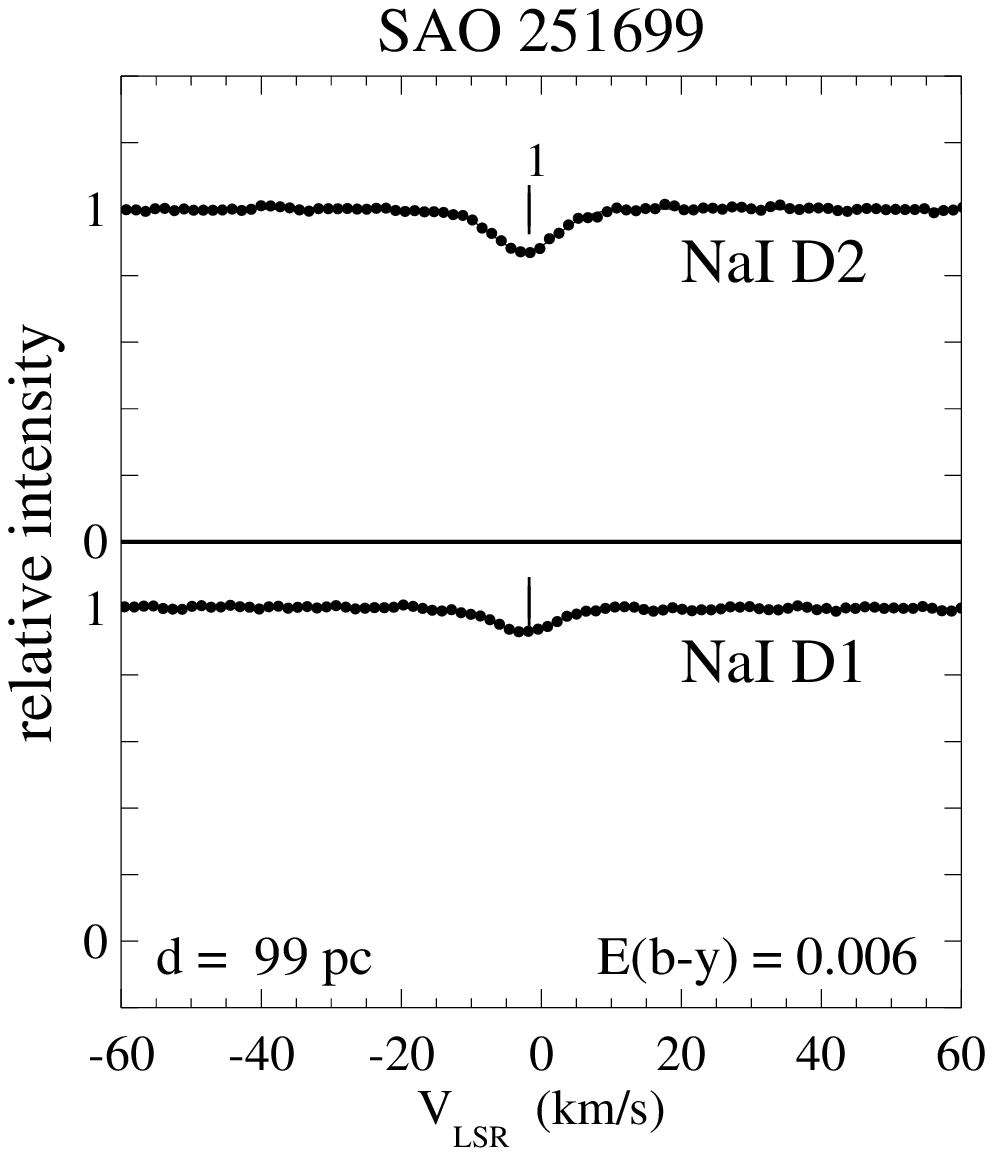} \hfill
\includegraphics[width=5.6cm]{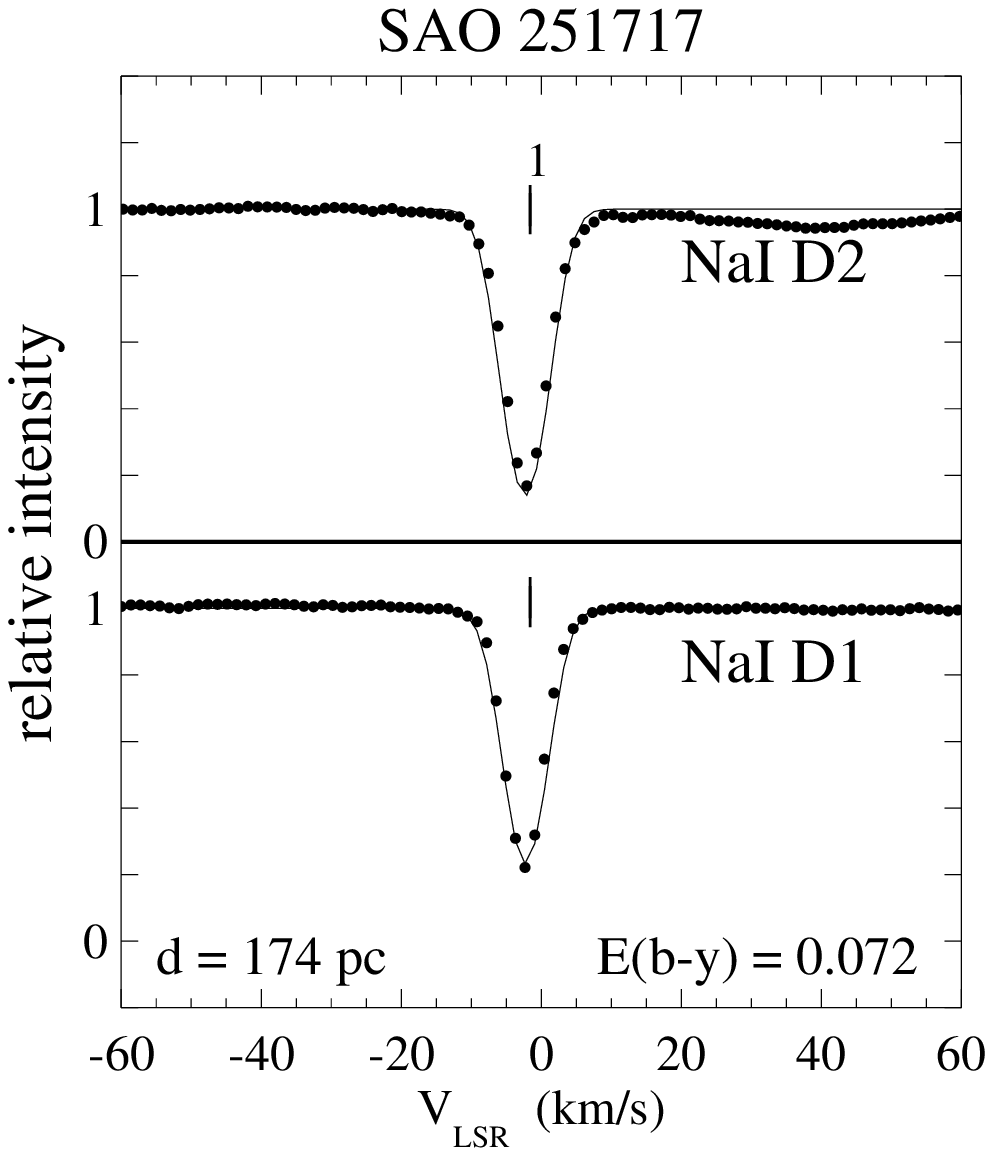} \hfill
\includegraphics[width=5.6cm]{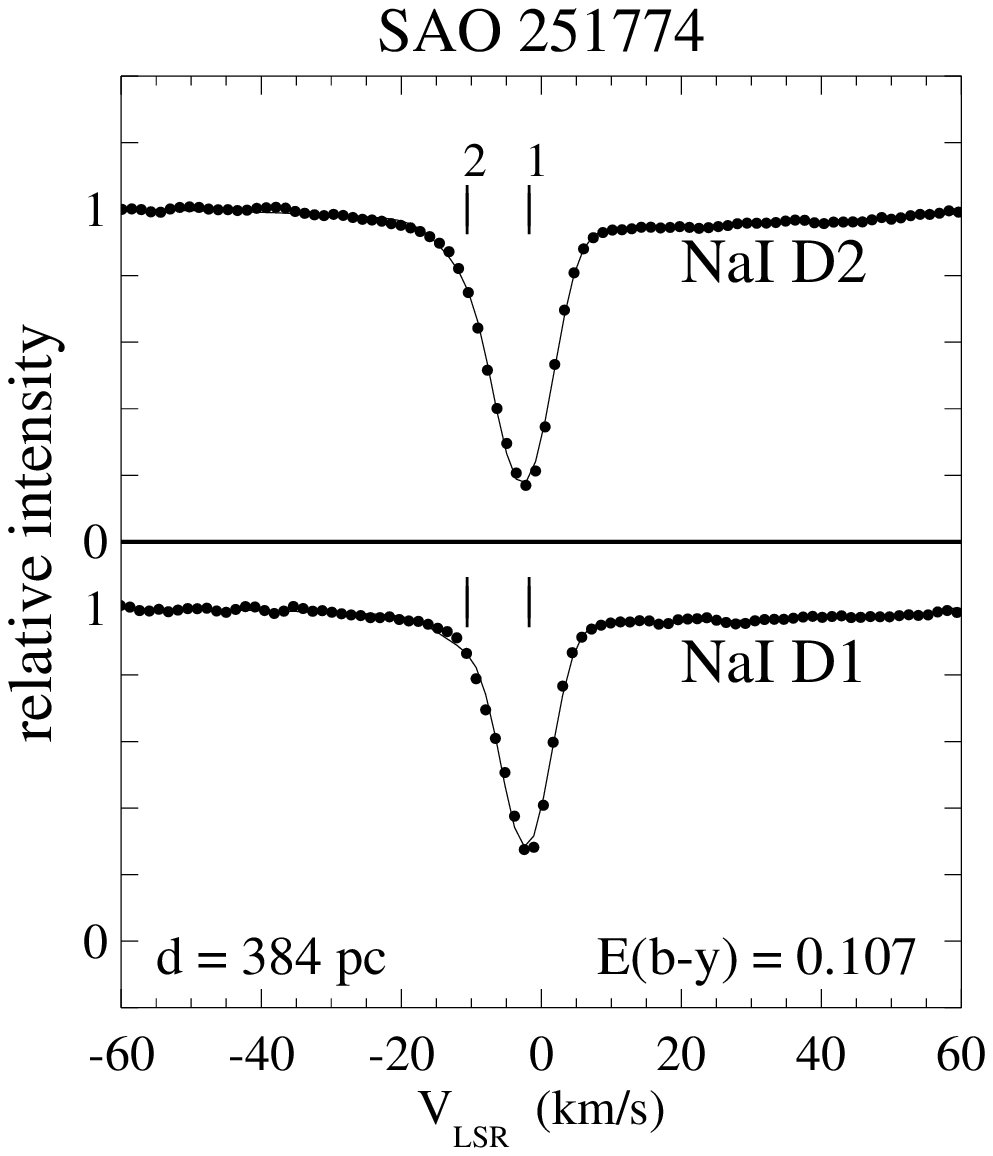} }
\vskip.3cm
\hbox{
\includegraphics[width=5.6cm]{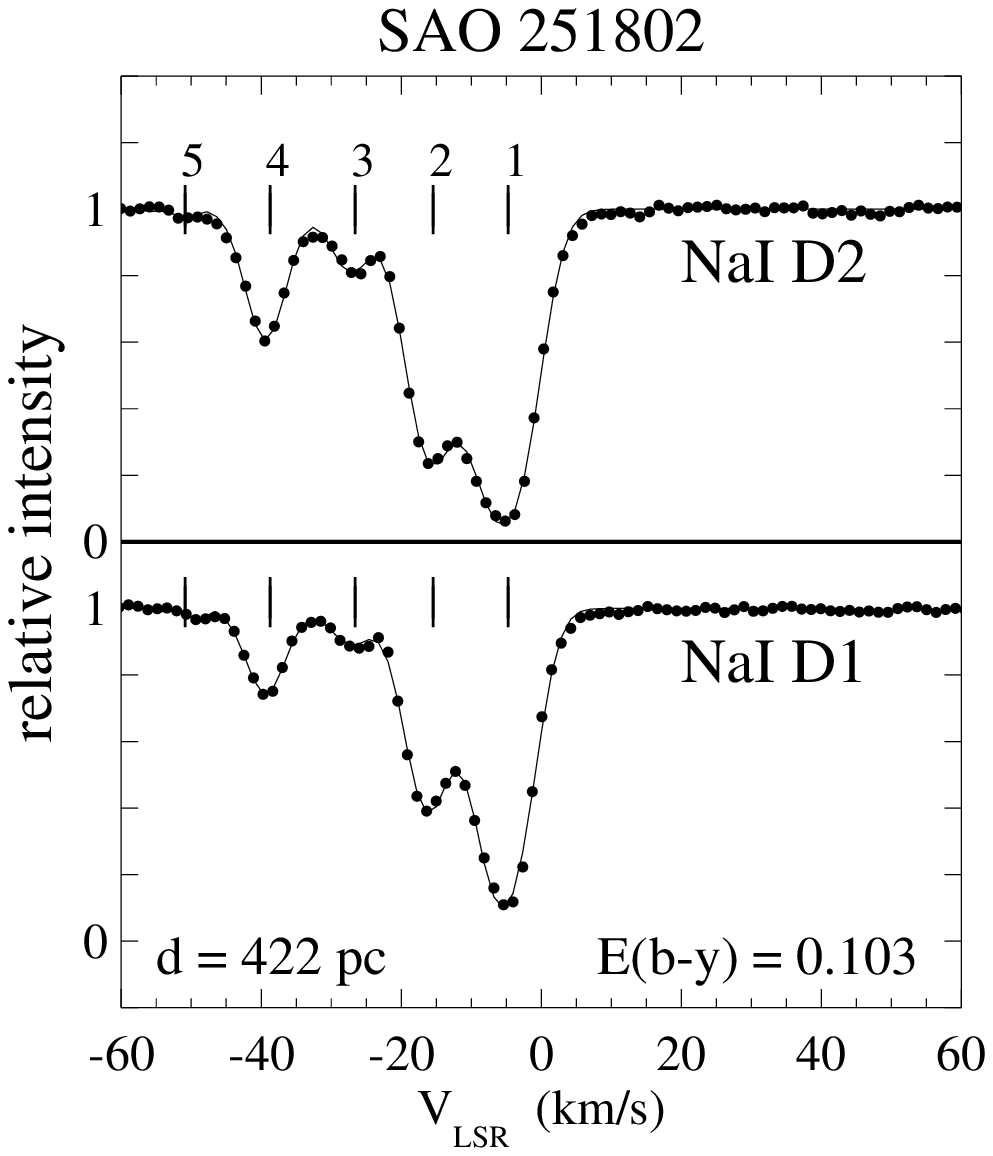} \hfill
\includegraphics[width=5.6cm]{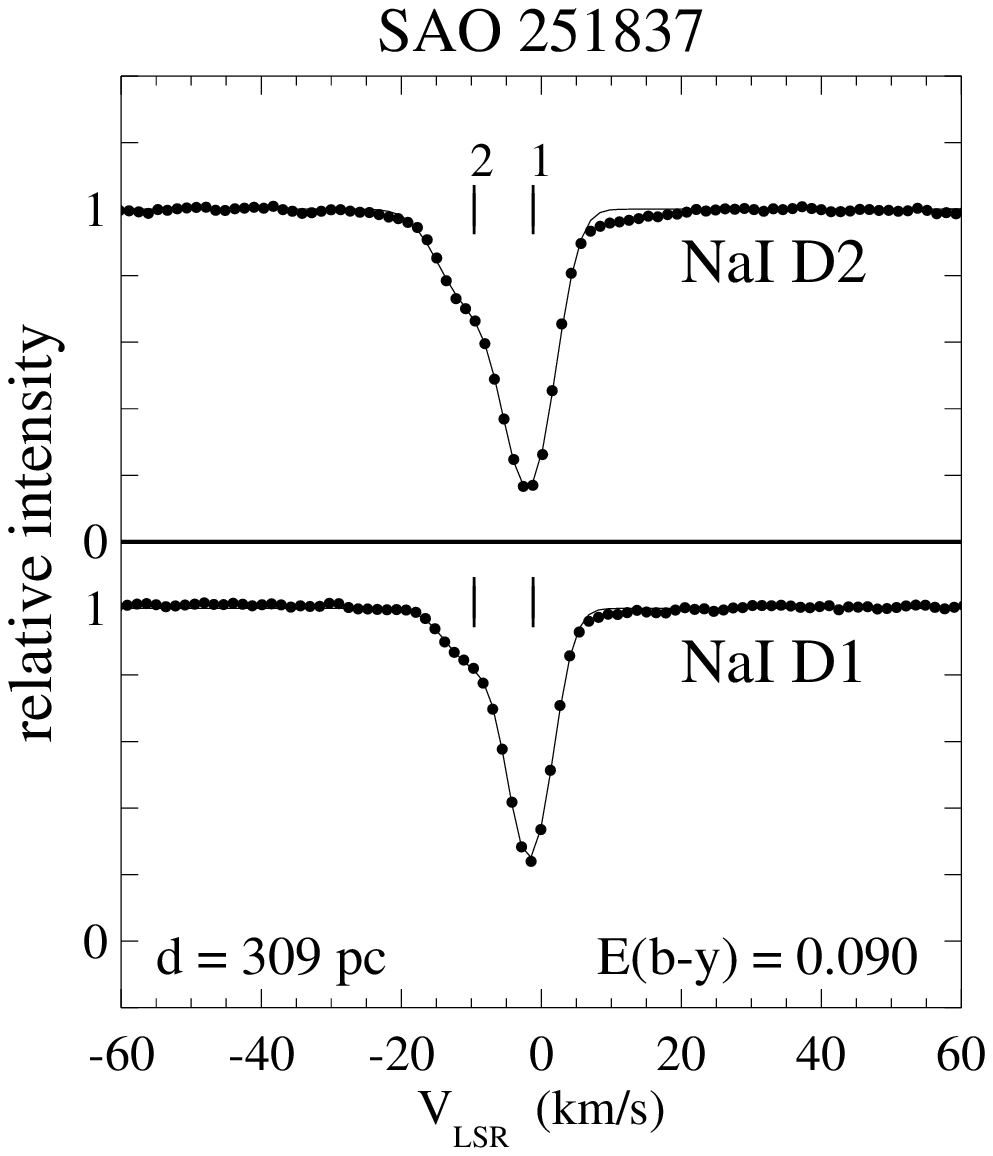} \hfill
\includegraphics[width=5.6cm]{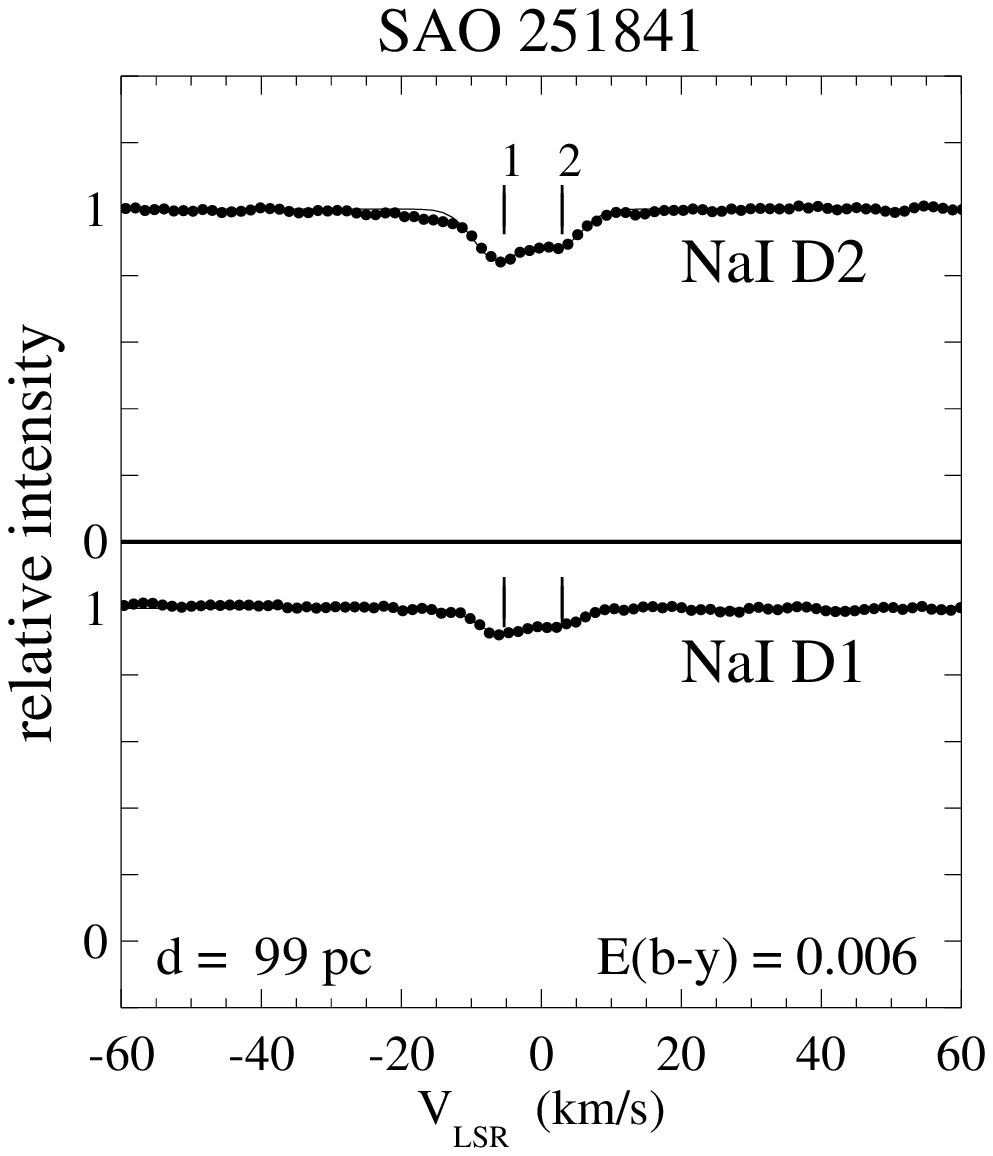} }
\vskip.3cm
\hbox{
\includegraphics[width=5.6cm]{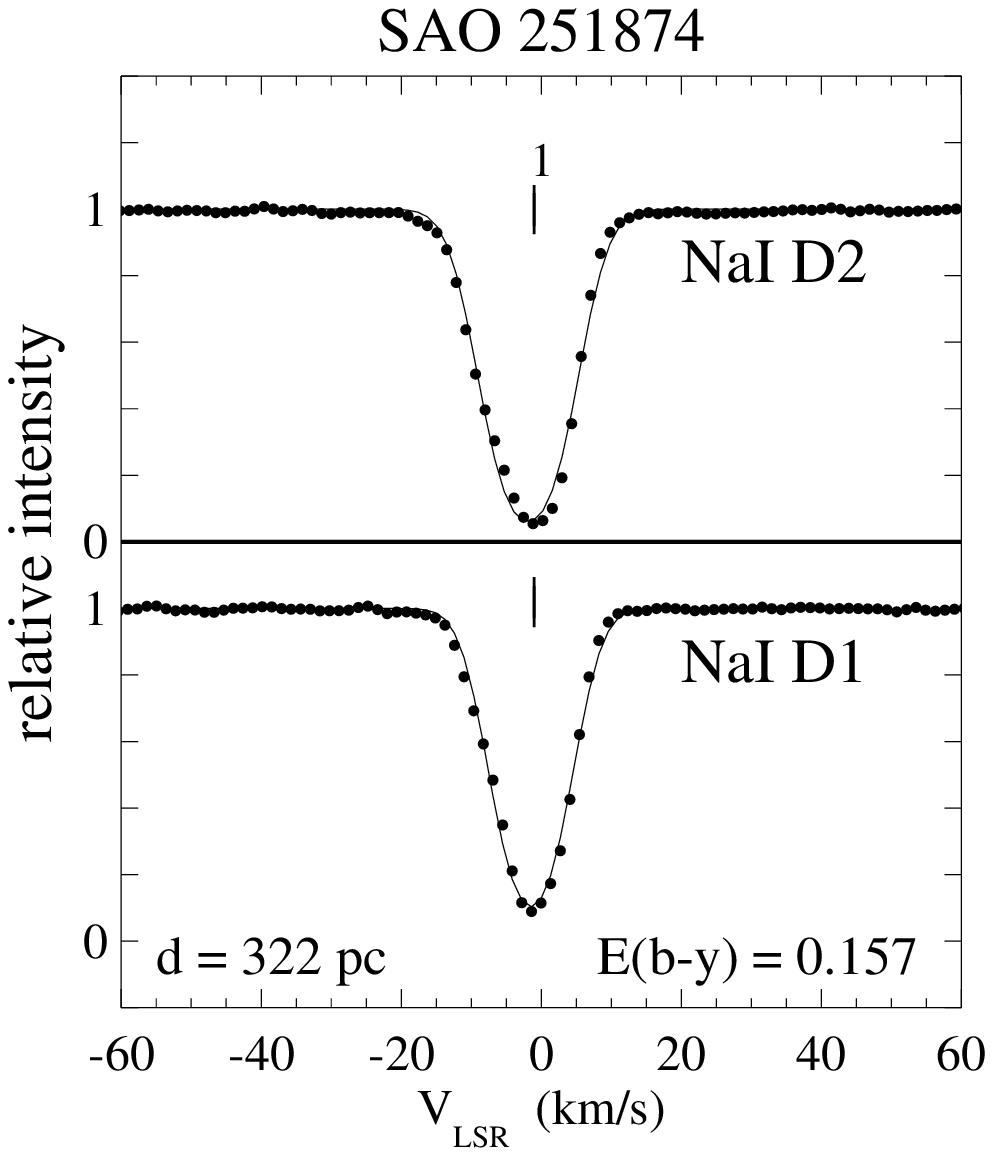} \hfill
\includegraphics[width=5.6cm]{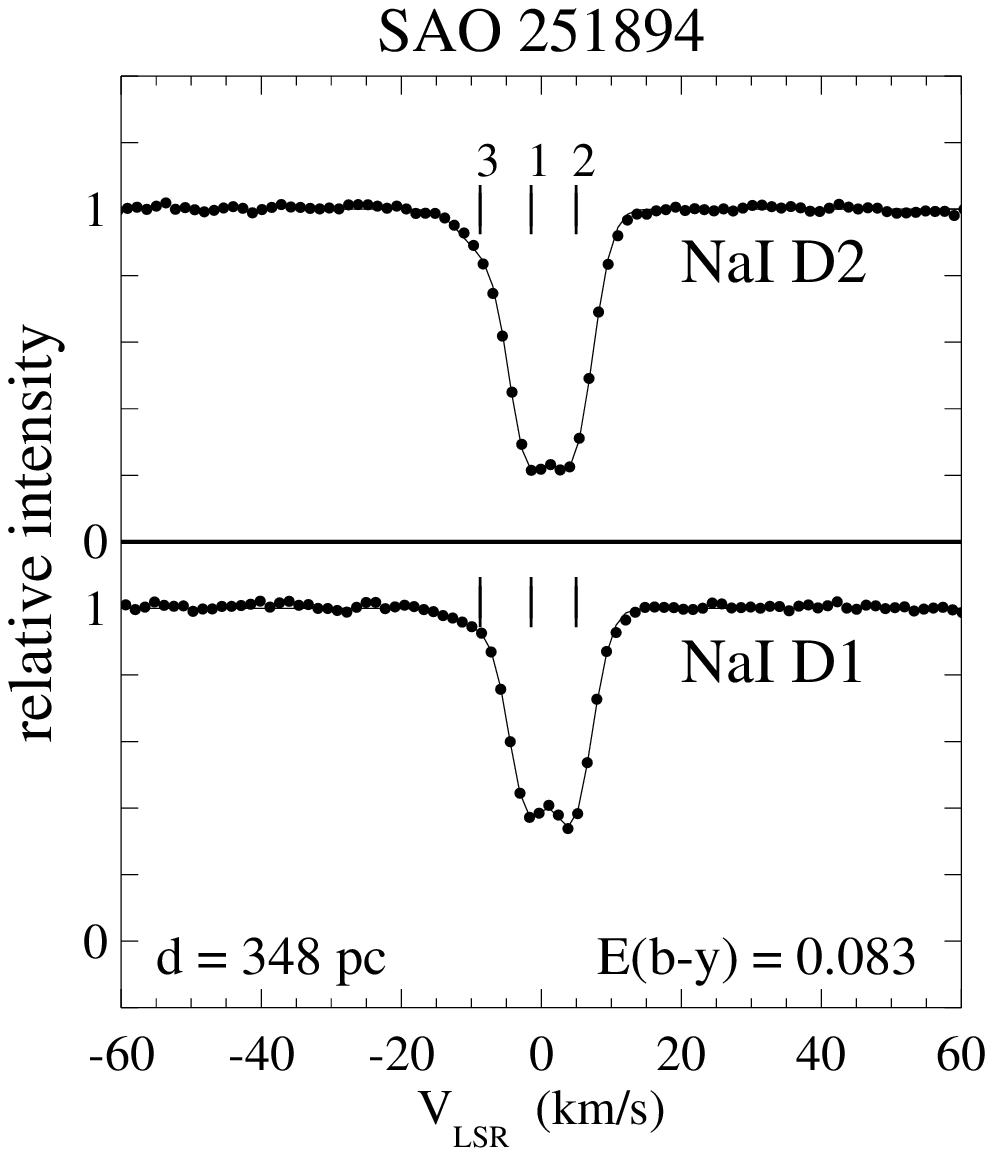} \hfill
\includegraphics[width=5.6cm]{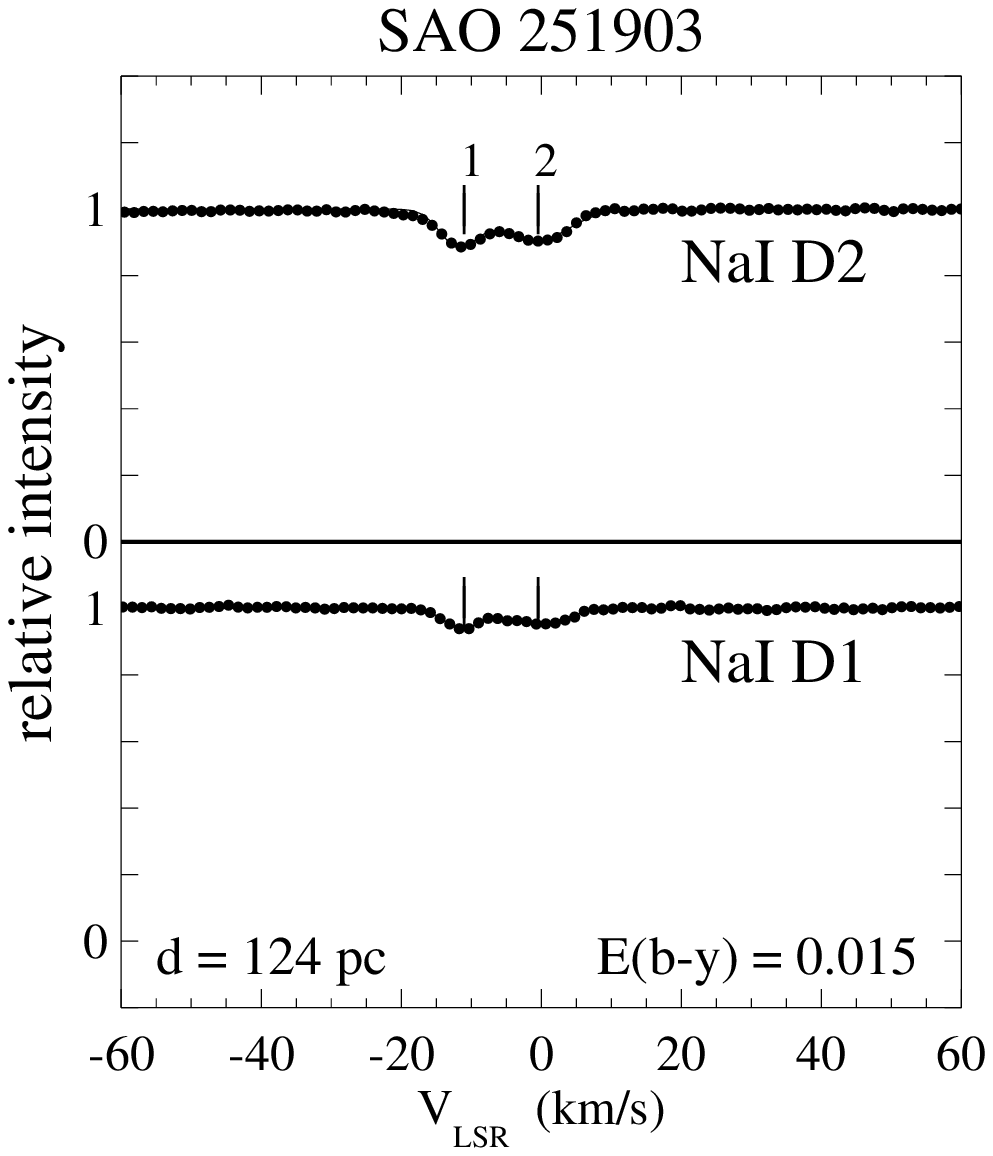} }
\caption{(Continued)}
\end{figure*}

\addtocounter{figure}{-1}
\begin{figure*}
\hbox{
\includegraphics[width=5.6cm]{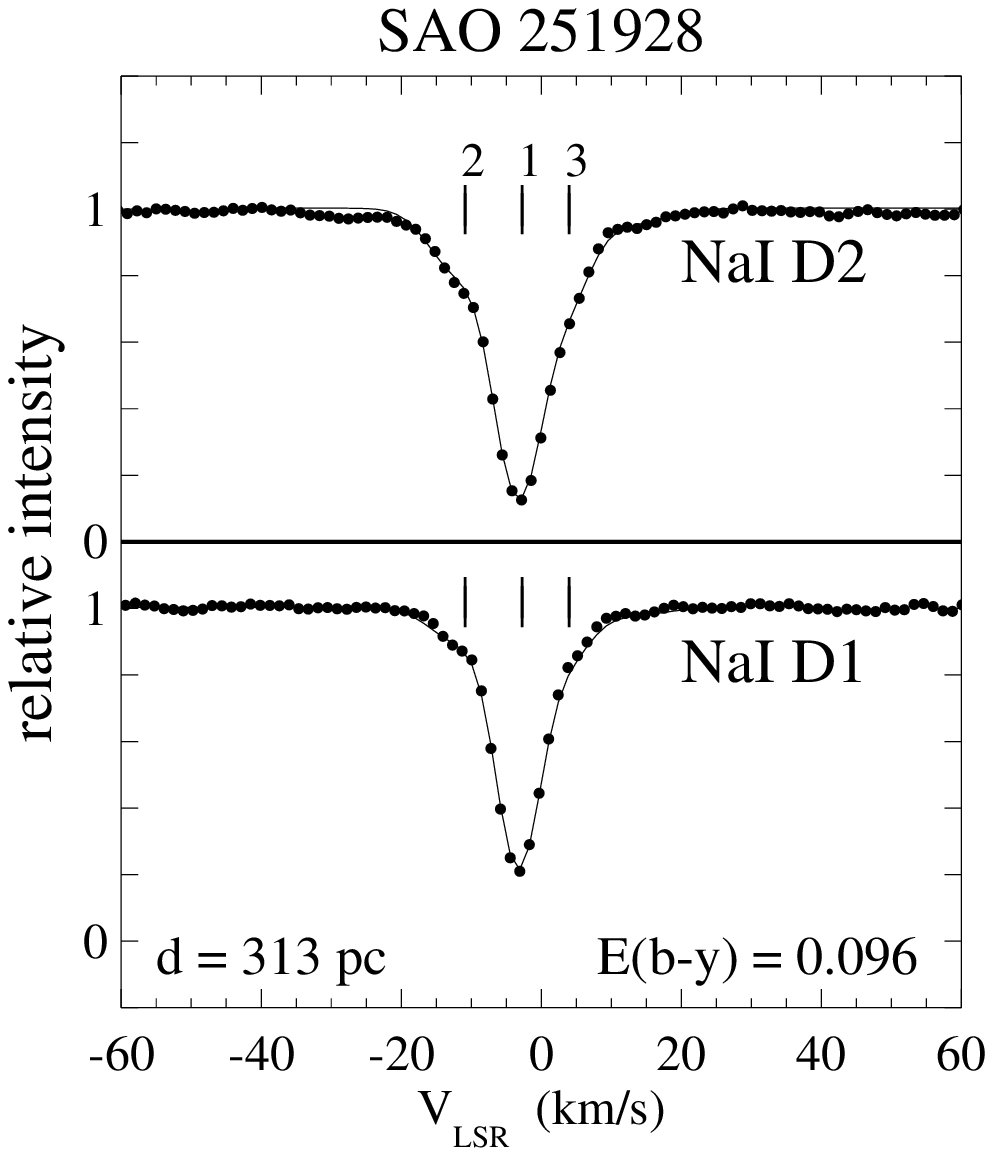} \hfill
\includegraphics[width=5.6cm]{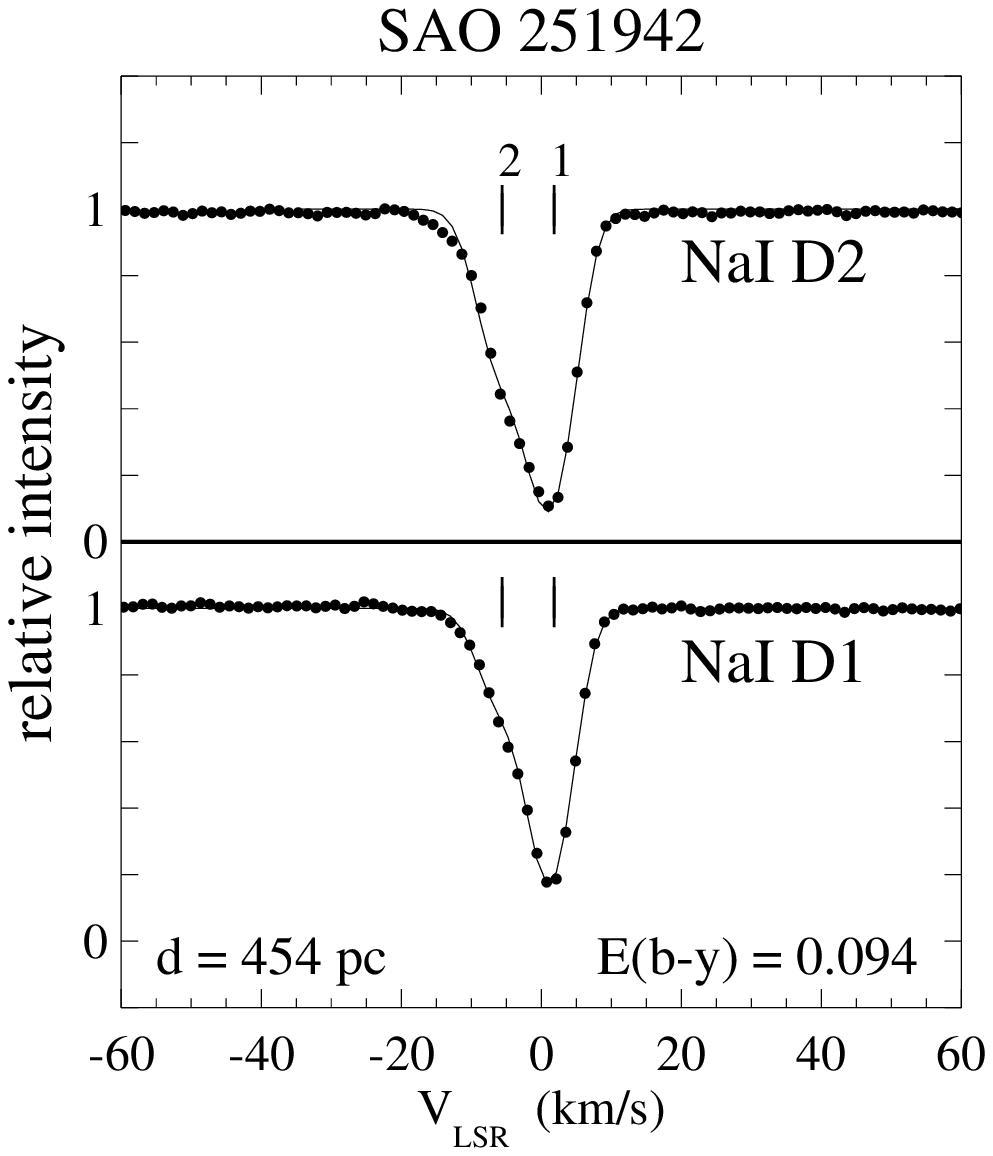} \hfill
\includegraphics[width=5.6cm]{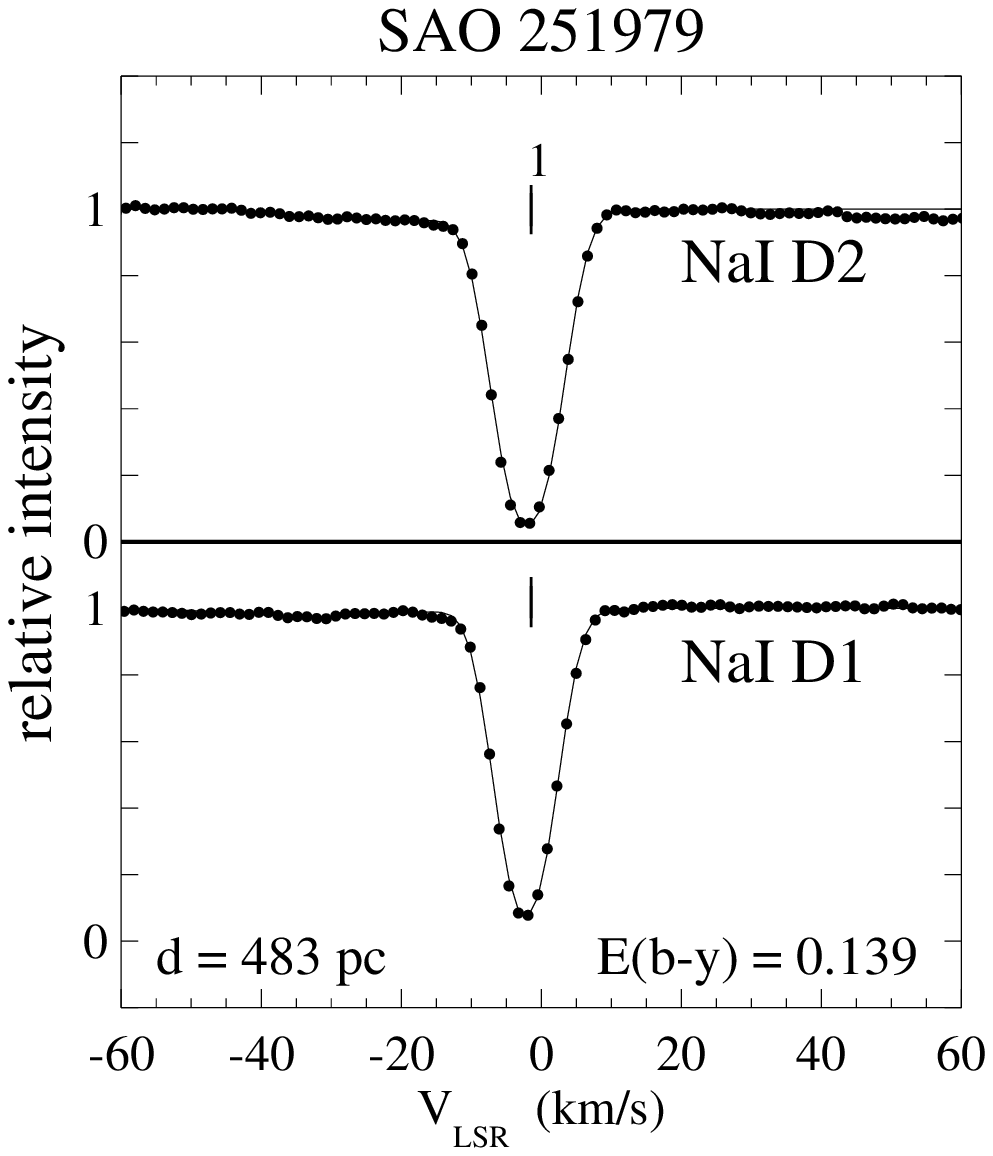} }
\vskip.3cm
\hbox{
\includegraphics[width=5.6cm]{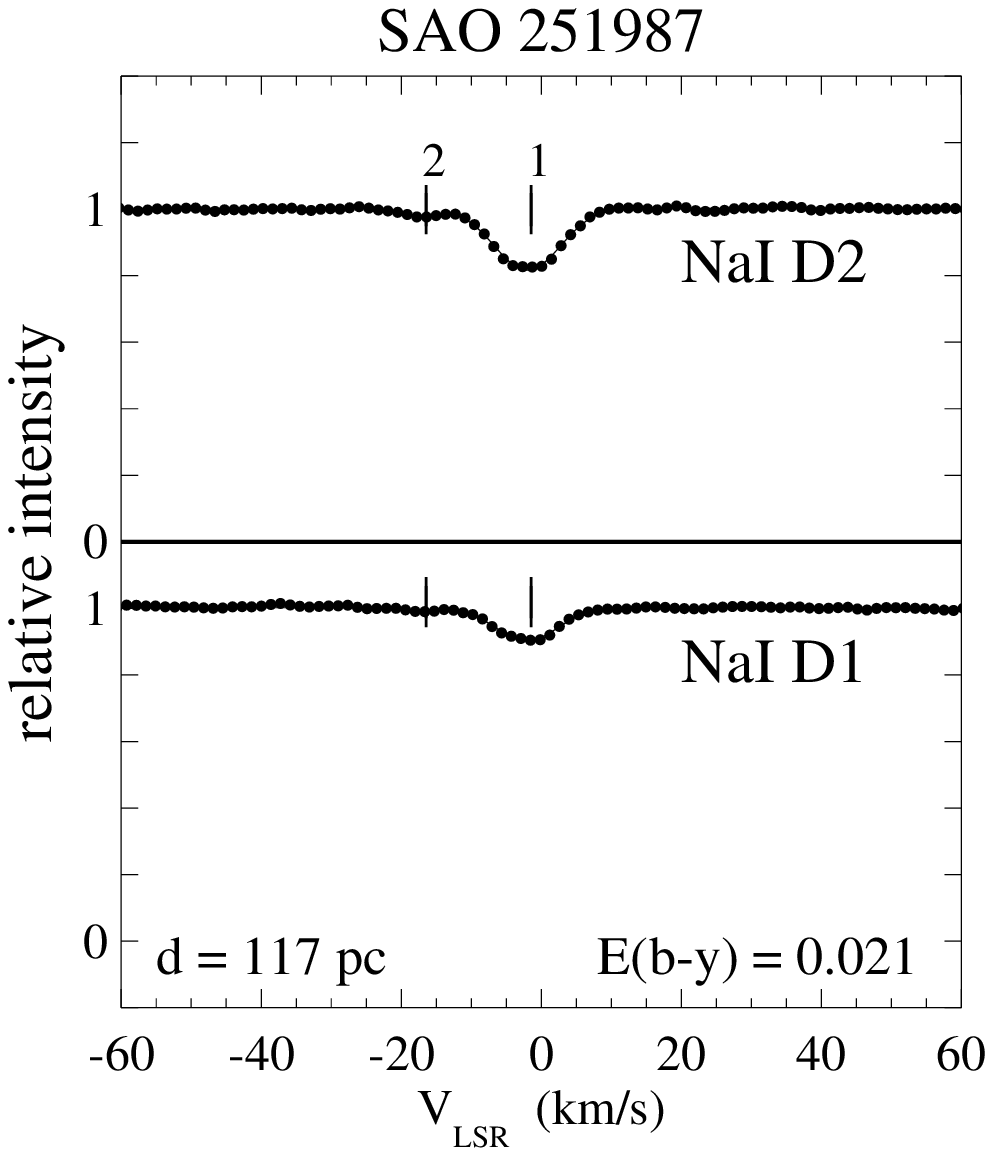} \hfill
\includegraphics[width=5.6cm]{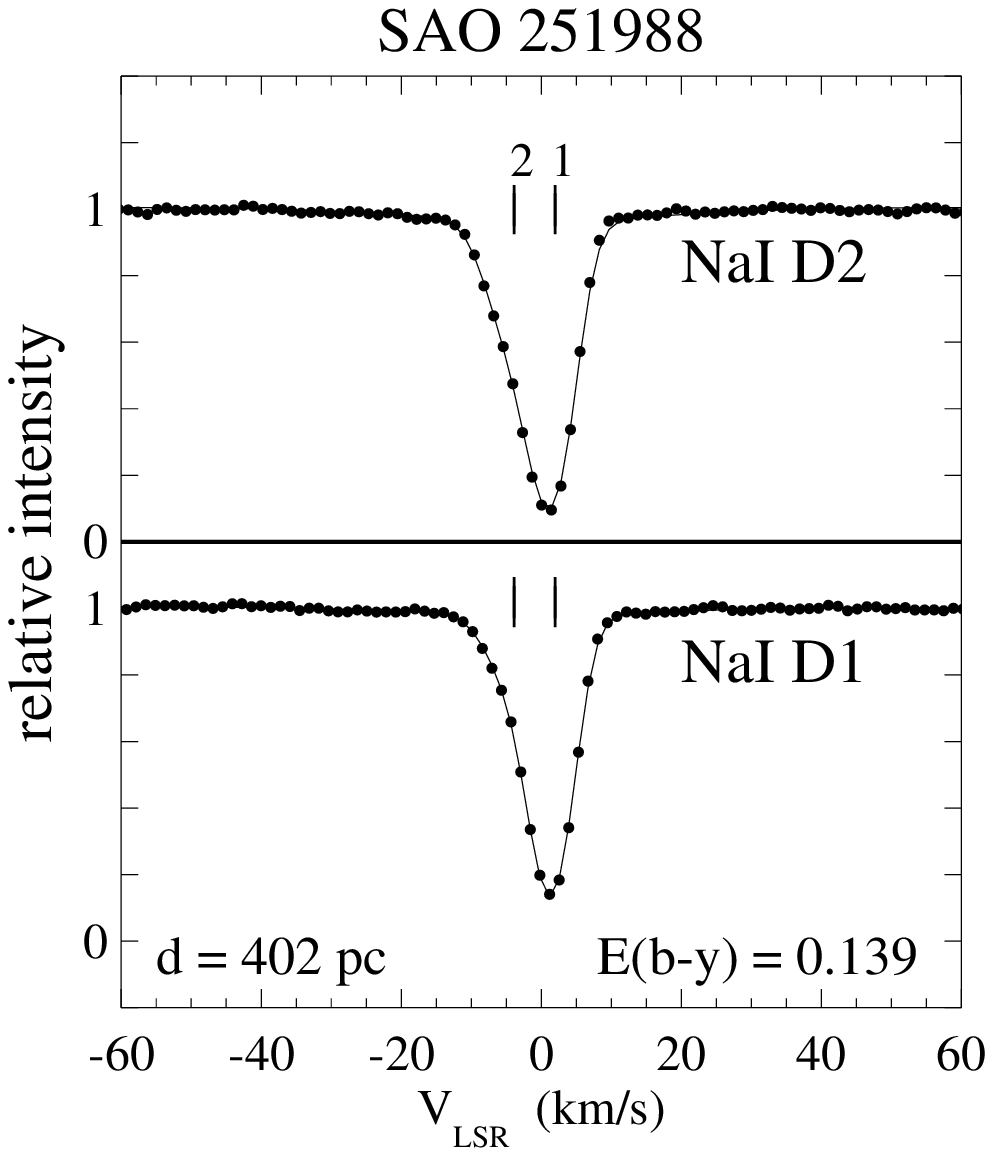} \hfill
\includegraphics[width=5.6cm]{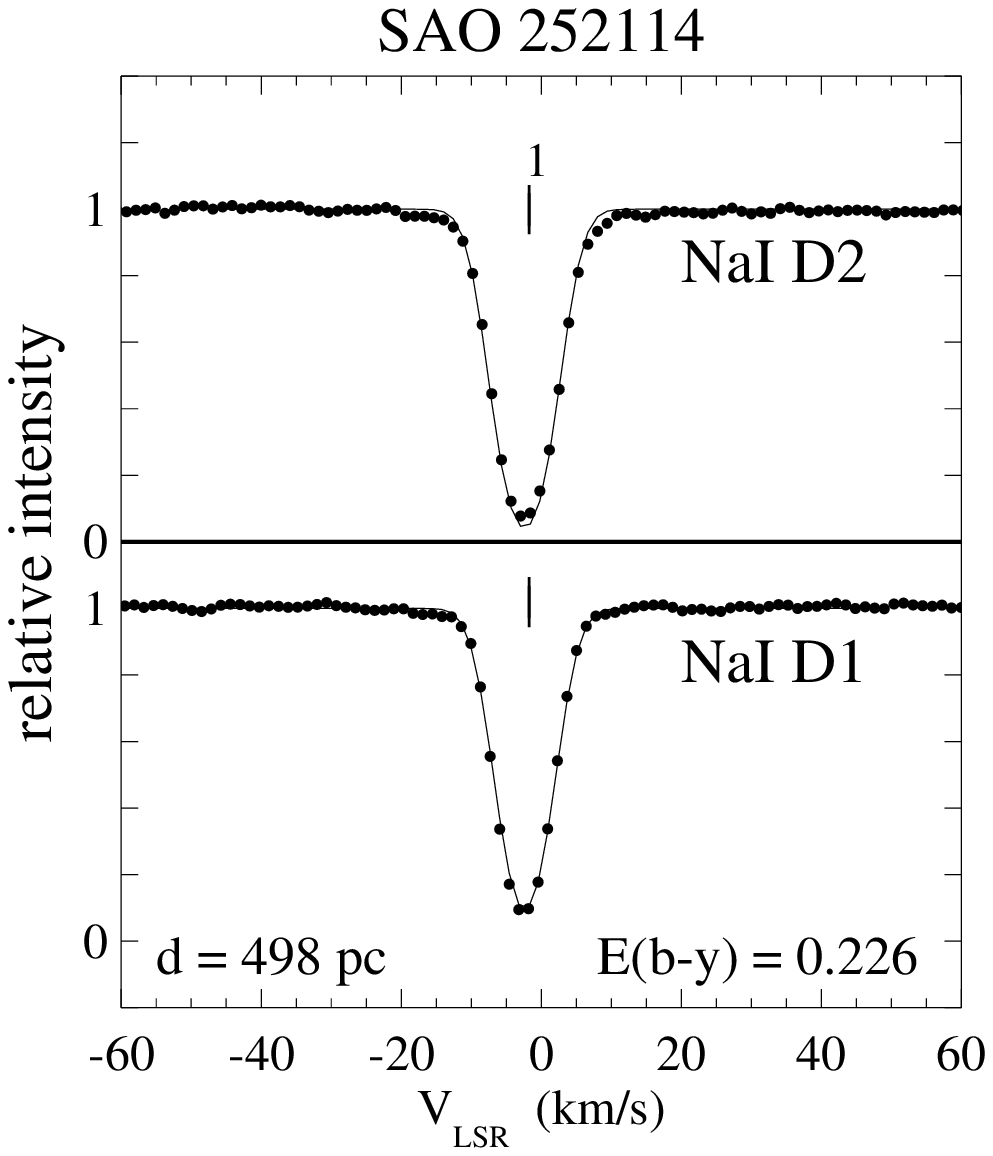} }
\vskip.3cm
\hbox{
\includegraphics[width=5.6cm]{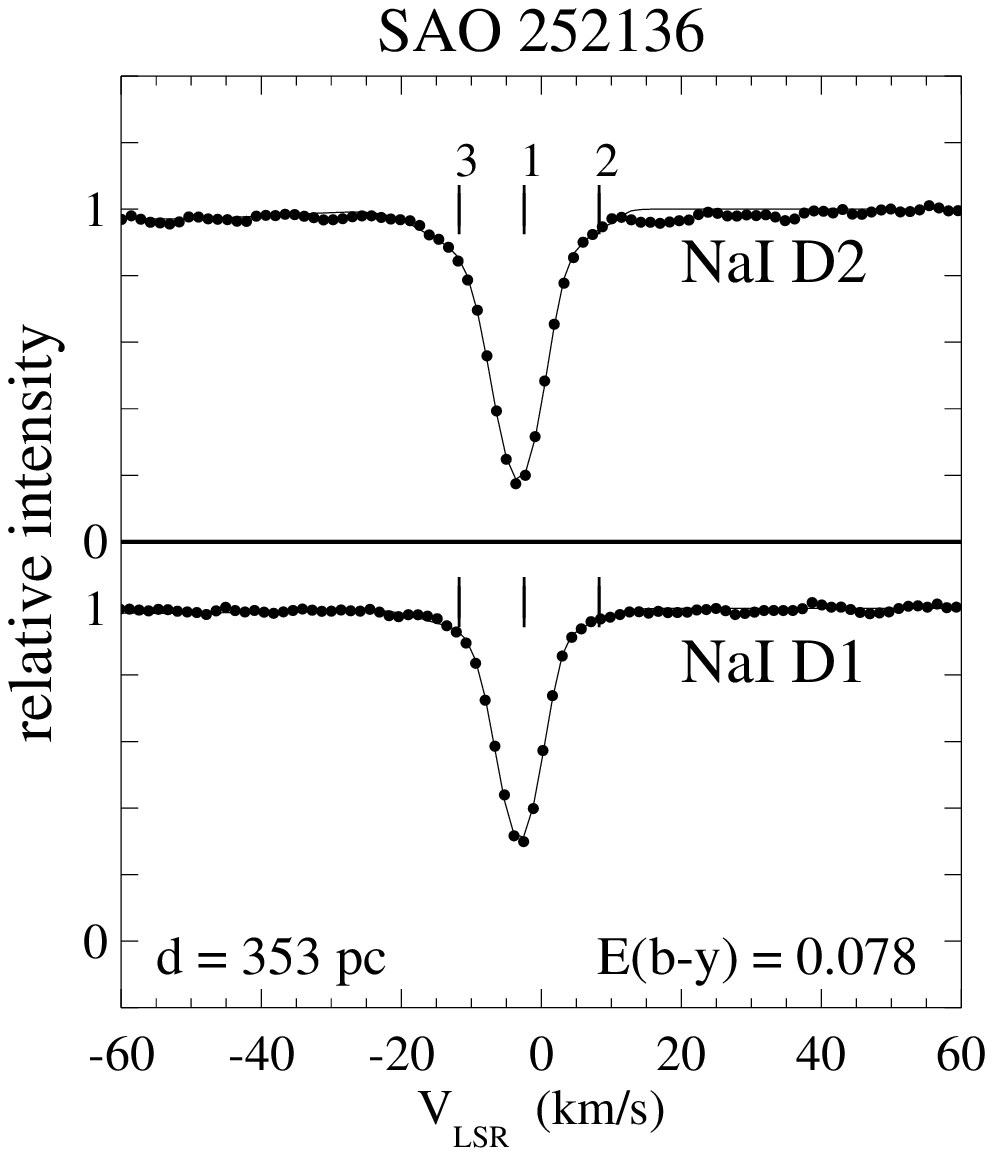} \hfill
\includegraphics[width=5.6cm]{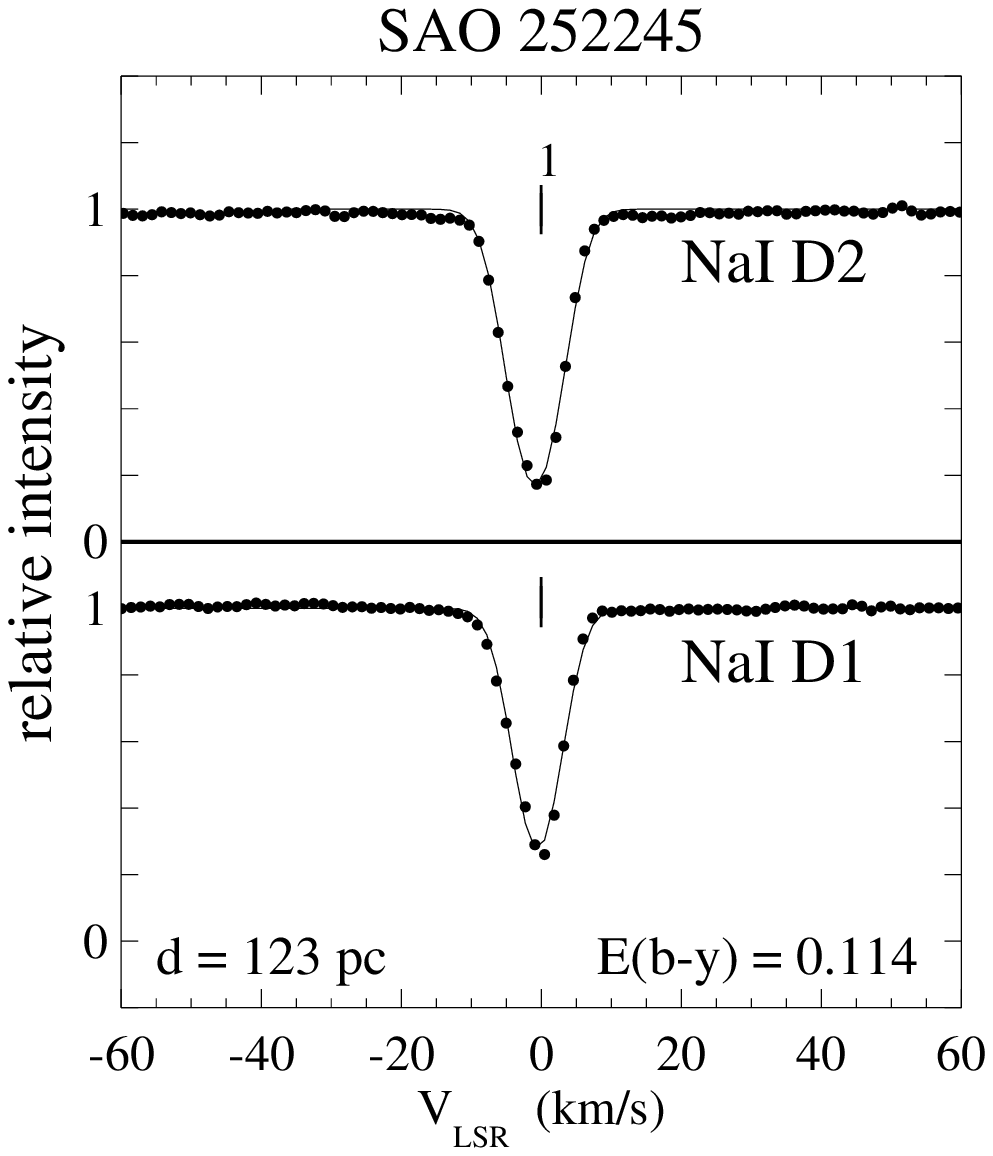} \hfill
\includegraphics[width=5.6cm]{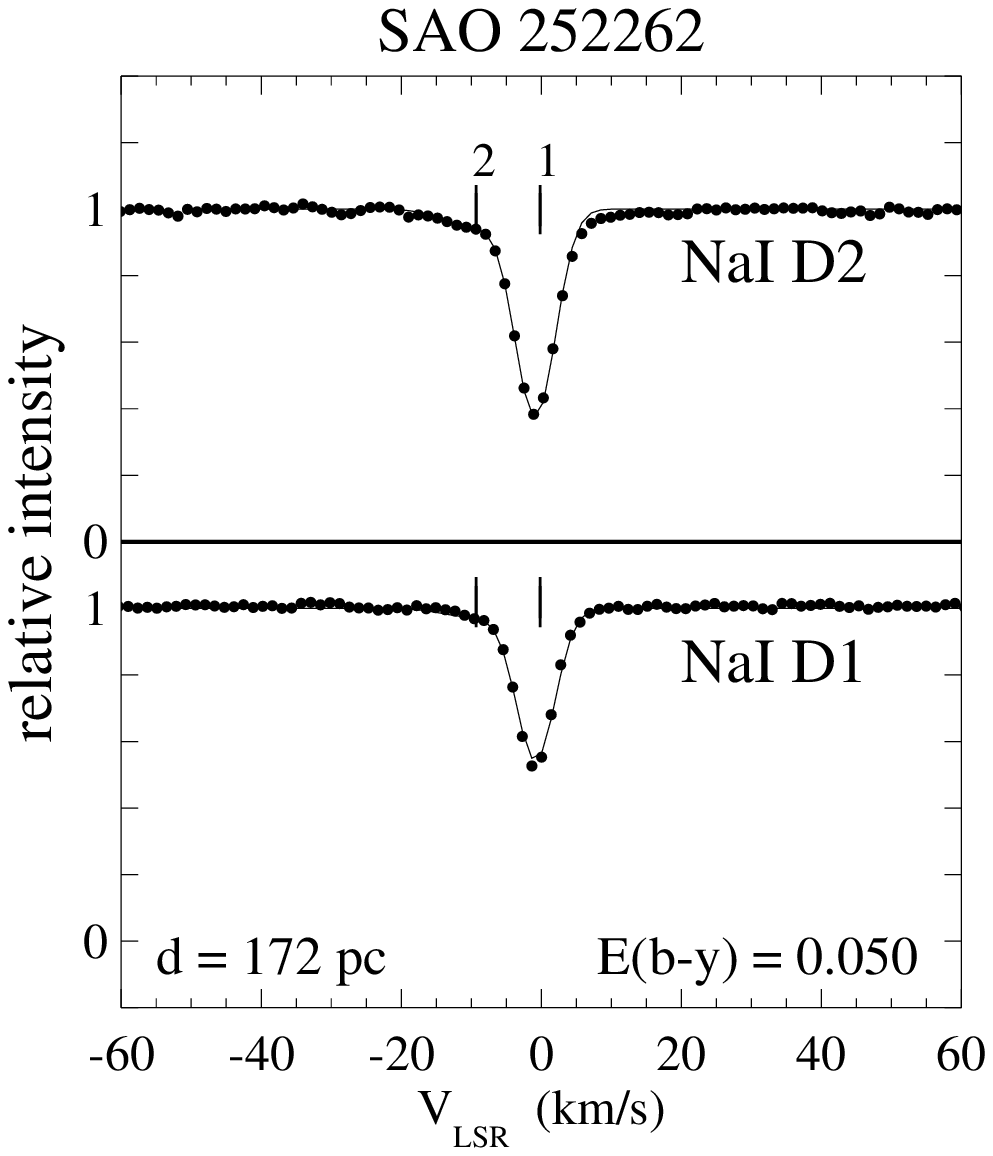}  }
\caption{(Continued)}
\end{figure*}

\addtocounter{figure}{-1}
\begin{figure*}
\hbox{
\includegraphics[width=5.6cm]{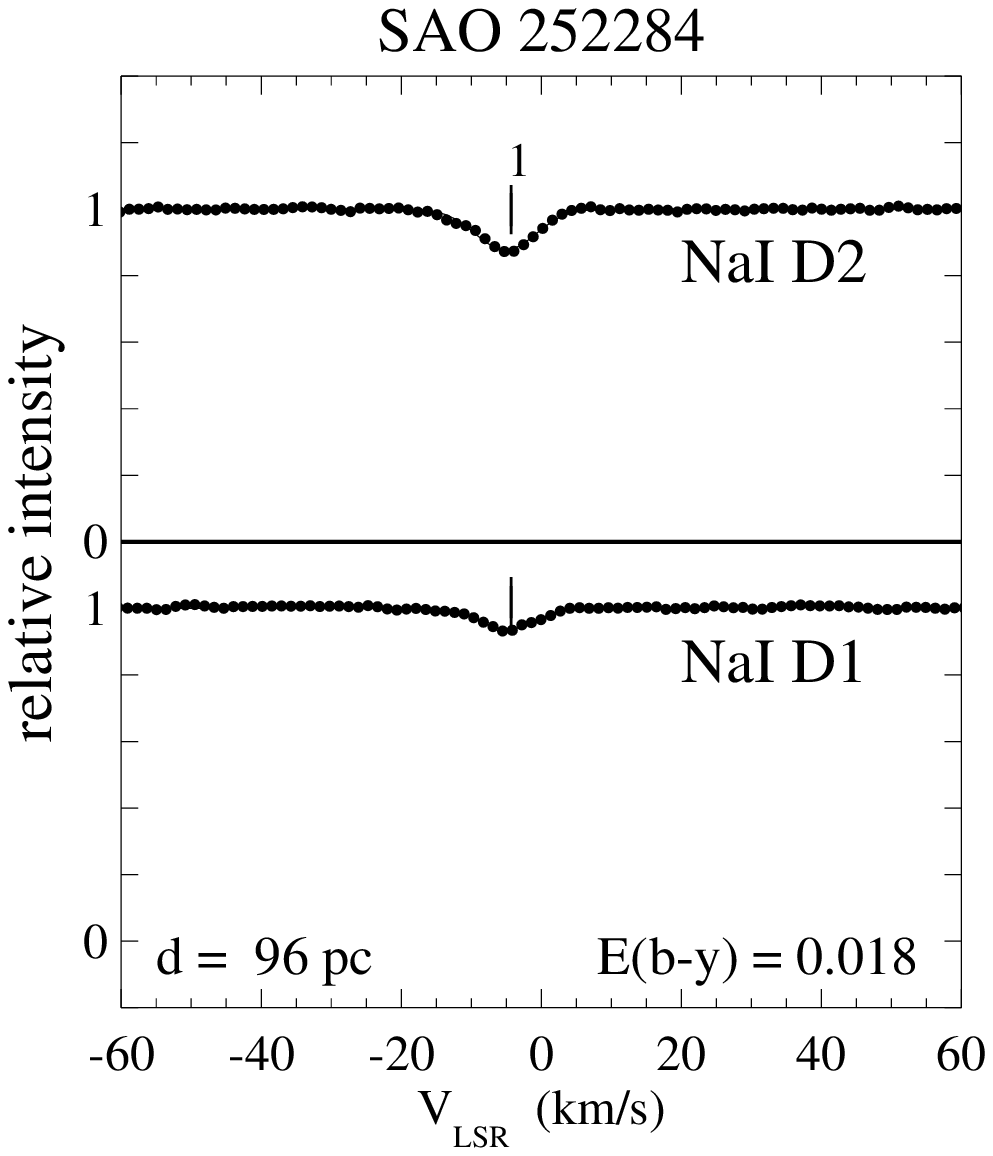} \hfill
\includegraphics[width=5.6cm]{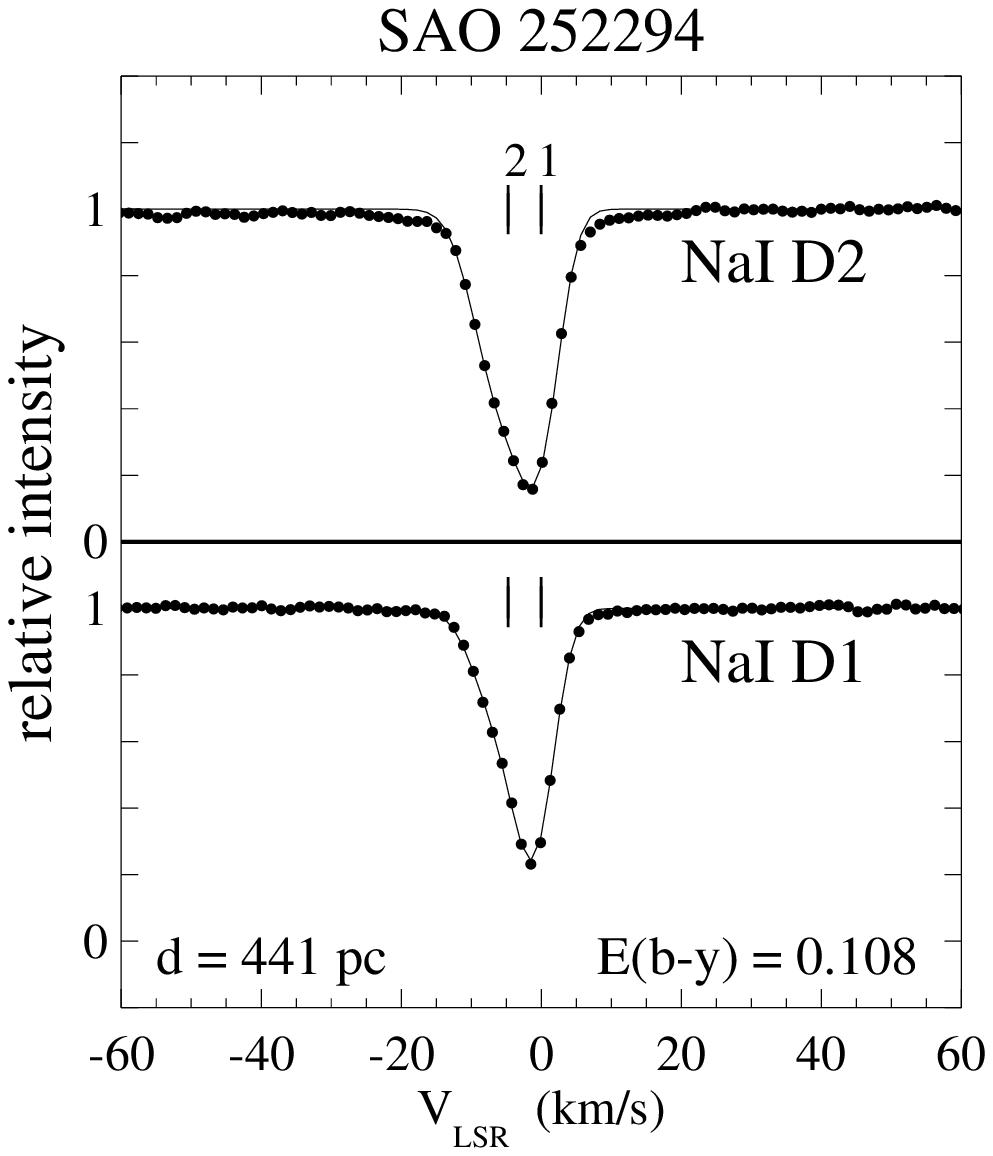} \hfill
\includegraphics[width=5.6cm]{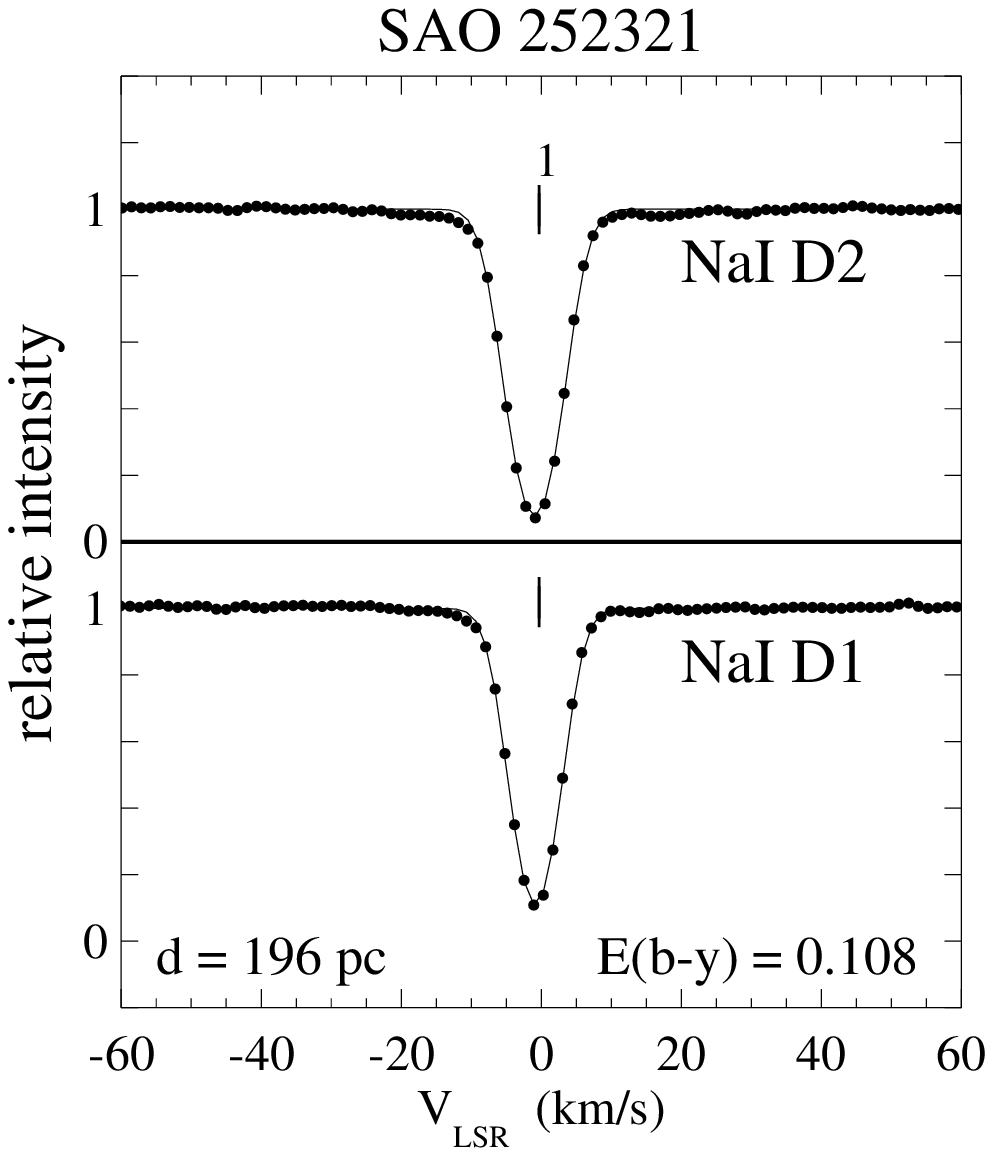} }
\vskip.3cm
\hbox{
\includegraphics[width=5.6cm]{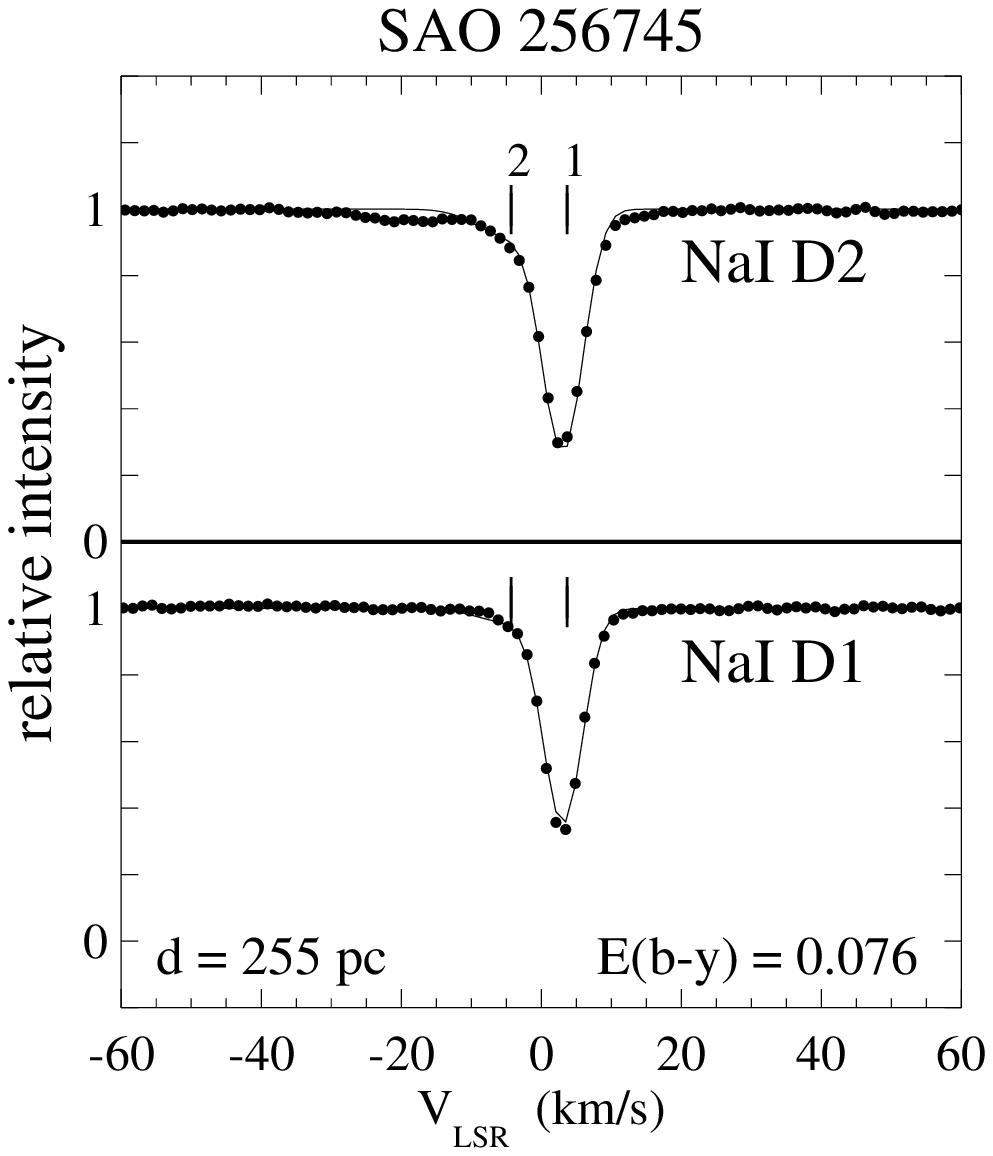} \hfill
\includegraphics[width=5.6cm]{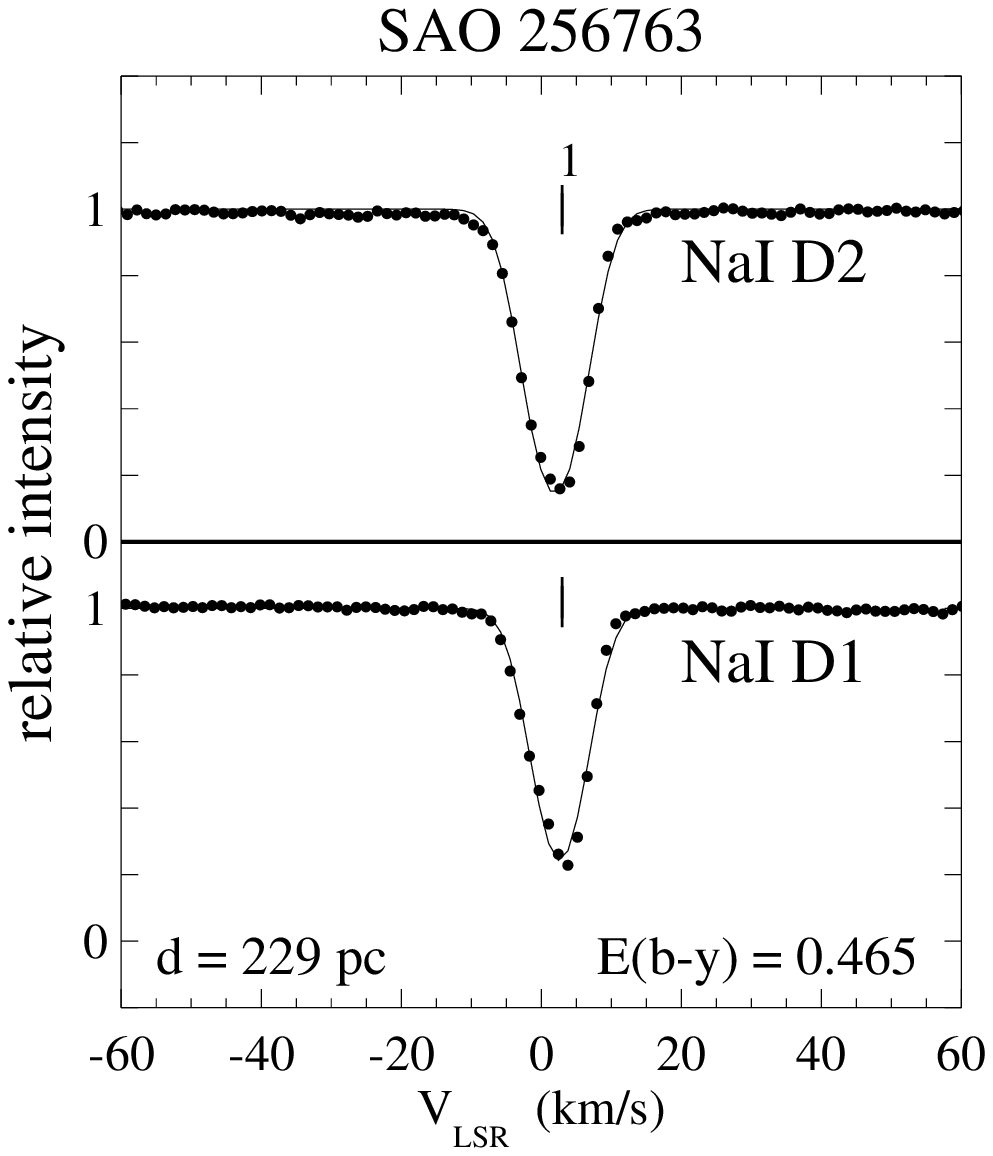} \hfill
\includegraphics[width=5.6cm]{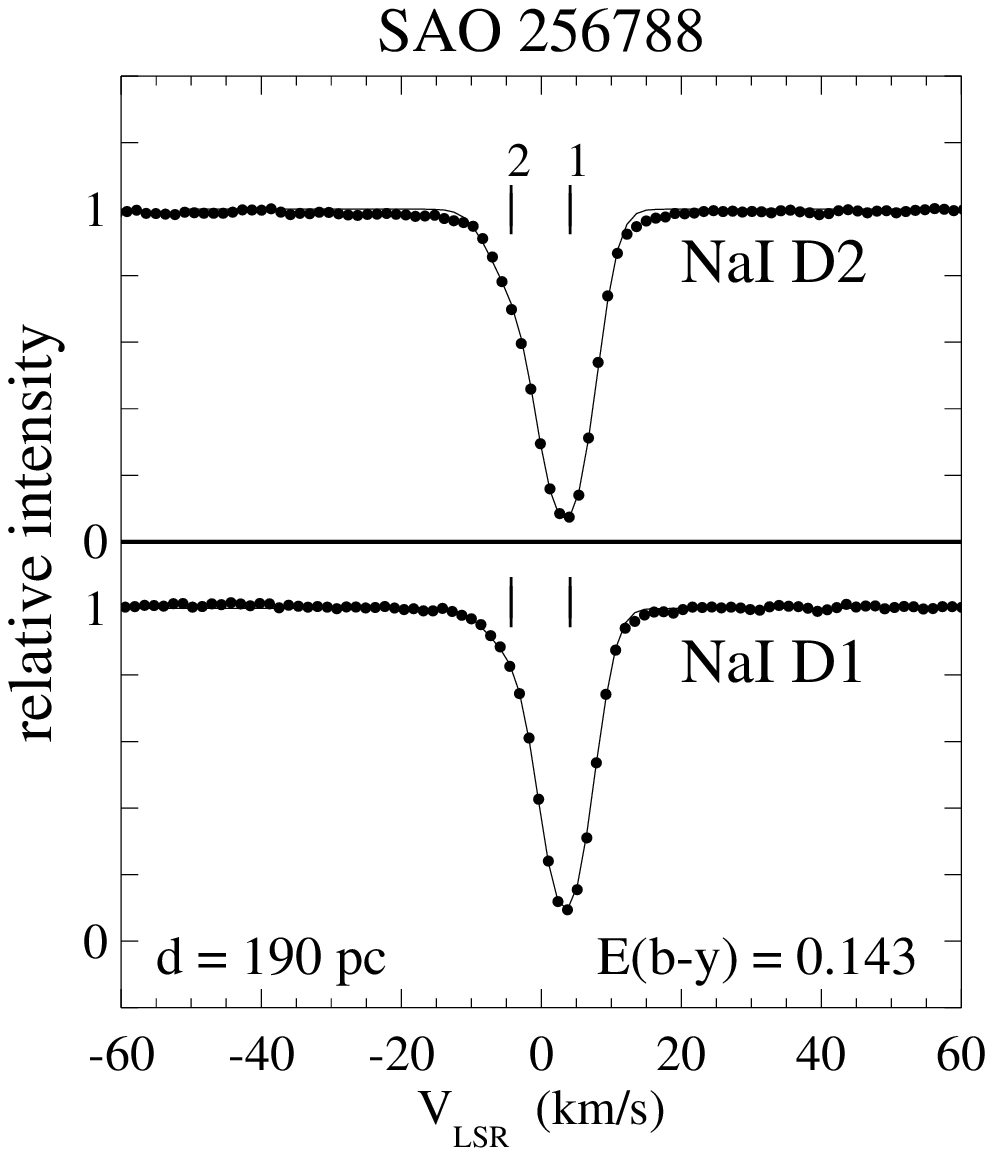} }
\vskip.3cm
\hbox{
\includegraphics[width=5.6cm]{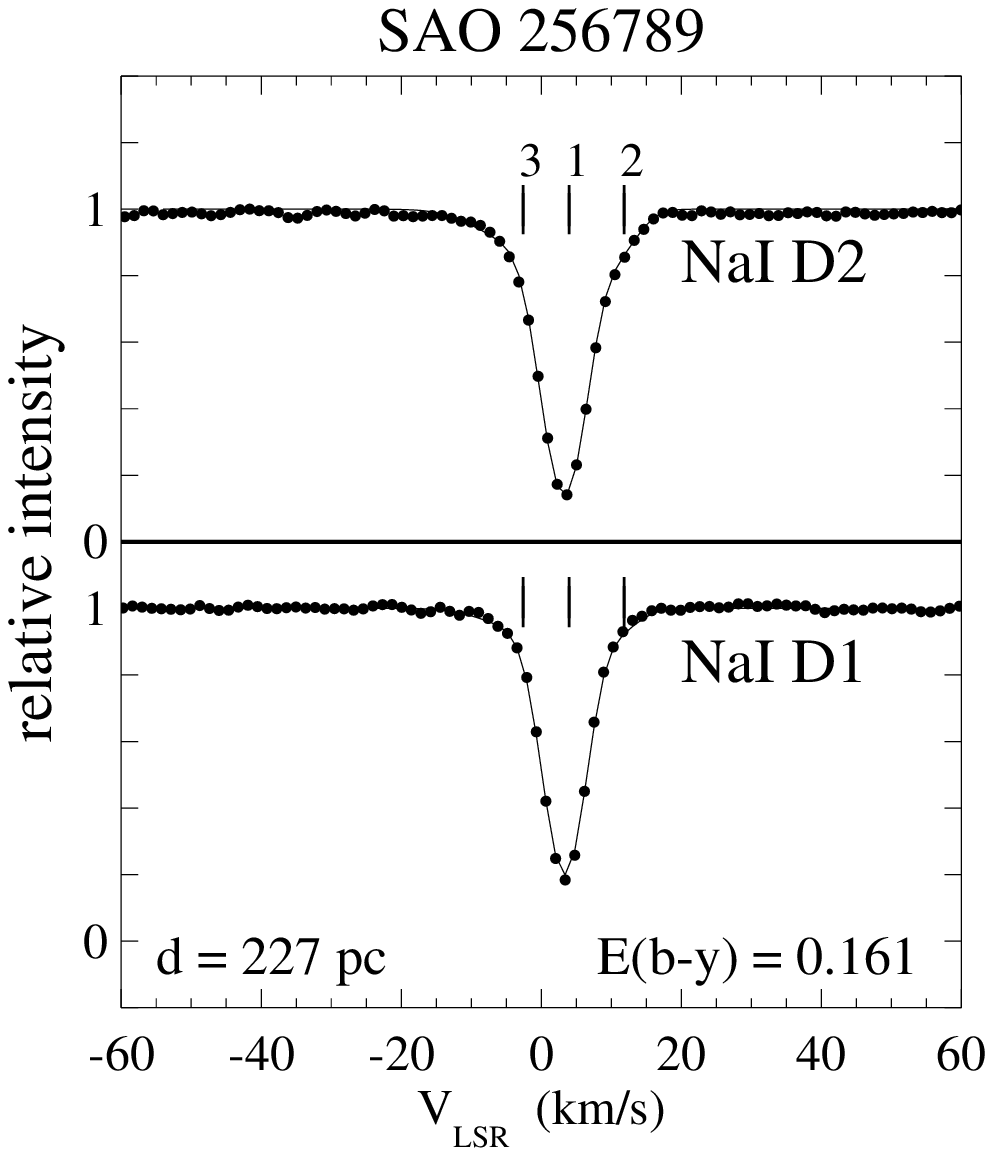} \hfill
\includegraphics[width=5.6cm]{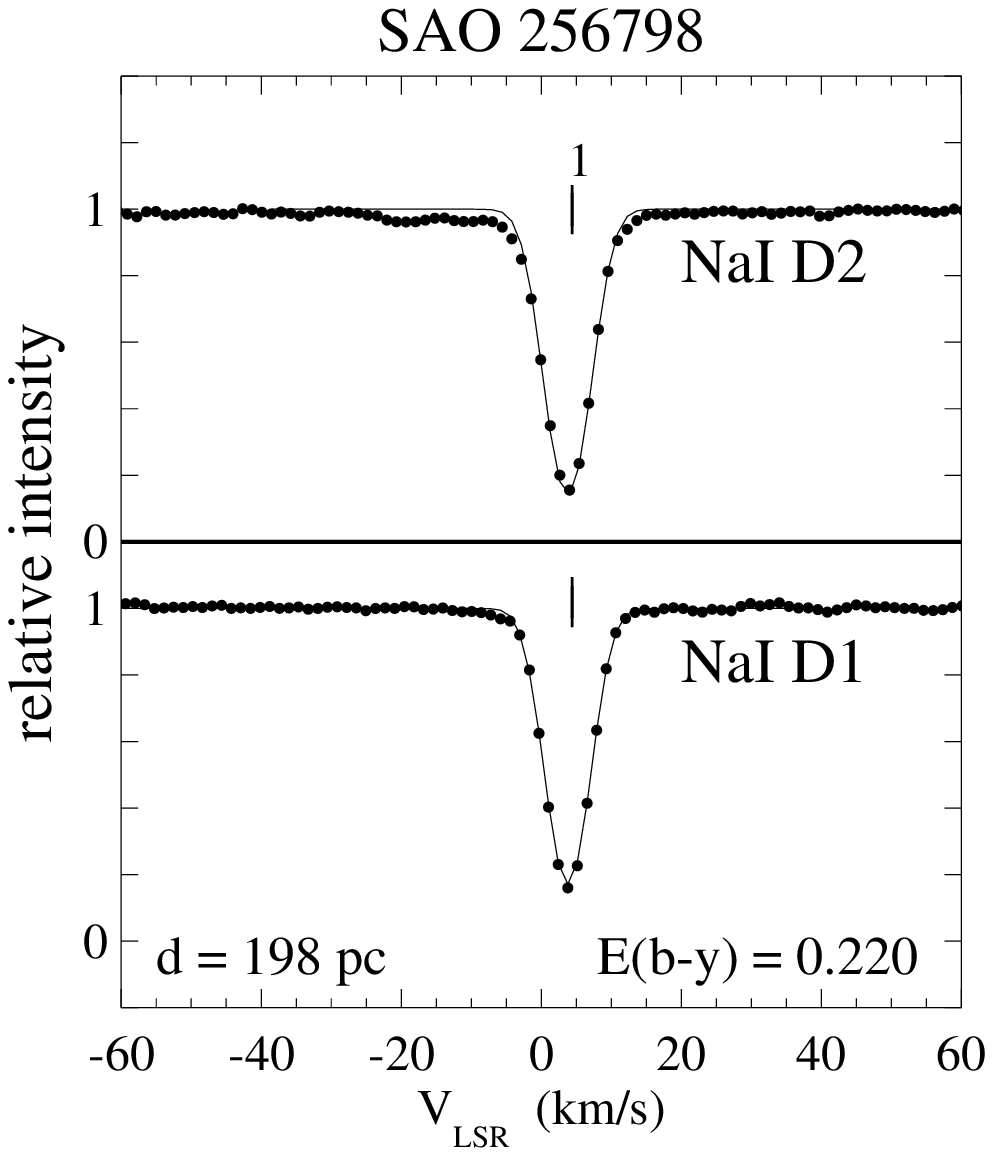} \hfill
\includegraphics[width=5.6cm]{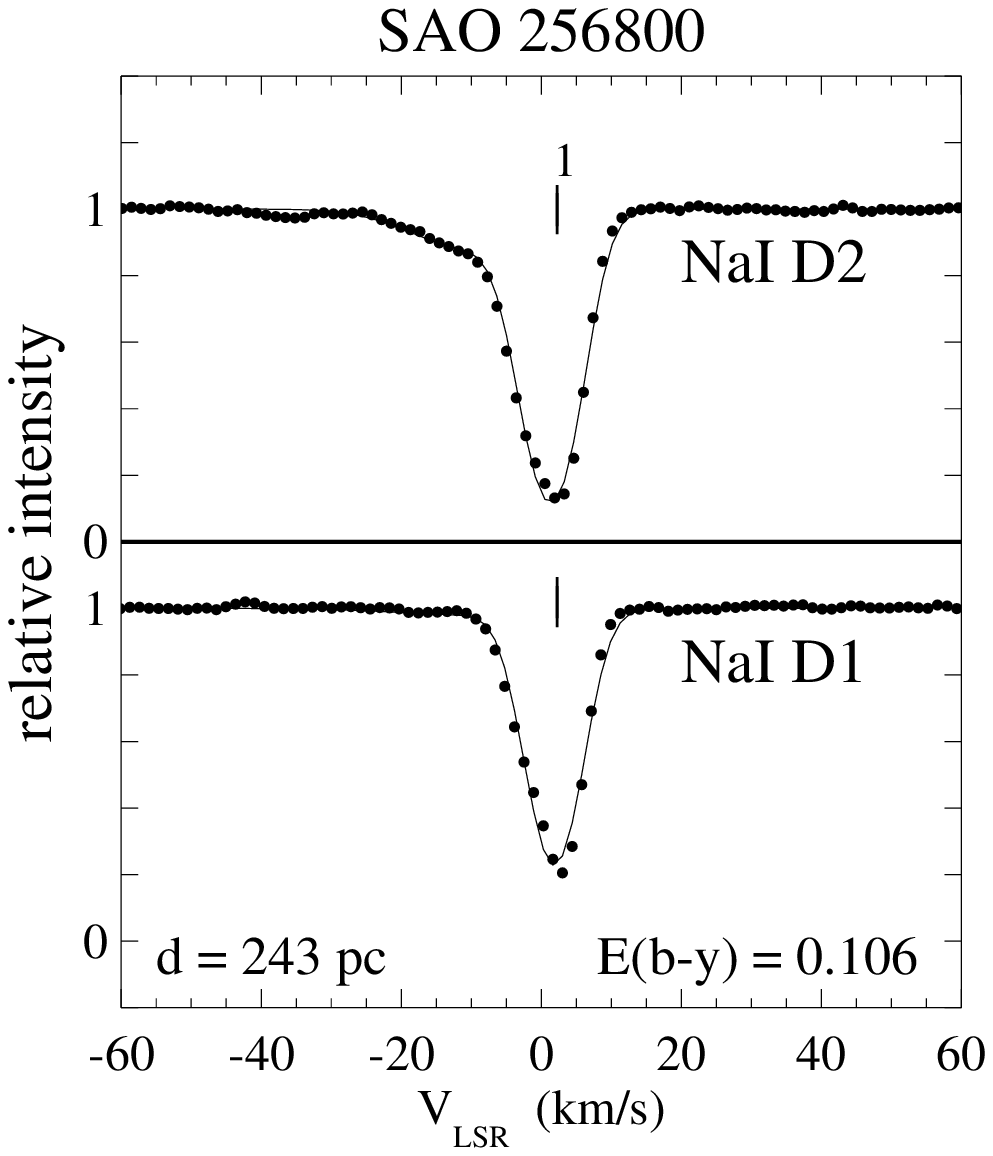} }
\caption{(Continued)}
\end{figure*}

\addtocounter{figure}{-1}
\begin{figure*}
\hbox{
\includegraphics[width=5.6cm]{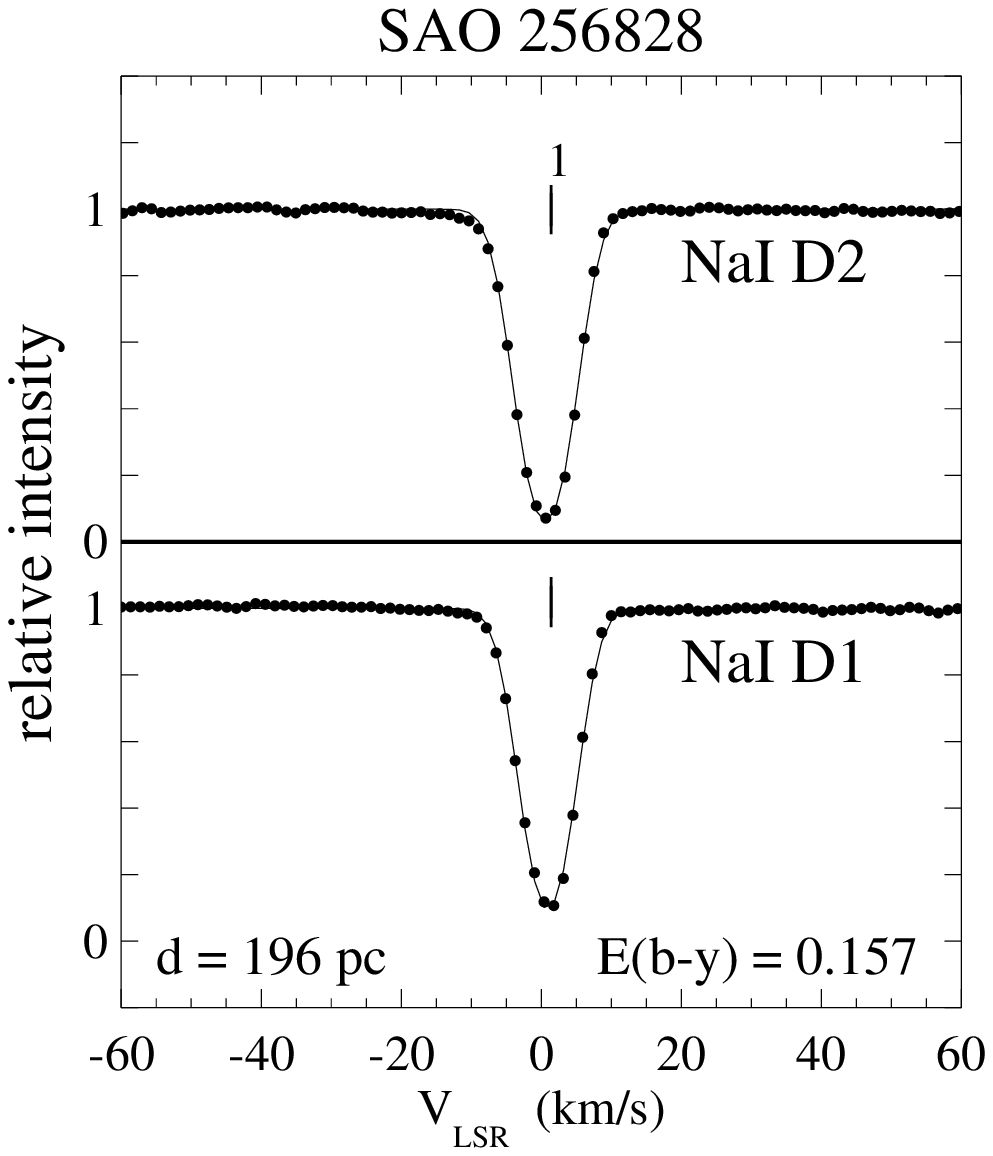} \hfill
\includegraphics[width=5.6cm]{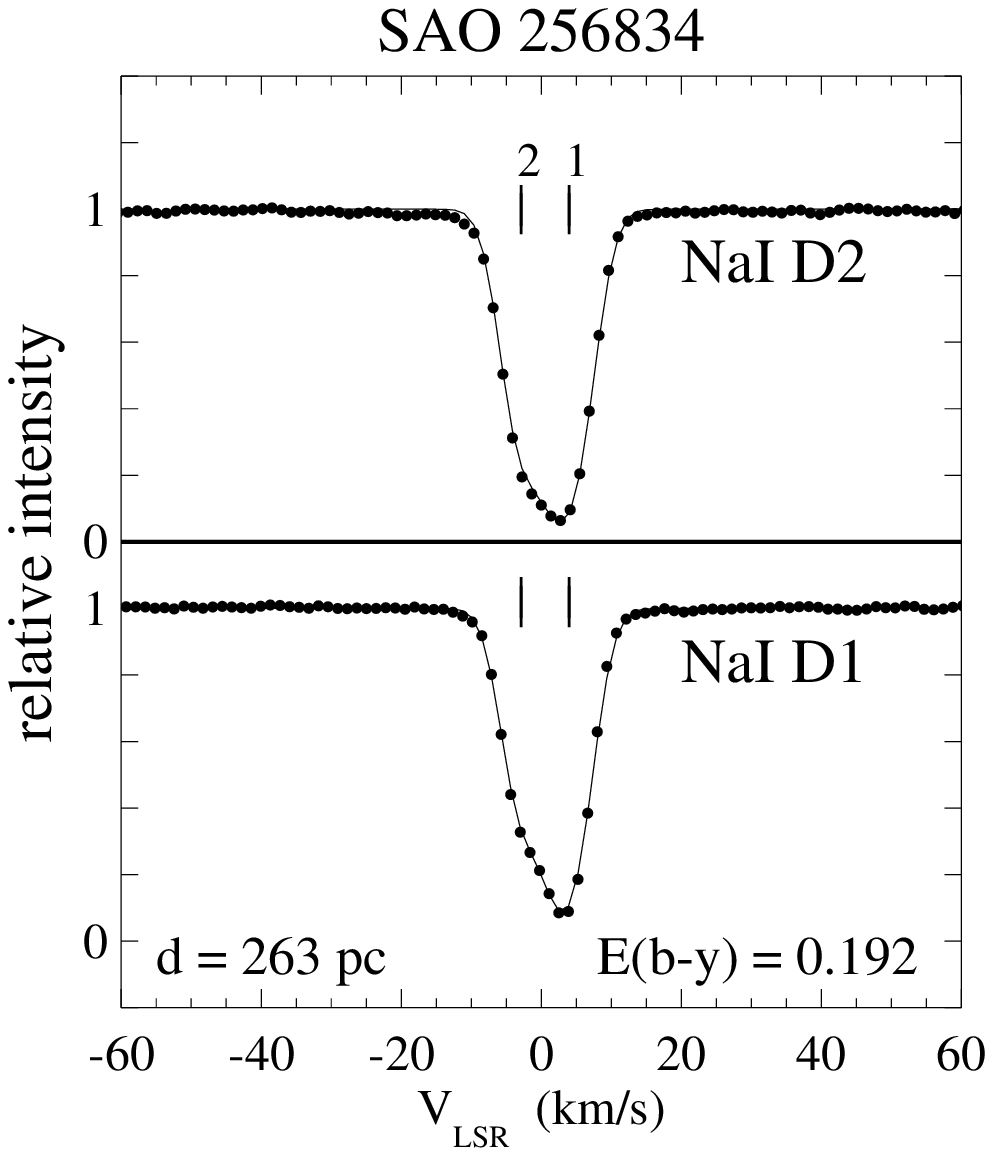} \hfill
\includegraphics[width=5.6cm]{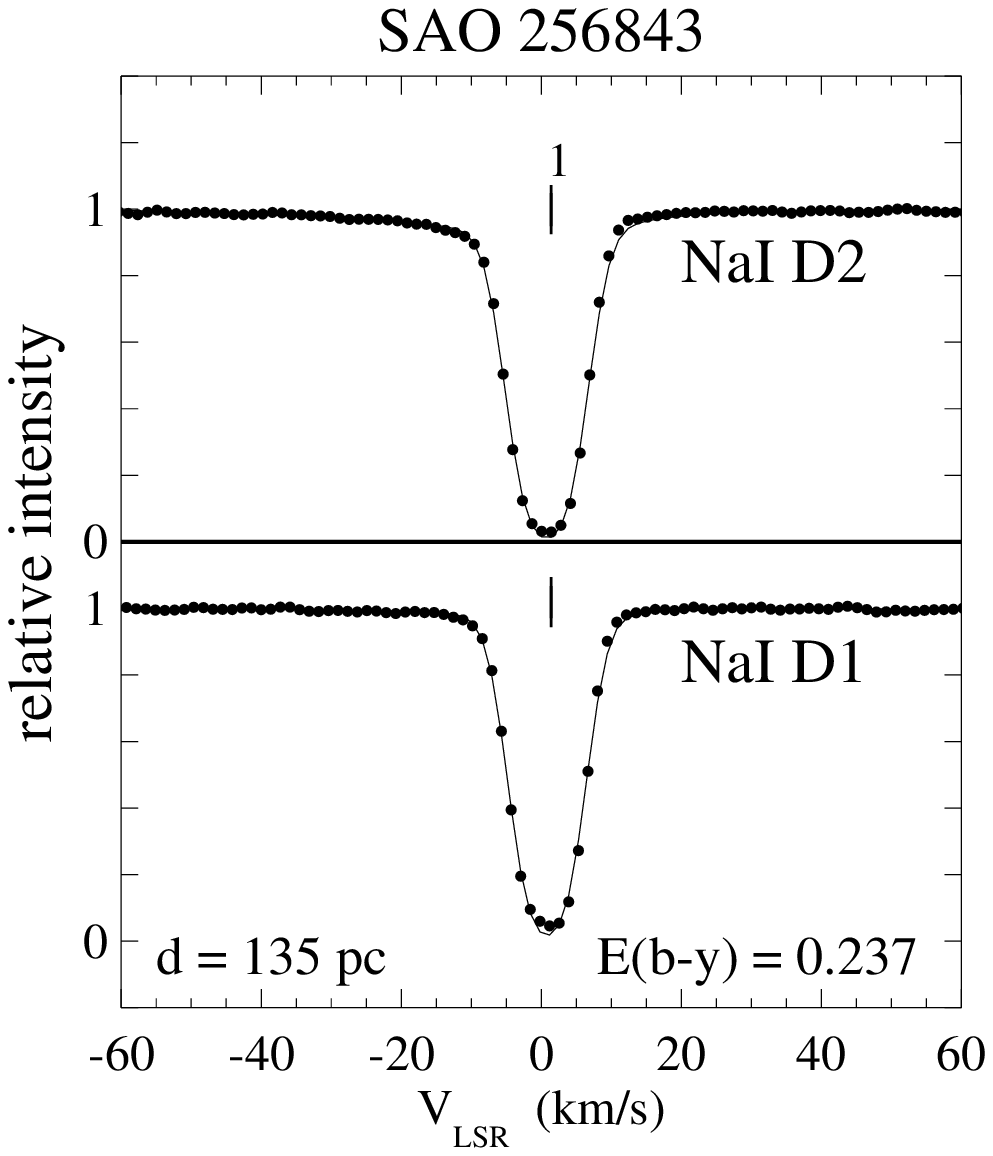} }
\vskip.3cm
\hbox{
\includegraphics[width=5.6cm]{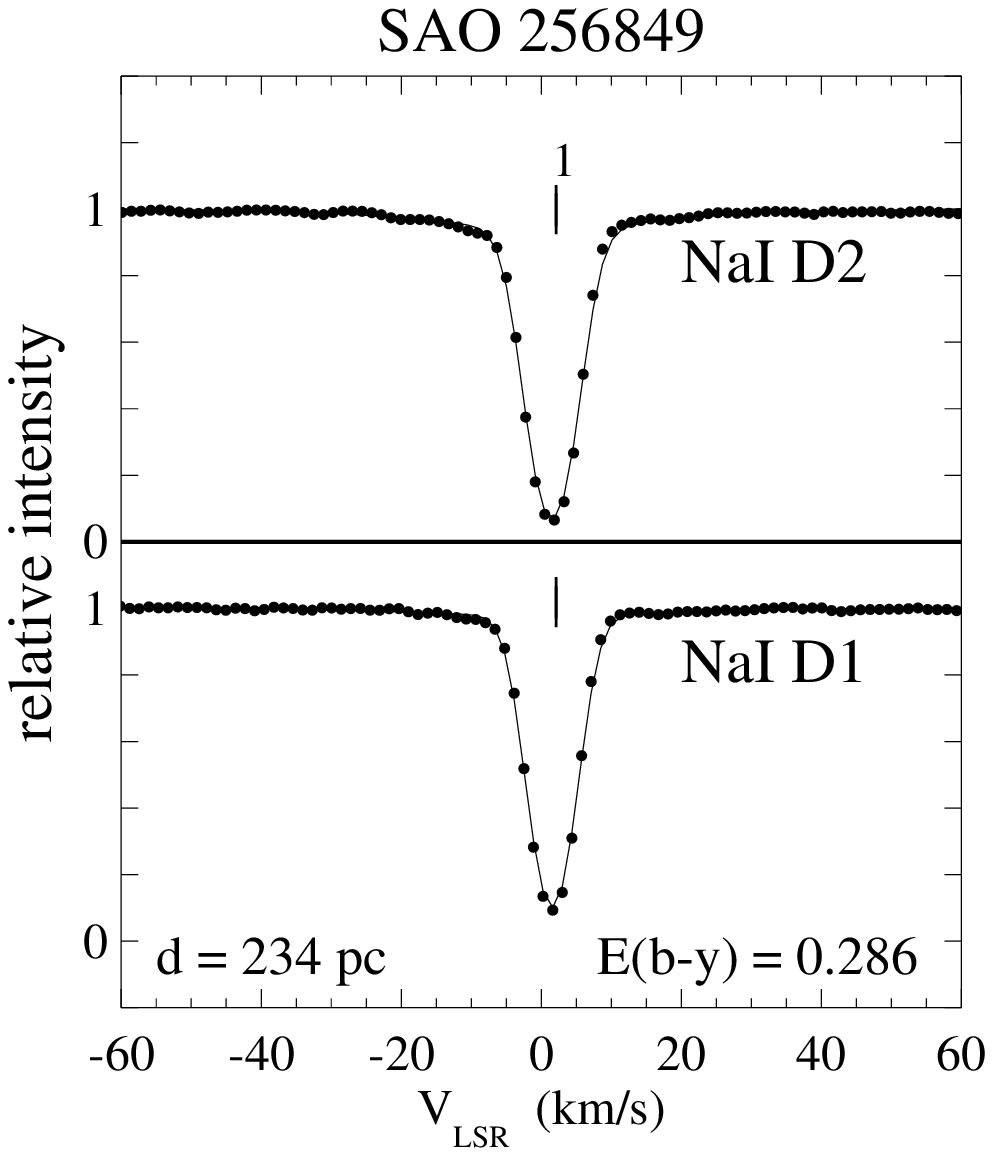} \hfill
\includegraphics[width=5.6cm]{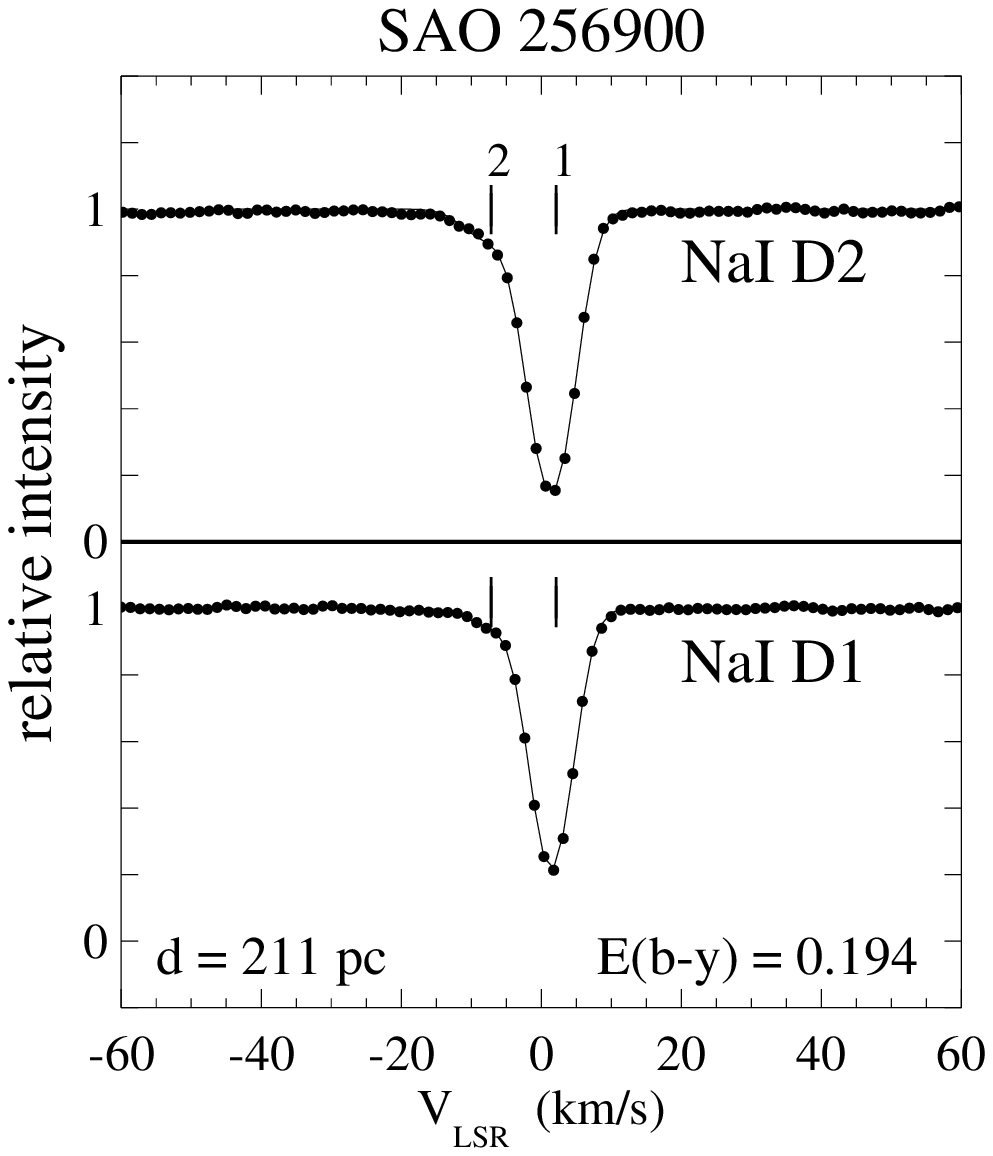} \hfill
\includegraphics[width=5.6cm]{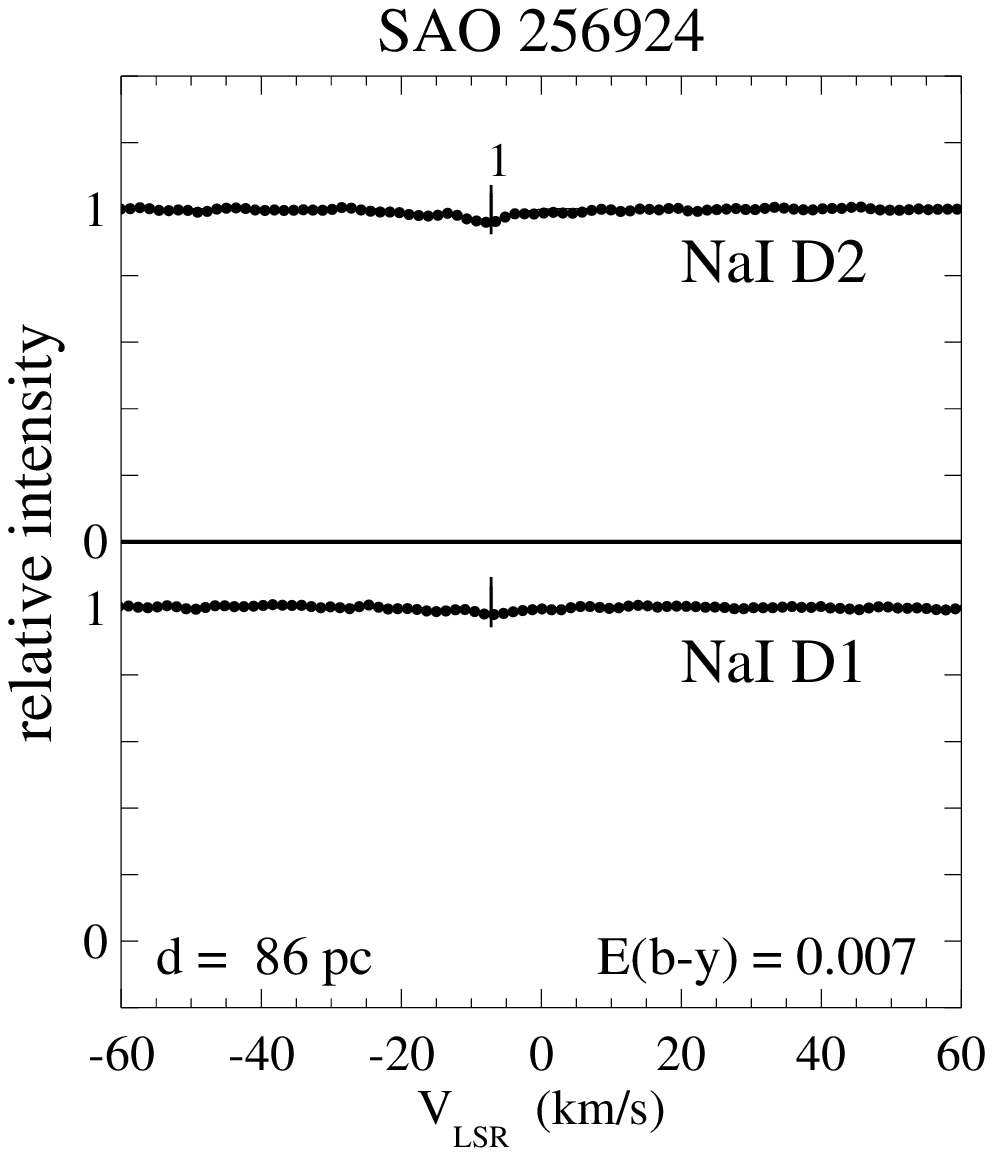} }
\vskip.3cm
\hbox{
\includegraphics[width=5.6cm]{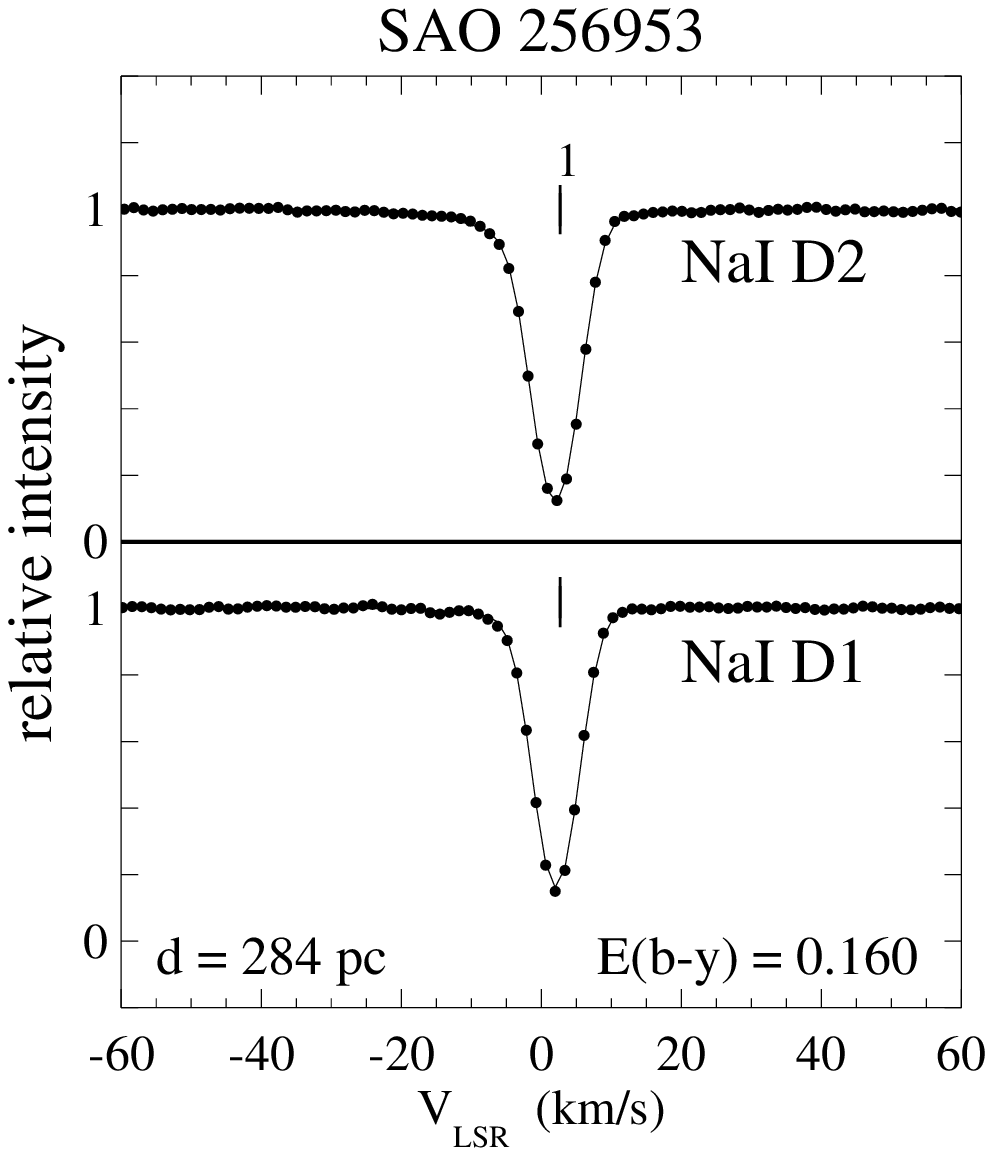} \hfill
\includegraphics[width=5.6cm]{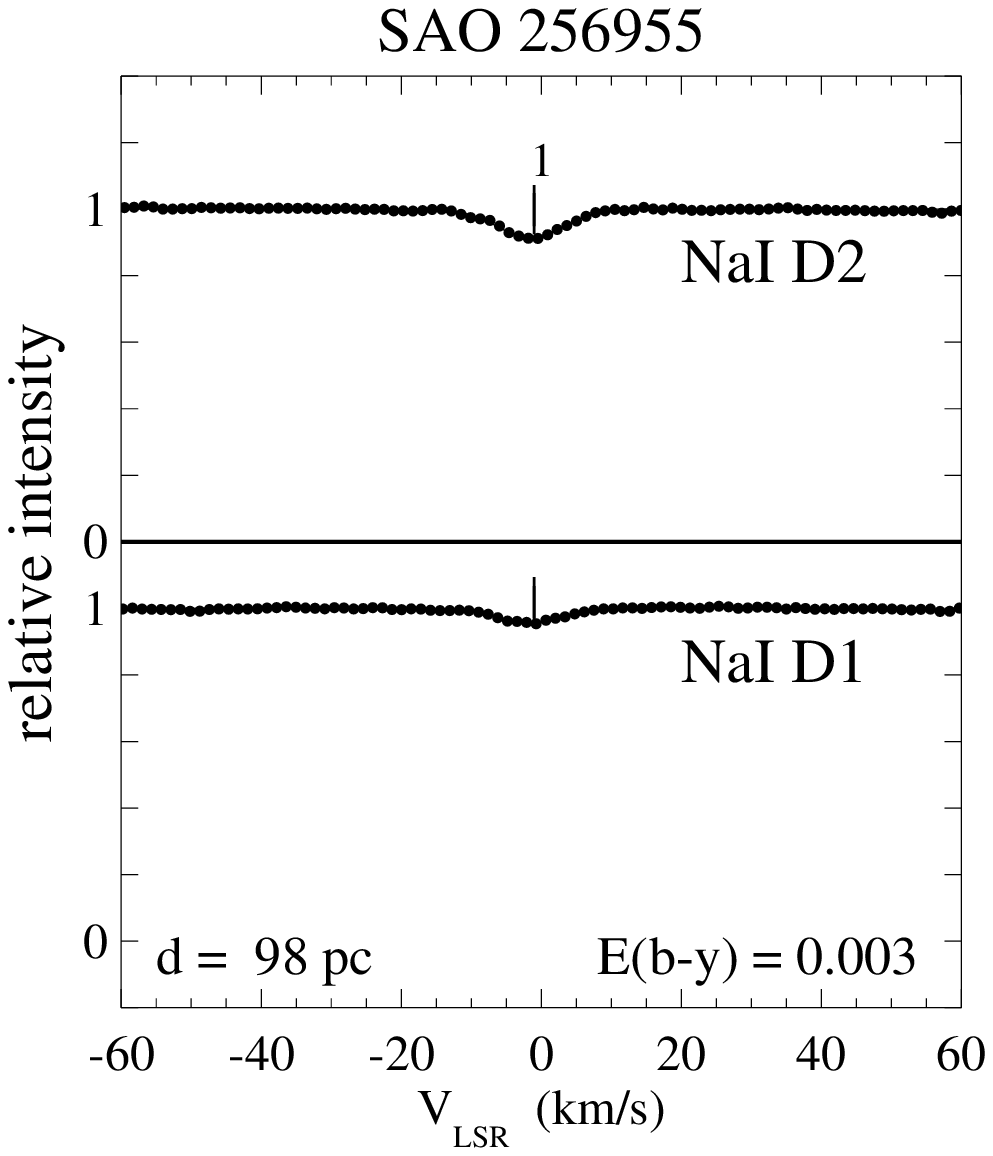} \hfill
\includegraphics[width=5.6cm]{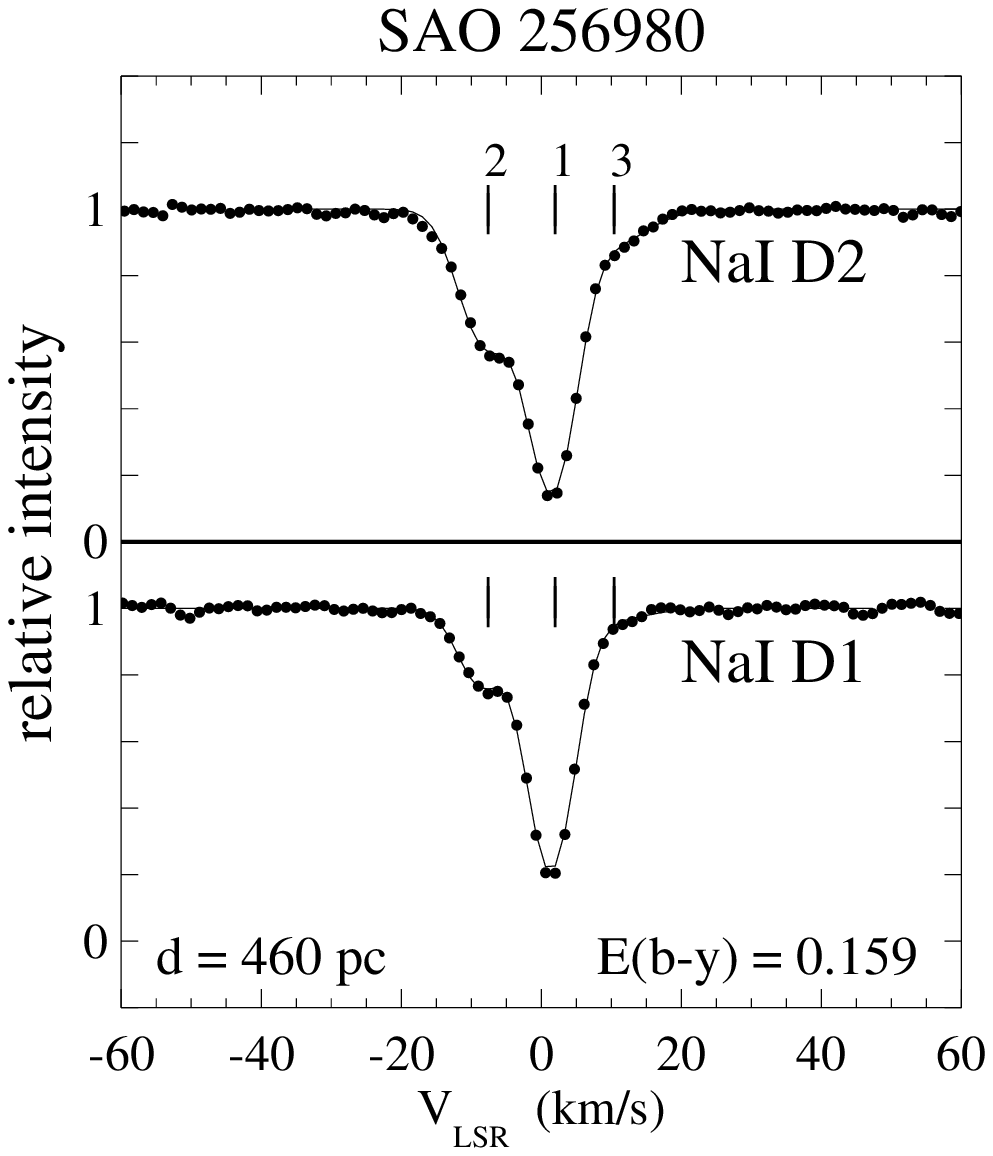} }
\caption{(Continued)}
\end{figure*}

\addtocounter{figure}{-1}
\begin{figure*}
\hbox{
\includegraphics[width=5.6cm]{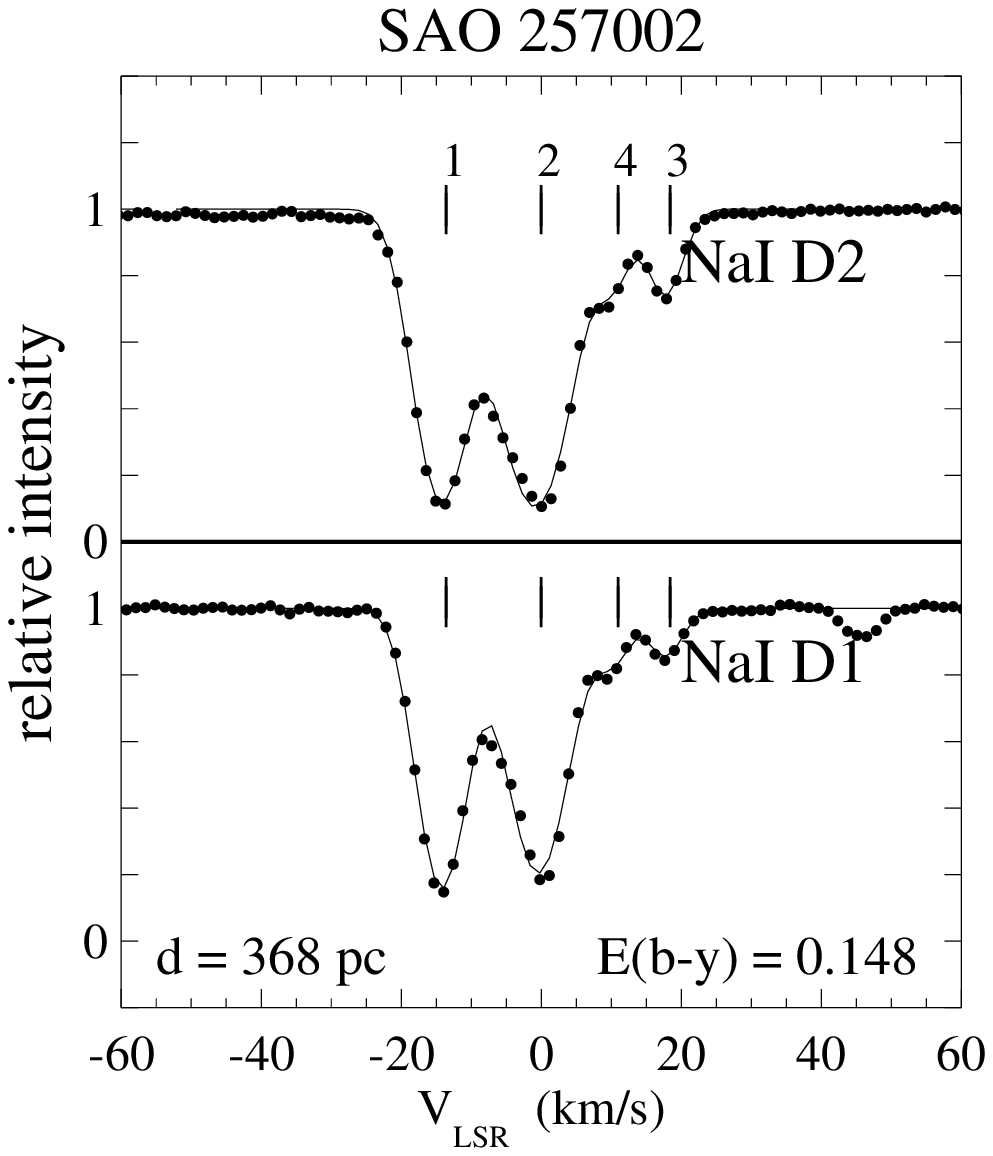} \hfill
\includegraphics[width=5.6cm]{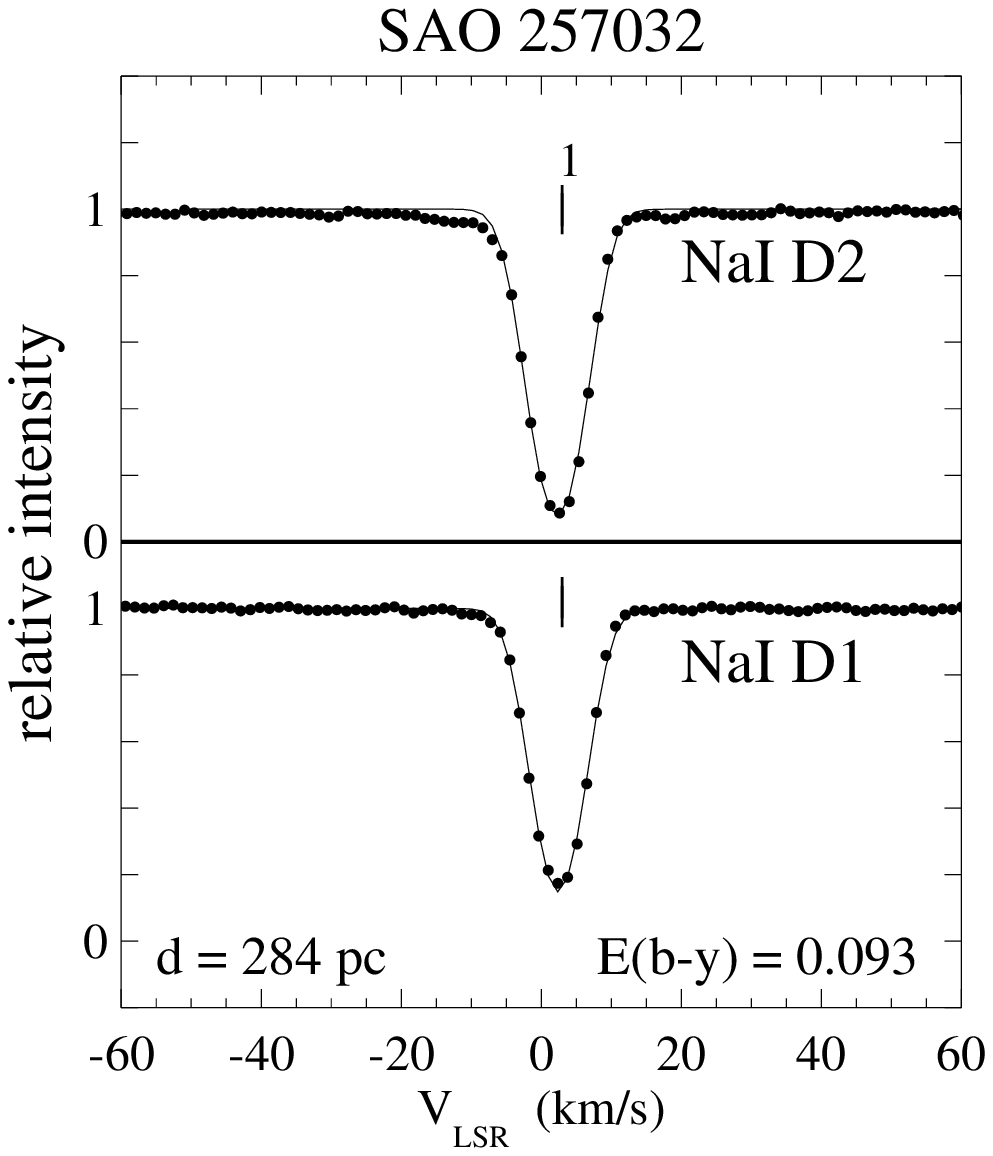} \hfill
\includegraphics[width=5.6cm]{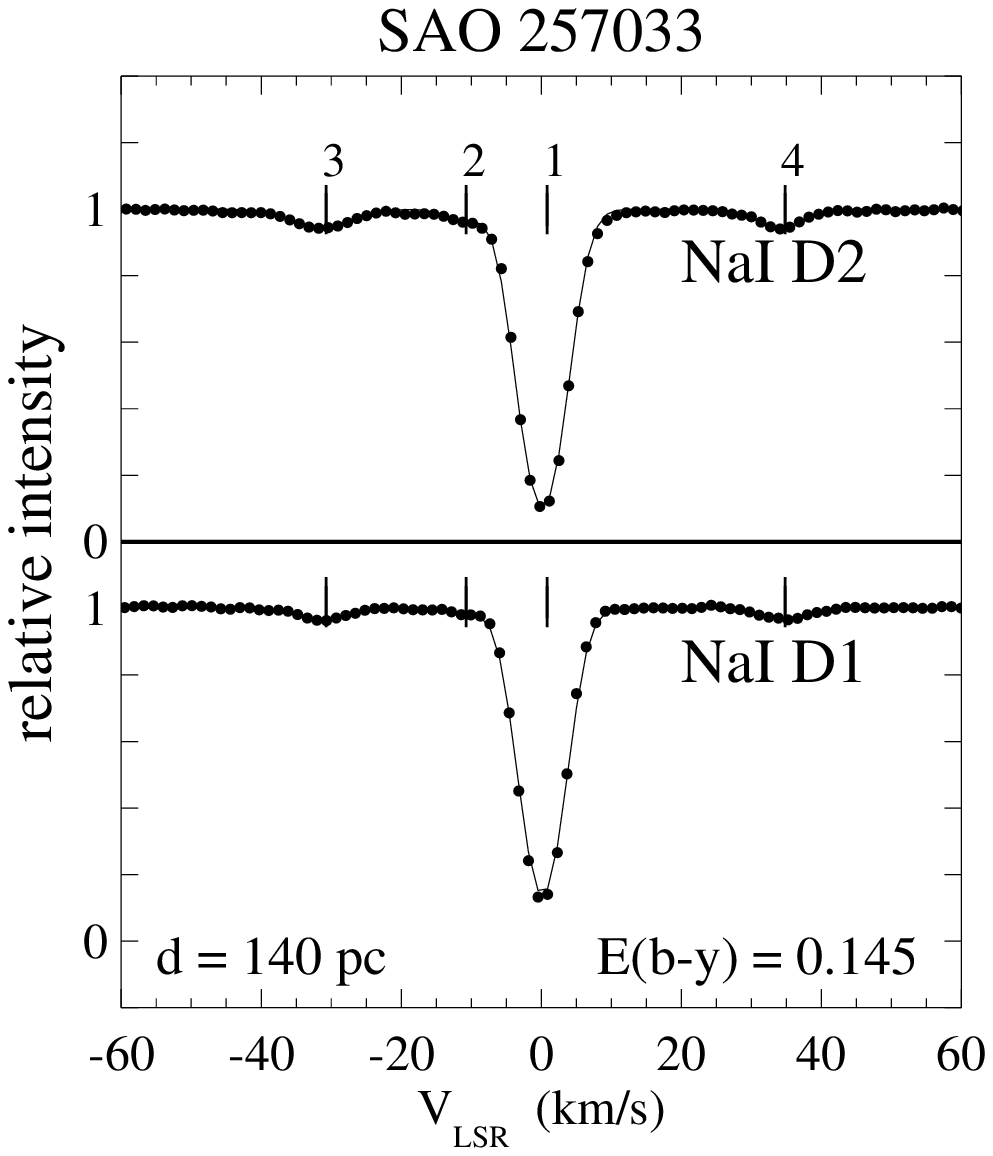} }
\vskip.3cm
\hbox{
\includegraphics[width=5.6cm]{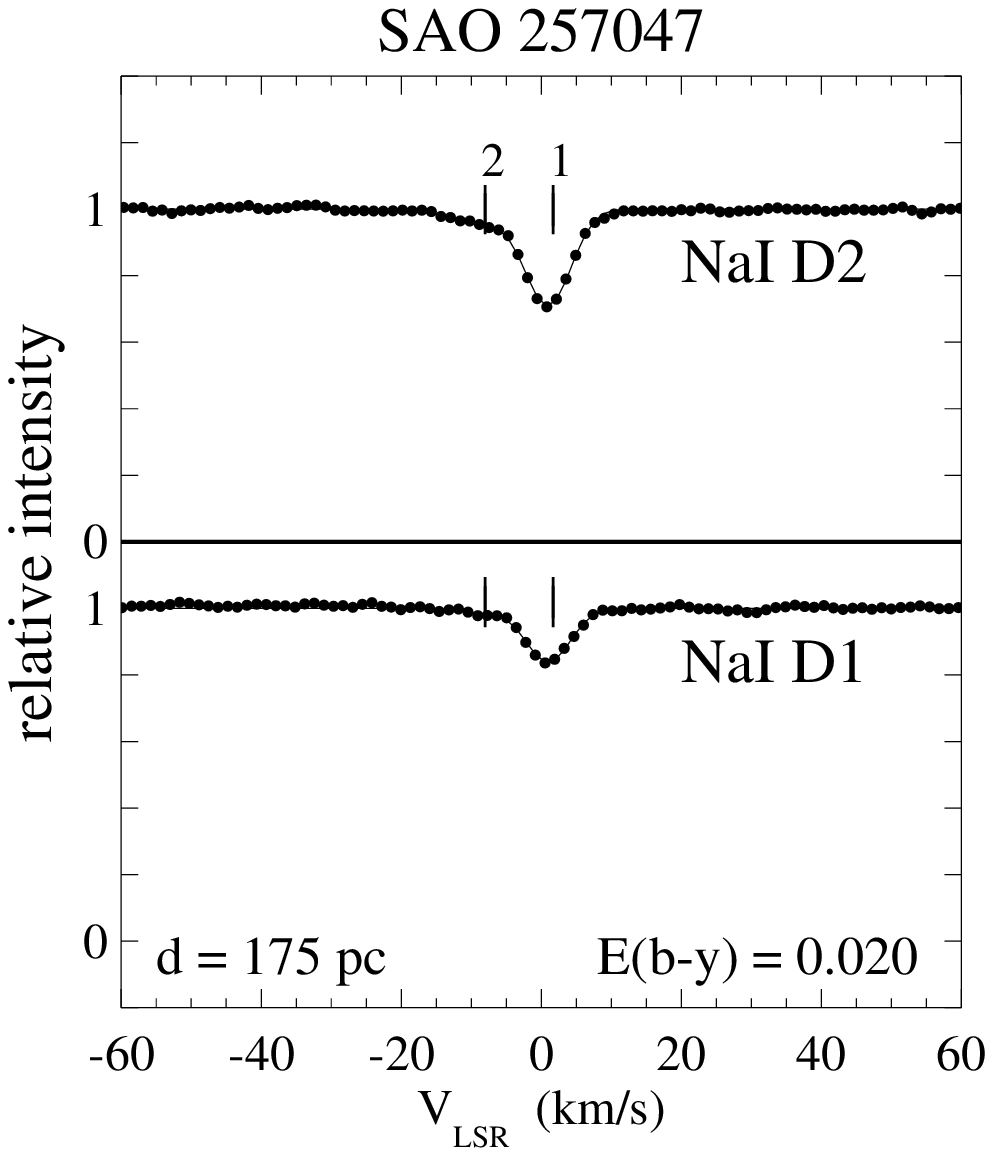} \hfill
\includegraphics[width=5.6cm]{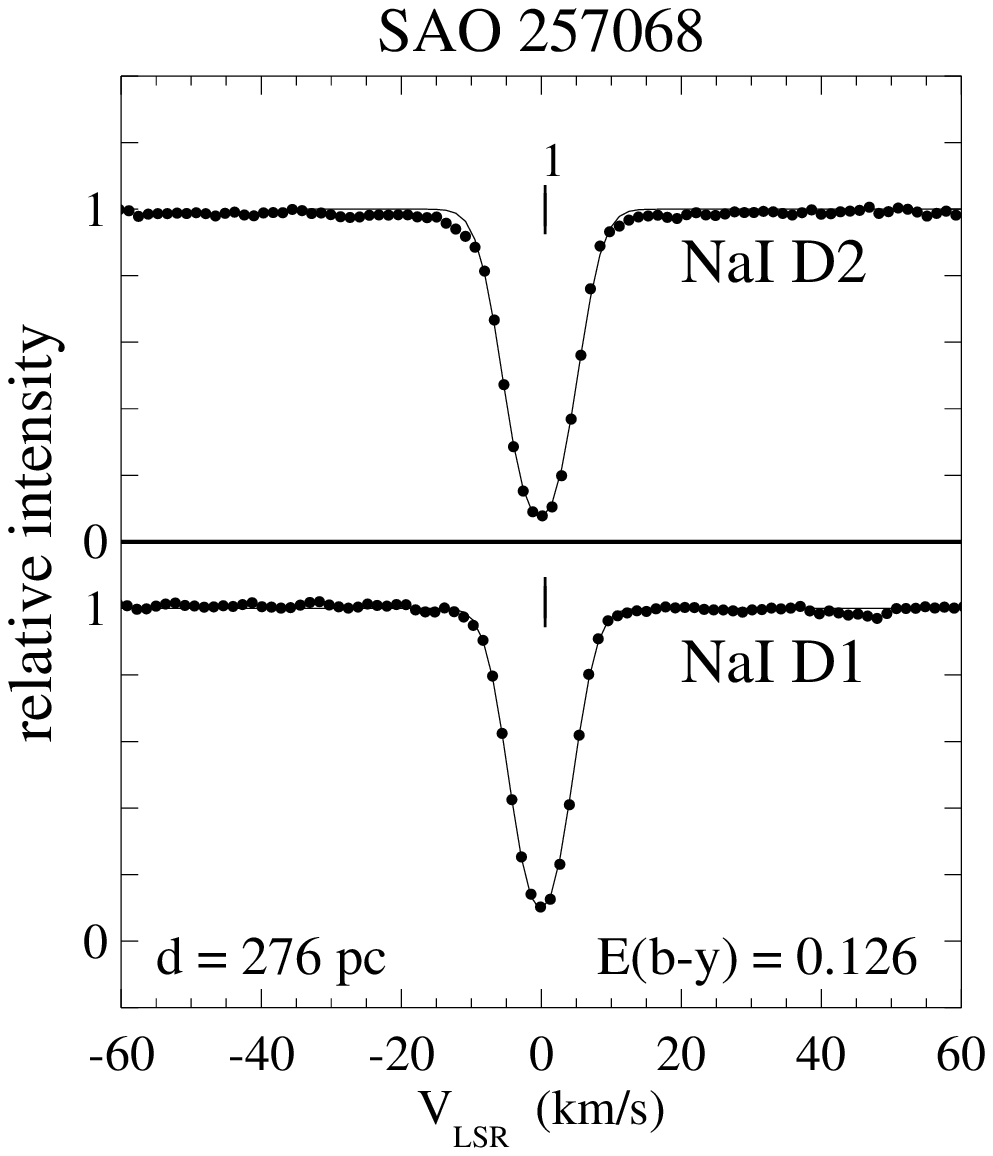} \hfill
\includegraphics[width=5.6cm]{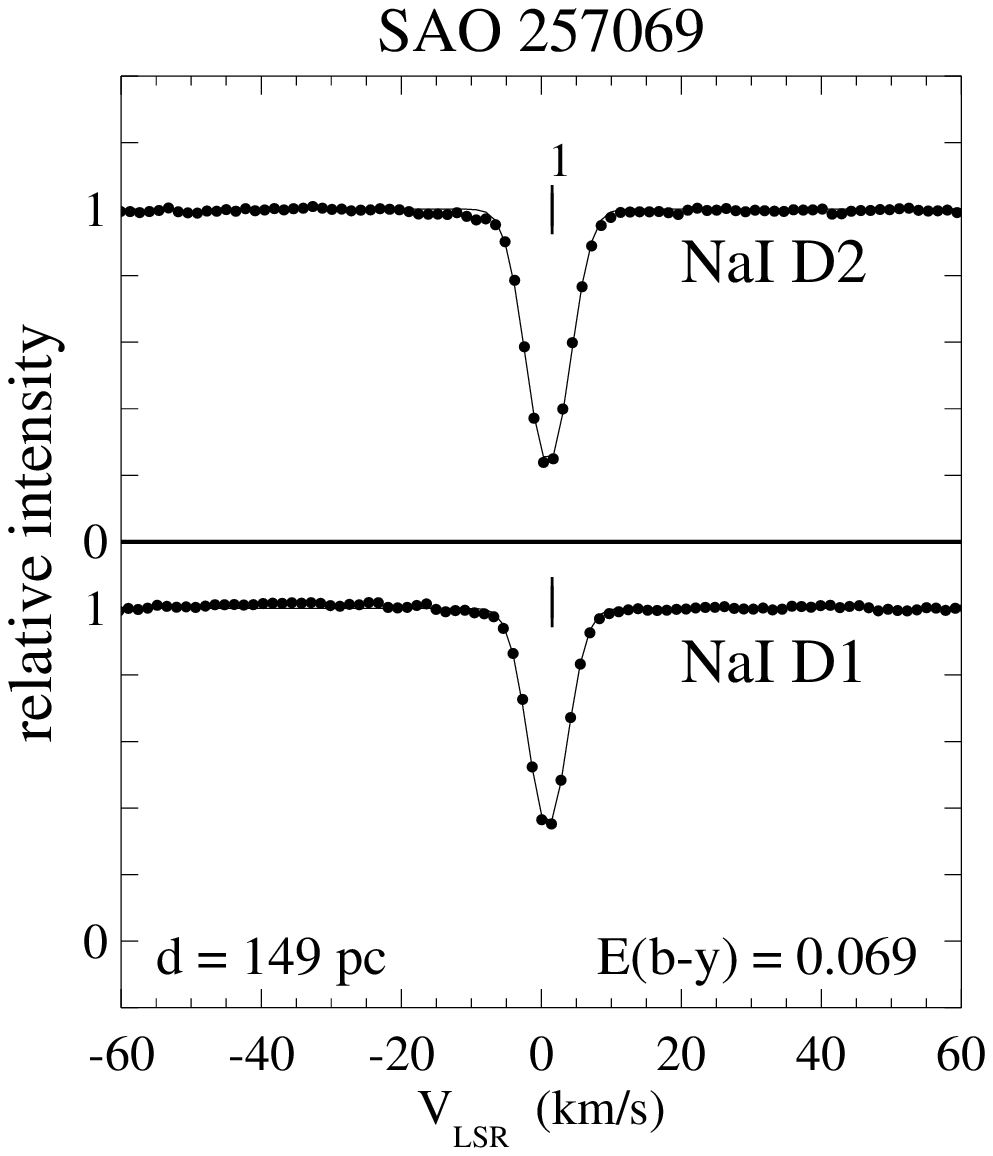} }
\vskip.3cm
\hbox{
\includegraphics[width=5.6cm]{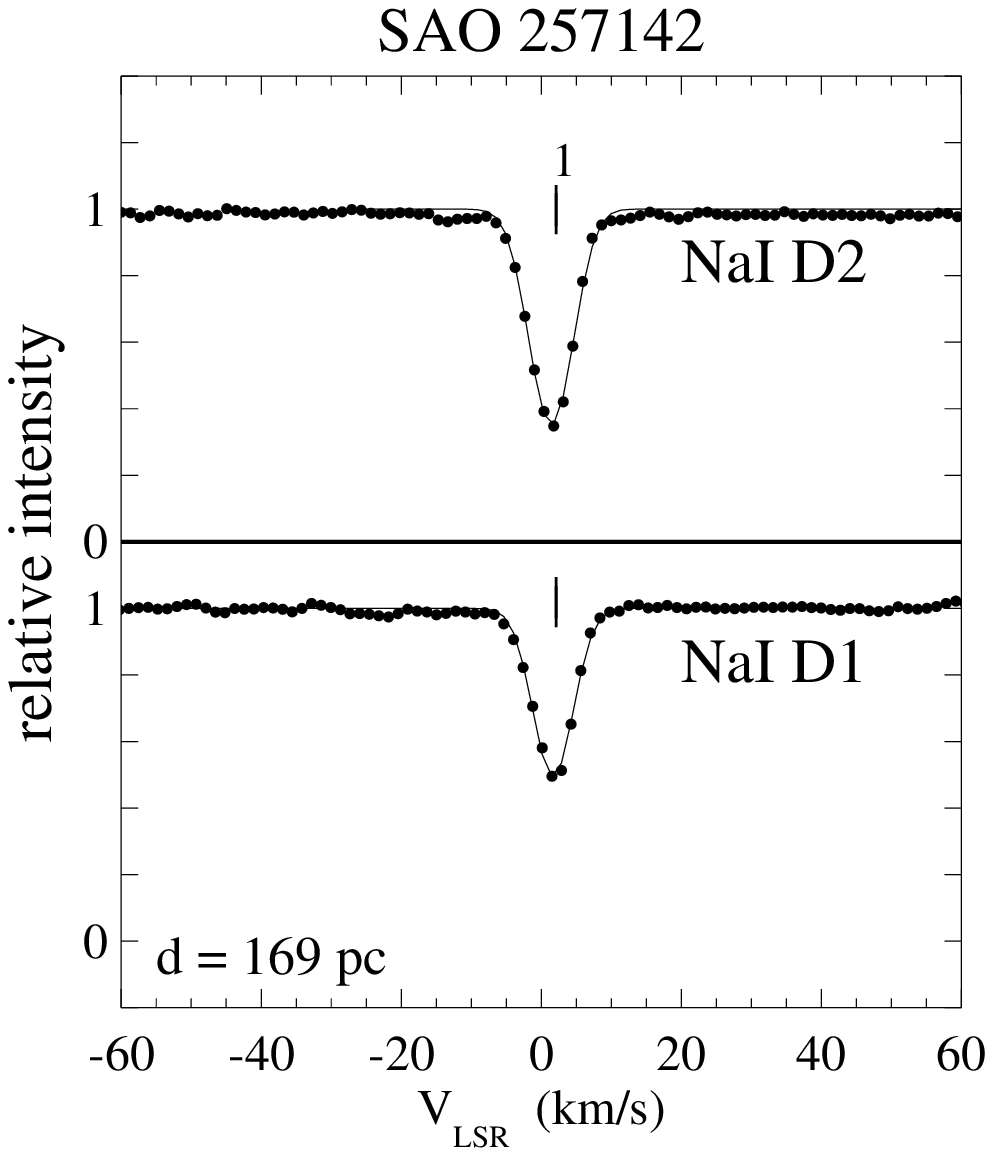} \hfill
\includegraphics[width=5.6cm]{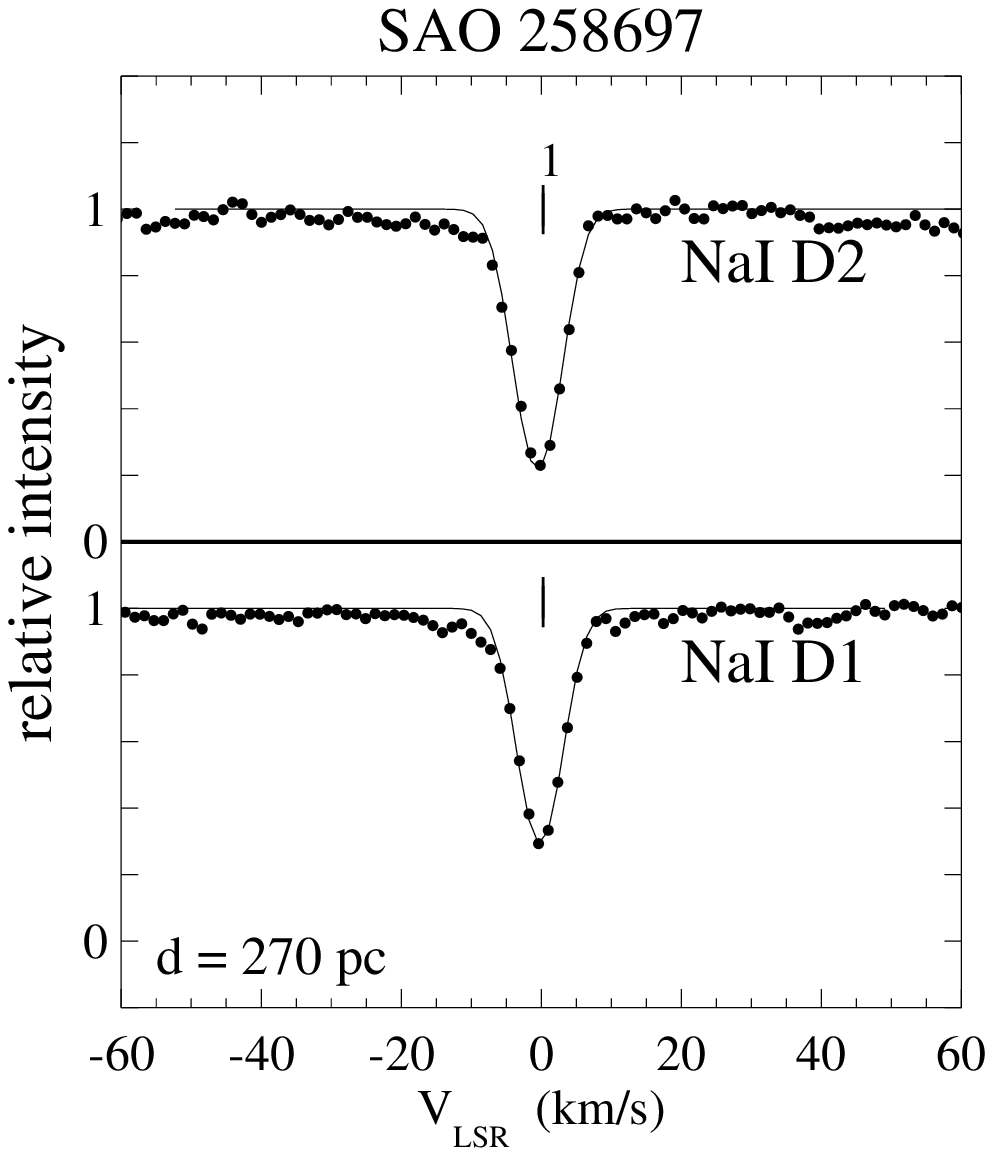} \hfill
\includegraphics[width=5.6cm]{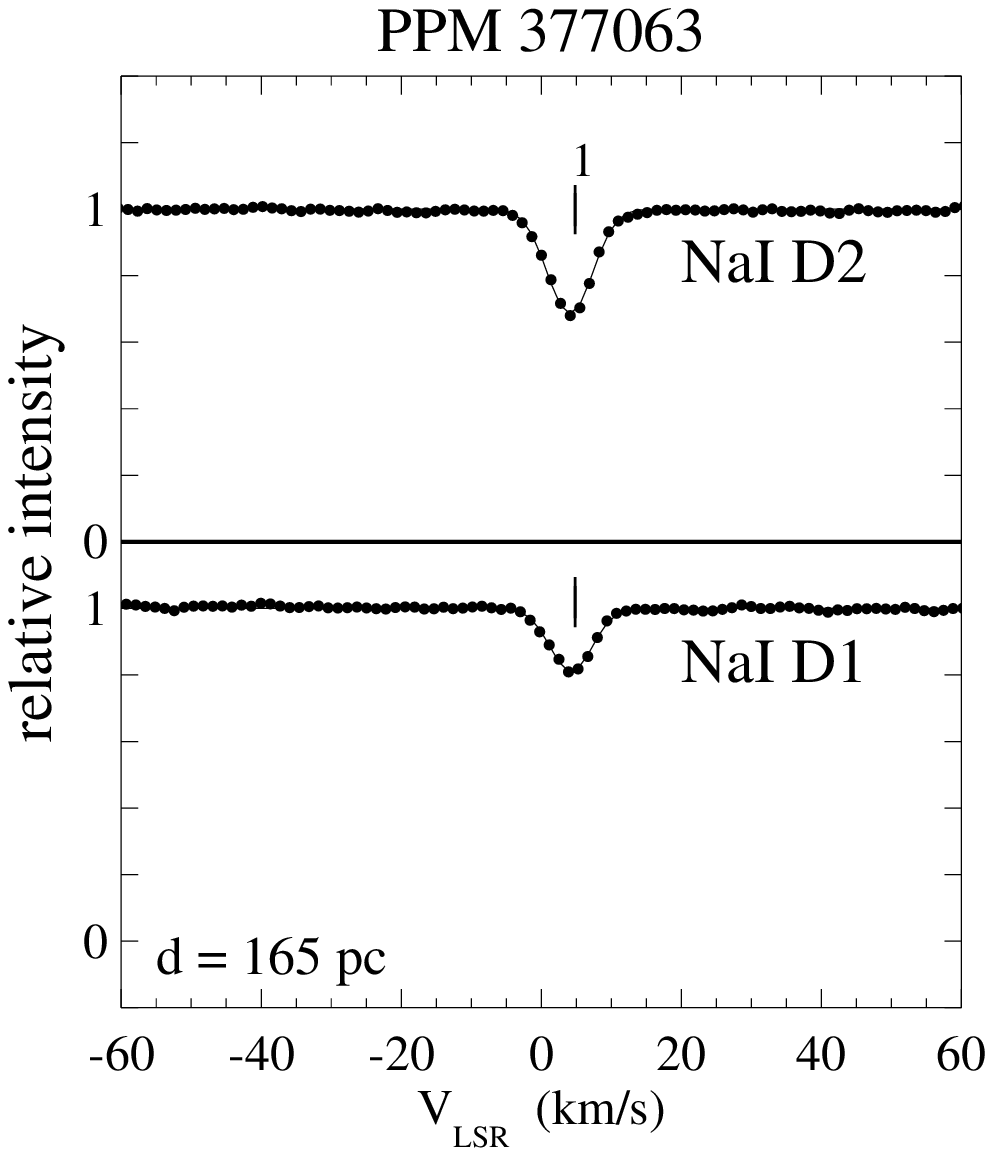} }
\caption{(Continued)}
\end{figure*}

\section{The absorption-line profiles}

All observed absorption lines show appreciable velocity structure, and thus
profile fitting provides the best method for determining accurate interstellar
column densities as well as to discern and determine the properties of the
individual interstellar clouds contributing to the line profile. The observed
profiles of the D$_2$ and D$_1$ absorption-lines are shown in 
Fig.~\ref{spectra}. The dots represent the object spectra and the solid line 
gives the adopted multicomponent fit, as described below. The line-fitting 
program (FITS6P) has been kindly provided to us by Dr. Welty. Further details 
of the employed method can be found in \citet{welty} and \citet{sembach}.
The zero point of the velocities in the upper and lower panels are with 
respect to the rest wavelength of the D$_2$\,(5889.9510 \AA) and 
D$_1$\,(5895.9242 \AA) lines, respectively. In each panel the tick marks and 
numbers above the profiles identifies the different components used to fit
the D$_2$ and D$_1$ lines.

Voigt line profiles parameterized by their velocity dispersion parameter b,
cloud component velocity \vlsr\ and cloud column density $N_{\rm \nai}$, were
convolved with a Gaussian instrumental profile of 5.3 km s$^{-1}$ width and
fitted to the observed data. Inspection of the calibration lamp spectra
supports our assumption of a Gaussian instrumental profile since no asymmetric
tails are observed in the thorium lines. Both the D$_2$ and D$_1$ lines
were fitted simultaneously, with the fewest components that seemed reasonable.
Inflections in line shape, as a result of additional components in the wings 
of the stronger lines, often indicated whether another component should be 
added to the model. The hyperfine structure splitting of $\sim$ 1 \kms\ of 
the individual D lines were not included in the adjustments since
the resolution of the data is sufficiently broad to be affected by the
splitting.

In most cases the assigned initial values of b were from 2.0 to 4.0 \kms\ to 
all components, but larger values are sometimes required, e.g. SAO\,239531. 
Also the three free parameters ($N_{\rm \nai}$, \vlsr, b) for each component 
were allowed to vary in an iterative nonlinear least-squares fit to
the observed profile. Some parameters occasionally were held fixed at
``reasonable'' values in order to facilitate convergence of the fits, but are
thought to be as well determined as those allowed to vary.

The average of the D$_2$ and D$_1$ best fit values of these parameters are 
listed in Table 2. A total of 125 entries are given. Each successive column 
gives the star's name, component number, LSR radial velocity \vlsr, 
heliocentric radial velocity $V_{\sun}$, logarithm of the column density 
\logn, velocity dispersion b and the equivalent widths $W_{\rm D2}$ and 
$W_{\rm D1}$, respectively.

The distribution of the absolute mean errors for \logn, \vlsr\ and b are shown
in Figs.~\ref{param}a-c. For narrow, isolated lines of moderate strength such 
errors are typically 1\%-10\% for \logn\ and few hundredths of a \kms\ for b 
and \vlsr. For stronger, wider and more blended lines the uncertainties in 
\logn\ are typically 5\%-20\%, and 0.1-0.4 \kms\ in b and \vlsr. For the 
strongest saturated lines, uncertainties in the actual component structure 
are likely to be the dominant error source.

The theoretical uncertainty in the placement of the continuum level is
approximately equal to the rms scatter divided by the square root of the number
of points used to define it \citep{howarth}. The statistical error on
line flux values are given by $\sigma_{\mbox{\sc f}}$ = $\frac{\Delta
\lambda\,\sqrt{n}}{\rm S/N}$, where $n$ is the number of observed data points
in the absorption profile, $\Delta\lambda$ is the resolution element in
\AA/pixel, and S/N is the signal-to-noise ratio of the exposure \citep{welsh90}.
In most cases the continua are very well defined and statistical errors
probably dominate the equivalent width errors. Typically for our observations,
$n = 16$, $\Delta \lambda$ = 0.027\AA/pix and S/N $\geq$ 150, so that
$\sigma_{\mbox{\sc f}} \leq$ 1 m\AA.

The star $\iota$\,Lup has been previously observed at similar resolution 
(4 \kms) by \citet{welsh94}. The parameters for (\logn, \vlsr, b) are
($10.9\pm 0.02$, $-13.8 \pm 0.3$, $4.6\pm 0.3$), and those obtained in this work
are ($11.0\pm 0.3$, $-13.42\pm 0.09$, $4.3 \pm0.7$). The remarkable aggreement
between the values is rather encouraging. Additional absorption features occur
in some of the spectra of low $v \sin{i}$ stars shown in Fig.~\ref{spectra}. 
The stellar \nad\ lines are expected to be very weak throghout the class of B 
stars (being $\leq$ 2 m\AA\ for stars earlier than B7), but the cases where 
stellar lines have been identified were removed from the multicomponent fit.

\section{The kinematical structure} \label{kinem}

\begin{table*}
\caption{Profile fit results for the Na\,{\sc i} D absorption-lines. Each
successive column gives the star's name, component number, LSR radial velocity
\vlsr, column density (\logn), velocity dispersion b and the equivalent width
of the D$_2$ and D$_1$ lines}
\scriptsize
\begin{tabular}{lcrrccrrclcrrccrr}
\hline
Star  & C& \multicolumn{1}{c}{V$_{\mbox{\sc LSR}}$} & \multicolumn{1}{c}
{V$_{\sun}$} & \multicolumn{1}{c}{\logn} & b & \multicolumn{1}{c}
{W$_{\rm D2}$} & \multicolumn{1}{c}{W$_{\rm D1}$}  & &
Star  & C& \multicolumn{1}{c}{V$_{\mbox{\sc LSR}}$} & \multicolumn{1}{c}
{V$_{\sun}$} & \multicolumn{1}{c}{\logn} & b & \multicolumn{1}{c}
{W$_{\rm D2}$} & \multicolumn{1}{c}{W$_{\rm D1}$}  \cr
\noalign{\smallskip} \cline{3-8} \cline{12-17} \noalign{\smallskip}
     &   & \multicolumn{2}{c}{km s$^{-1}$} &
\multicolumn{1}{c}{cm$^{-2}$} & km s$^{-1}$ & \multicolumn{2}{c}{m\AA} & &
     &   & \multicolumn{2}{c}{km s$^{-1}$} &
\multicolumn{1}{c}{cm$^{-2}$} & km s$^{-1}$ & \multicolumn{2}{c}{m\AA} \cr\hline
239327&1&  0.5& 10.0& 12.3&  3.2& 162.9& 122.4&&251928&1&  3.5& 11.4& 11.5&  3.5&  48.5&  26.2\cr
      &2& $-$7.4&  2.0& 11.5&  8.0&  60.6&  34.5&&      &2& $-$3.3&  4.6& 12.6&  2.1& 139.5& 118.9\cr
239370&1&  0.6&  9.8& 13.0&  2.5& 188.2& 168.9&&      &3&$-$11.5& $-$3.5& 11.4&  4.8&  41.7&  21.9\cr
      &2& $-$3.2&  6.0& 11.9&  7.1& 137.1&  76.7&&251942&1&  1.2&  8.9& 12.6&  2.4& 157.9& 134.6\cr
239423&1& $-$2.2&  6.8& 11.7&  5.0&  88.5&  48.8&&      &2& $-$6.2&  1.5& 11.7&  2.7&  68.3&  39.8\cr
      &2& $-$9.4& $-$0.4& 11.2&  1.4&  27.0&  14.8&&251979&1& $-$2.1&  5.5& 12.7&  3.4& 215.4& 188.5\cr
239531&1& $-$1.3&  7.5& 11.8&  2.7&  89.5&  52.3&&251987&1& $-$2.0&  6.1& 11.4&  5.4&  43.0&  21.7\cr
      &2& $-$5.9&  3.0& 11.8&  9.5& 108.5&  56.9&&      &2&$-$17.0& $-$8.9& 10.2&  0.3&   3.3&   1.2\cr
239702&1& $-$0.5&  7.9& 12.6&  2.4& 154.5& 132.0&&251988&1&  1.4&  9.9& 13.9&  1.4& 150.1& 138.3\cr
      &2&$-$11.8& $-$3.4& 11.4&  5.9&  50.5&  26.5&&      &2& $-$4.5&  3.9& 11.5&  3.3&  56.2&  30.5\cr
239779&1&  1.0&  9.2& 12.5&  2.2& 133.9& 111.0&&252114&1& $-$2.4&  4.8& 12.8&  3.1& 205.3& 176.4\cr
      &2& $-$7.7&  0.5& 11.6&  3.2&  65.8&  37.1&&252136&1&  7.7& 15.0& 10.9&  0.2&   9.3&   6.2\cr
240041&1& $-$1.9&  5.7& 12.9&  2.4& 176.4& 156.5&&      &2& $-$3.1&  4.3& 12.3&  3.1& 152.2& 115.4\cr
      &2& $-$6.0&  1.6& 11.6&  6.2&  67.4&  35.8&&      &3&$-$12.3& $-$5.0& 11.1&  5.8&  23.8&  11.5\cr
240265&1&  0.5&  7.5& 12.5&  1.9& 121.4& 105.2&&252245&1& $-$0.7&  6.6& 12.3&  3.2& 162.6& 125.1\cr
      &2& $-$6.3&  0.7& 11.0&  1.5&  19.3&  10.4&&252262&1& $-$0.8&  6.2& 12.0&  1.9&  88.7&  62.7\cr
      &3&$-$13.9& $-$7.0& 11.2&  6.0&  29.1&  14.9&&      &2& $-$9.9& $-$2.8& 10.7&  4.7&  10.3&   5.2\cr
240368&1& $-$0.4&  6.3& 10.7&  1.9&  10.1&   5.0&&252284&1& $-$4.9&  1.0& 11.1&  4.7&  26.1&  12.3\cr
      &2& $-$9.7& $-$3.1& 10.9&  4.3&  16.1&   8.4&&252294&1& $-$0.6&  6.3& 12.5&  1.6& 101.2&  95.1\cr
240621&1&  0.8&  6.9& 12.6&  2.9& 178.5& 133.9&&      &2& $-$5.3&  1.6& 11.9&  4.0& 113.8&  65.9\cr
      &2&$-$11.8& $-$5.6& 12.7&  5.9& 312.8& 269.5&&252321&1& $-$0.9&  6.3& 12.8&  2.7& 182.2& 156.9\cr
      &3&$-$23.3&$-$17.2& 12.6&  5.3& 274.4& 225.2&&256745&1&  3.1& 14.0& 12.4&  1.6& 103.5&  88.4\cr
      &4&$-$32.2&$-$26.0& 12.3&  2.3& 134.3& 102.4&&      &2& $-$4.9&  6.0& 10.9&  5.0&  16.8&   8.5\cr
      &5&$-$38.8&$-$32.7& 15.3&  0.5& 277.4& 223.2&&256763&1&  2.3& 13.3& 12.4&  3.9& 186.5& 148.7\cr
      &6&$-$49.6&$-$43.5& 12.5&  2.6& 160.7& 131.7&&256788&1&  3.5& 14.4& 12.9&  2.6& 179.0& 165.9\cr
      &7&$-$58.1&$-$52.0& 12.6&  3.0& 188.9& 149.6&&      &2& $-$4.9&  6.0& 11.2&  2.7&  29.8&  15.1\cr
240645&1& $-$3.8&  2.3& 11.2&  5.1&  28.0&  13.4&&256789&1& 11.2& 22.0& 10.9&  2.1&  16.5&   8.6\cr
251460&2& $-$2.9&  7.1& 13.0&  1.8& 144.6& 130.6&&      &2&  3.4& 14.2& 12.7&  2.0& 138.7& 122.0\cr
      &3&$-$10.4& $-$0.4& 11.3&  8.2&  38.8&  19.9&&      &3& $-$3.2&  7.7& 11.2&  6.5&  27.3&  13.9\cr
      &1&  3.3& 13.3& 11.2&  8.6&  27.6&  14.0&&256798&1&  3.8& 14.5& 12.8&  2.0& 141.3& 131.1\cr
251530&1& $-$0.3&  9.4& 13.0&  2.0& 152.4& 145.2&&256800&1&  1.7& 12.5& 12.4&  3.9& 189.1& 153.8\cr
      &2& $-$5.6&  4.1& 11.7&  7.3&  86.4&  47.8&&256828&1&  0.8& 11.3& 12.7&  3.0& 191.6& 171.6\cr
251545&1& $-$2.0&  7.7& 14.9&  1.1& 214.4& 167.8&&256834&1&  3.4& 13.8& 12.9&  2.4& 168.2& 157.9\cr
      &2&$-$13.2& $-$3.5& 10.1&  0.2&   2.8&   0.8&&      &2& $-$3.5&  7.0& 12.1&  1.5&  80.6&  66.8\cr
251554&1& 36.7& 46.5& 11.0&  6.9&  17.7&   9.0&&256843&1&  0.8& 11.2& 14.6&  2.1& 258.6& 235.9\cr
      &2& $-$1.6&  8.2& 12.9&  2.7& 190.6& 174.7&&256849&1&  1.6& 11.8& 14.2&  1.6& 182.4& 156.5\cr
      &3&$-$11.8& $-$2.0& 10.9&  4.3&  16.1&   8.2&&256900&1&  1.4& 11.0& 12.6&  2.2& 144.2& 120.6\cr
251582&1& $-$2.6&  7.0& 12.7&  2.3& 161.1& 140.5&&      &2& $-$7.8&  1.7& 10.9&  2.8&  13.4&   7.3\cr
251625&1& $-$1.8&  7.6& 12.9&  2.6& 188.4& 166.8&&256924&1& $-$7.7&  1.9& 10.5&  2.4&   6.1&   2.6\cr
      &2&$-$14.0& $-$4.5& 11.2&  4.8&  29.9&  15.4&&256953&1&  2.1& 11.4& 13.1&  1.8& 145.3& 131.9\cr
251629&1& 11.8& 21.0& 10.9&  0.3&   9.1&   4.9&&256955&1& $-$1.6&  7.3& 11.0&  5.7&  20.4&  10.0\cr
      &2& $-$1.8&  7.4& 12.5&  3.7& 213.2& 162.9&&256980&1&  9.8& 18.2& 11.1&  6.5&  27.6&   9.0\cr
      &3& $-$8.8&  0.4& 12.6&  1.2&  83.4&  86.7&&      &2&  1.4&  9.9& 12.4&  2.7& 147.1& 133.0\cr
      &4&$-$18.1& $-$8.9& 11.6&  3.2&  59.1&  34.0&&      &3& $-$8.2&  0.2& 11.7&  3.5&  77.8&  36.9\cr
251699&1& $-$2.4&  7.0& 11.2&  5.4&  29.2&  15.0&&257002&1& 17.8& 26.1& 11.4&  0.8&  31.6&  16.9\cr
251717&1& $-$2.2&  6.7& 12.5&  2.3& 145.4& 121.1&&      &2& 10.4& 18.7& 11.4&  0.6&  24.3&  18.3\cr
251774&1& $-$2.4&  6.2& 12.3&  3.4& 162.5& 122.4&&      &3& $-$0.6&  7.7& 12.5&  4.4& 223.1& 169.4\cr
      &2&$-$11.2& $-$2.6& 11.1&  4.0&  20.1&  14.4&&      &4&$-$14.2& $-$5.9& 12.5&  3.1& 170.1& 150.4\cr
251802&1& $-$5.4&  4.2& 12.7&  3.5& 212.2& 178.6&&257032&1&  2.4&  9.9& 12.6&  3.3& 188.3& 162.5\cr
      &2&$-$16.1& $-$6.5& 12.2&  2.9& 128.7&  98.6&&257033&1& 34.3& 42.0& 10.7&  3.5&   9.8&   5.0\cr
      &3&$-$27.2&$-$17.6& 11.2&  2.1&  26.0&  15.0&&      &2&  0.2&  7.9& 13.1&  1.9& 155.7& 141.1\cr
      &4&$-$39.4&$-$29.8& 11.6&  2.1&  56.5&  35.8&&      &3&$-$11.4& $-$3.7& 10.4&  0.3&   4.2&   2.3\cr
      &5&$-$51.4&$-$41.8& 10.4&  0.2&   2.6&   3.1&&      &4&$-$31.3&$-$23.6& 10.8&  5.1&  12.4&   6.3\cr
251837&1& $-$1.7&  6.5& 12.4&  2.7& 149.8& 117.7&&257047&1&  1.0&  9.0& 11.4&  2.9&  46.3&  25.0\cr
      &2&$-$10.2& $-$2.0& 11.5&  4.3&  56.7&  28.5&&      &2& $-$8.7& $-$0.6& 10.6&  5.2&   9.0&   2.8\cr
251841&1&  2.4& 10.9& 11.0&  3.1&  18.2&   8.1&&257068&1& $-$0.1&  7.4& 12.6&  3.6& 211.5& 183.2\cr
      &2& $-$6.0&  2.5& 11.1&  3.4&  26.0&  12.8&&257069&1&  1.0&  9.0& 12.3&  1.9& 116.3&  90.6\cr
251874&1& $-$1.6&  7.1& 12.6&  5.3& 281.9& 232.4&&257142&1&  1.6&  9.7& 12.0&  2.3& 100.8&  72.0\cr
251894&1&  4.4& 13.2& 12.3&  1.8& 103.7&  86.6&&258697&1& $-$0.3&  8.4& 12.3&  2.7& 132.7& 114.5\cr
      &2& $-$2.0&  6.8& 12.5&  1.4&  98.7&  74.7&&377063&1&  4.2& 15.8& 11.5&  2.8&  50.3&  28.3\cr
      &3& $-$9.3& $-$0.5& 10.9&  2.6&  14.1&   7.6&&                                              \cr
251903&1& $-$1.0&  7.2& 11.1&  5.5&  23.3&  10.2&&                                              \cr
      &2&$-$11.6& $-$3.5& 10.9&  1.2&  14.3&   7.2&&                                              \cr \hline
\end{tabular}
\end{table*}

\begin{figure*}
\includegraphics[width=\hsize]{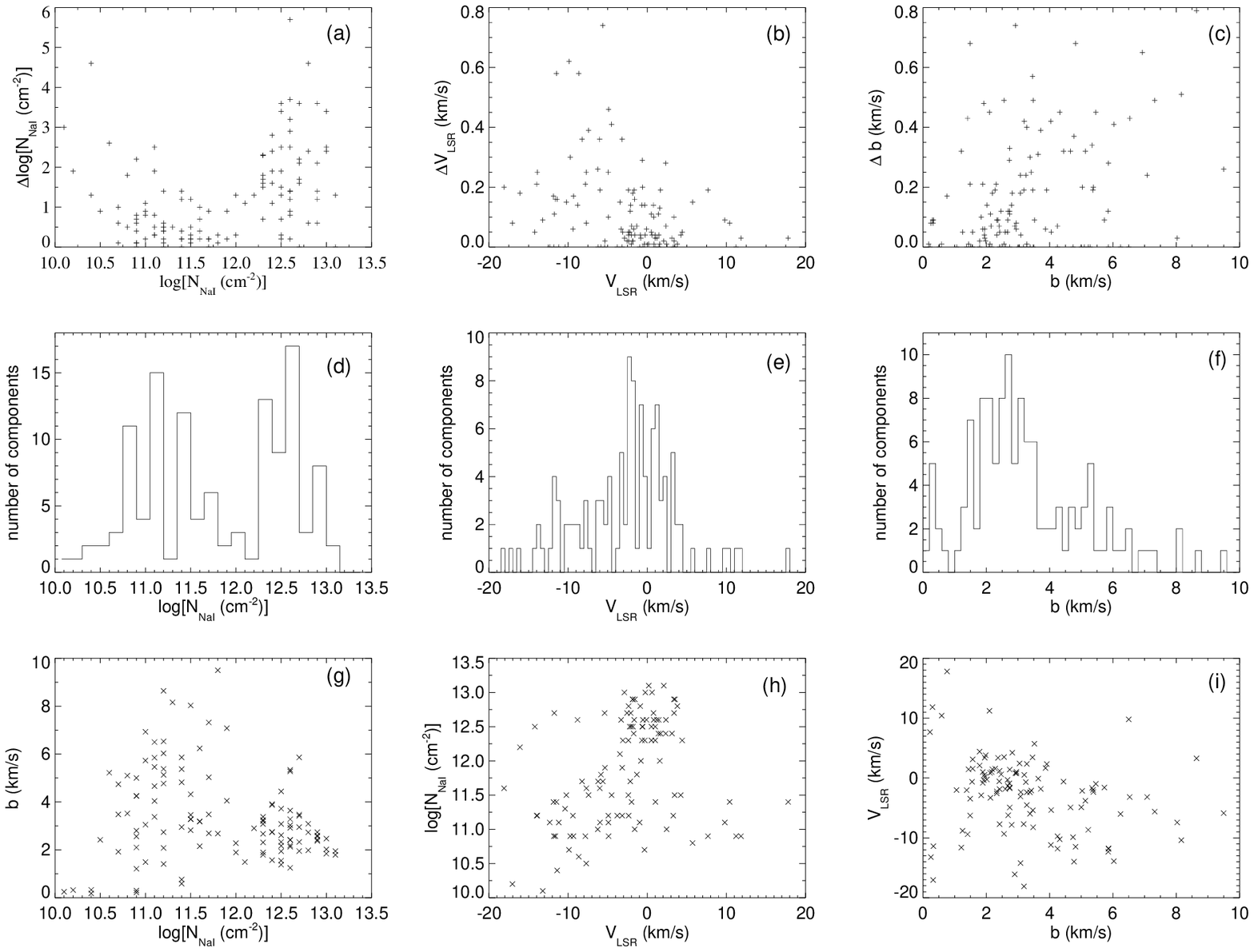}
\caption{Distribution of the free parameters determined in the profile fits and
their absolute mean errors. {\bf a-c} Distribution of the absolute mean errors
in the fitted values of \logn, \vlsr\ and b, respectively. For narrow, isolated
lines of moderate strength such errors are typically 1\%-10\% for \logn\ and
few hundredths of a \kms\ for b and \vlsr. For stronger, wider and more blended
lines the uncertainties in \logn\ are typically 5\%-20\%, and 0.1-0.4 \kms\ in
b and \vlsr. {\bf d-f} Histogram of \logn, \vlsr and b, respectively. There is a
clear division of the column densities around \logn\ $\approx$ 12.0 \cm2. The
LSR velocities, although centered around 0 \kms, are skewed to the negative
side. Also the velocity dispersion parameter b shows a concentration around 2.5
\kms\ and another around 5.5 \kms. {\bf (g)} Distribution of \logn\ {\it vs.} b.
{\bf (h)} Distribution of \logn\ {\it vs.} \vlsr. {\bf (i)} Distribution of
\vlsr {\it vs.} b. The lower column densities span the whole velocity 
dispersion range, and are concentrated mostly on the negative velocity range. 
On the other hand, the higher column densities are concentrated around b 
$\approx$ 3 \kms, and are essentialy at rest in the LSR frame; although a 
$\pm$ 2 \kms\ spread may be present. For the sake of clarity few points of 
very high negative and positive LSR velocities have been excluded from these 
plots. Further details can be found in the text}
\label{param}
\end{figure*}

Histograms of the individual distributions of \logn, \vlsr\ and b are given
in Figs.~\ref{param}d to \ref{param}f. Inspection of Fig.~\ref{param}d shows 
a clear division of the column densities around \logn\ $\approx$ 12.0 \cm2. 
The average column density of the sub-samples is 11.2 \cm2\ and 12.7 \cm2, 
respectively, both with small dispersion of $\pm$ 0.6 \cm2. 
Figure~\ref{param}e shows that the distribution of the LSR
velocities, although centered around 0 \kms, is skewed to the negative side.
The velocity dispersion parameter b shows a concentration around 2.5 \kms\
and barely suggests another around 5.5 \kms. These trends are somewhat
confirmed by Figs.~\ref{param}g-i, that shows the plots of b $vs.$ \logn, 
\logn\ $vs.$ \vlsr\ and \vlsr\ $vs.$ b, respectively. For the sake of 
clarity few points of very high negative and positive LSR velocities have 
been excluded from these plots.

As one can see the lower column densities span the whole
velocity dispersion range, and are concentrated mostly on the negative velocity
range. On the other hand, the higher column densities are concentrated around b
$\approx$ 3 \kms, and have velocities spread by $\pm$ 3 \kms\ around zero \kms\
in the LSR frame. The apparent lack of points with \logn\ $\leq$
10.5 \cm2\ for all b may be due to our detection limits. The lack of points with
b $\leq$ 1.5 \kms\ for \logn\ $\geq$ 11.5 \cm2 may be due to our inability to
resolve closely blended components.

\begin{figure*}
\centering
\includegraphics[width=14.6cm]{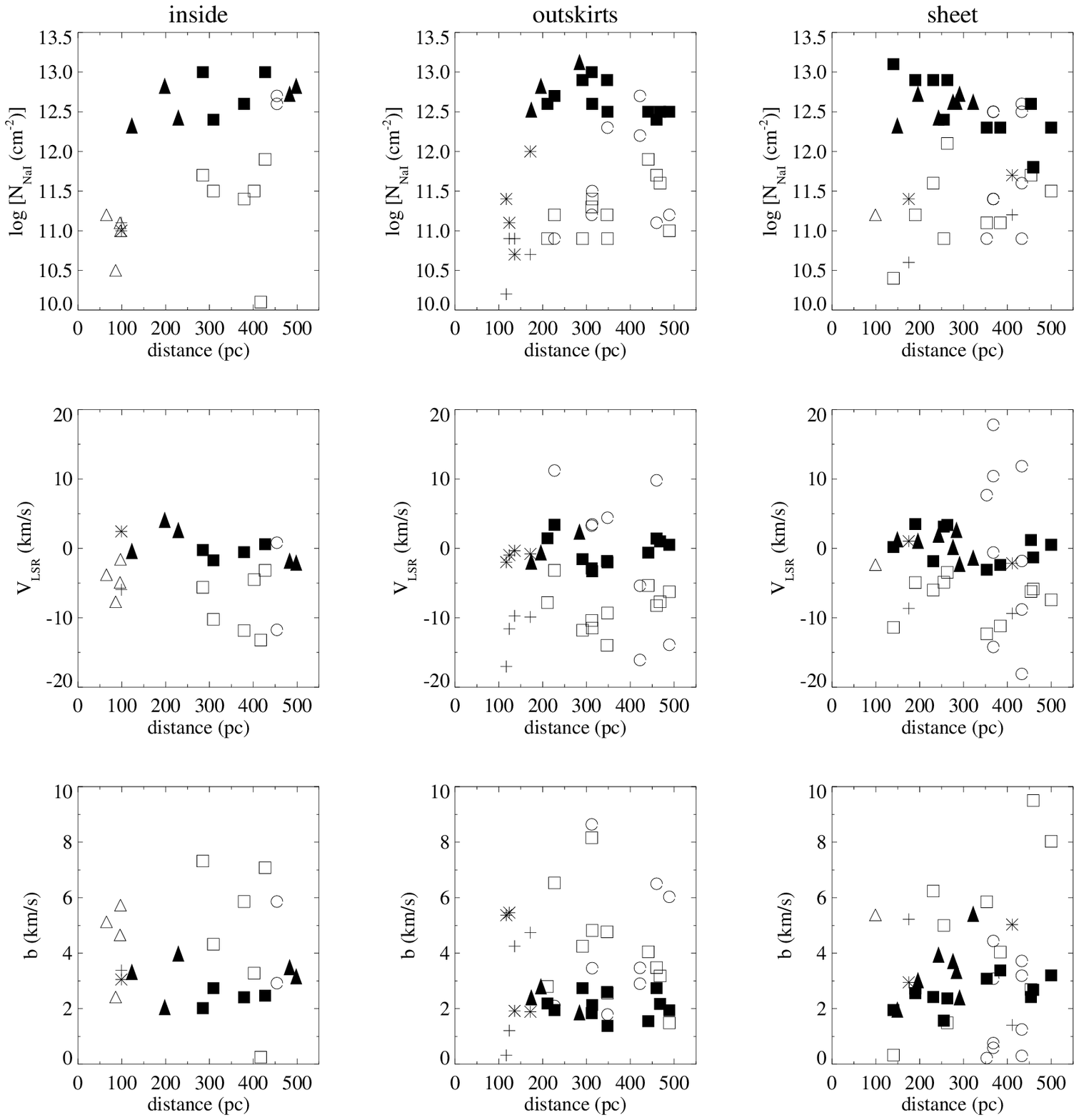}
\caption{Plots of \logn, \vlsr\ \ and b against the stellar distances. From
{\em left} to {\em right} the columns show the stars with line-of-sight {\em
inside} the clouds' contours, in their {\em outskirts}, and towards the {\em
sheet}. The line profiles with a {\em single weak} feature are represented by
the open triangles, while the profiles with a {\em single strong} feature are
represented by the solid triangles. The profiles with {\em two weak} features
are represented by the plusses and crosses, which indicate the features with
velocity similar to the single weak and strong components, respectively. The
profiles with {\em two or more} features are represented by the squares and
circles. The open and solid squares indicate those features similar to the
single weak and strong components. The open circles indicate the other ones,
i.e., the third, fourth, etc. The same structure for the {\em weak}
and the {\em strong} components suggests one interstellar gas at $d \leq$ 60 pc
and other $\approx$ 120-150 pc. The {\em strong} component (\logn\ $\approx$
12.7 \cm2) is spread $\pm$ 3 \kms around the zero velocities, and has low
velocity dispersion b $\approx$ 2.5 \kms. The {\em weak} component
(\logn\ $\approx$ 11.2 \cm2) is approaching to the Sun with average \vlsr\
around $-$7 \kms\ and have larger velocity dispersion b = 5 \kms. Further details
in the text}
\label{dis_nbv}
\end{figure*}

\subsection{Distribution of the interstellar gas components}

To investigate the distribution and kinematics of the interstellar gas
components along the line-of-sight plots of \logn, \vlsr\ and b as a function of
the stellar distance are given in Fig.~\ref{dis_nbv}. The {\em left} column 
shows those stars with line-of-sight {\em inside} the clouds' contours; the 
{\em middle} column those in their {\em outskirts}; and the {\em right} 
column those towards the {\em sheet}.

The plotting simbols indicate a further division of the components
according to the line profile characteristics. The line profiles with a {\em
single weak} feature are represented by the open triangles, while those profiles
with a {\em single strong} feature are represented by the solid triangles. The
profiles with {\em two weak} features are represented by the plusses and
asterisks, which indicate the features with velocity similar to the single weak
and single strong components, respectively. The profiles with {\em two or more}
features are represented by the squares and circles. The open and solid squares
indicate those features similar to the single weak and to the single strong
components. The open circles indicate the other components, i.e., the third,
fourth and so on.

The structure of the column density $vs.$ distance diagrams for the stars
{\em inside} the clouds' contours, in their {\em outskirts} and towards the
{\em sheet} is fairly similar. {\em Single weak} components are seen for the
stars closer than 60-80 pc from Sun (open triangles and plusses), and for most
of the more distant stars along the distance range (open squares). On the other
hand, the {\em single strong} component is seen only for stars more distant than
120-150 pc from the Sun (solid triangles, asterisks and solid squares). The open
circles tend to appear at distances greater than 300--350 pc, but without
clear correlation with the column densities or the radial velocities.

\begin{figure*}
\centering
\includegraphics[width=15.cm]{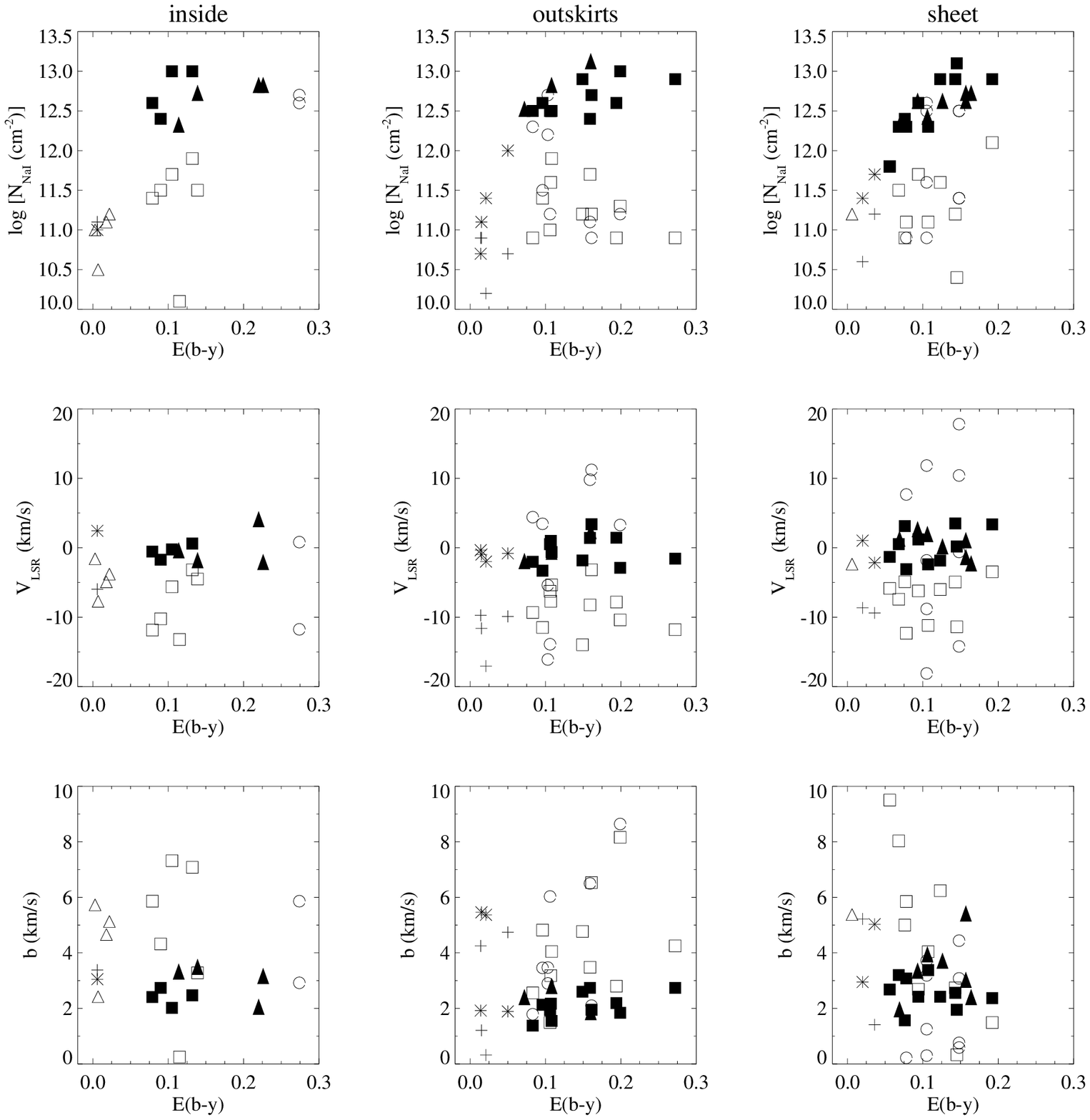}
\caption{Plots of \logn, \vlsr\ and b against the $E(b-y)$ colour excess. From
{\em left} to {\em right} the columns show the stars with line-of-sight {\em
inside} the clouds' contours, in their {\em outskirts}, and towards the {\em
sheet}. The plotting symbols have the same meaning as in Fig.~\ref{dis_nbv}.
The {\em strong} component is only picked up by the reddened stars, whereas 
the {\em weak} component is picked up by both reddened and unreddened stars. 
Such diagrams suggest that dust and gas apparently have the same distribution 
along the line of sight}
\label{eby_nbv}
\end{figure*}

The velocities of the components identified along the line of sight also have a
similar structure for the stars {\em inside} the clouds' contours, in their
{\em outskirts} and towards the {\em sheet}. The {\em single weak} component
(open triangles, open squares and plusses) has velocity more negative than $-$5
\kms, while the {\em single strong} component (solid triangle, solid square and
asterisks) is centered at 0 \kms\ in the LSR frame. The fact that the plusses
and asterisks show kinematic structure consistent with those of the {\em single
weak} and the {\em single strong} components, respectively, suggests that the
material have some inhomogeneities. The open circles appear in the whole range
of radial velocities.

The velocity dispersions of the components also have a similar
structure for the stars {\em inside} the clouds' contours, in their {\em
outskirts} and towards the {\em sheet}. The {\em single weak} component has b
$\approx$ 5.5 \kms. On the other side the {\em single strong} component has b
$\approx$ 2.5 \kms.

The b value is a measure of both thermal and turbulent velocities in an
interstellar cloud. We have b $= \sqrt{\frac{2kT}{m} + 2 v_{\rm t}^2}$, where
$v_{\rm t}$ is the rms turbulent velocity, $k$ is the Boltzmann's constant, $T$
is the kinetic temperature and $m$ is the mass of the atom. There is no
independent way to determine the degree of turbulence within the material, but
some qualitative considerations can be done.
The typical kinetic temperature of interstellar diffuse clouds is generally
taken to be about 80 -- 115 K, from observations of rotational excitation of
H$_2$ \citep{savage}. For \nai\ this corresponds to 0.24 $\leq$ b $\leq$
0.29 \kms. In this case the upper limits to $v_{\rm t}$ for the {\em weak} and
{\em strong} components would be 1.8 and 3.9 \kms.
Assuming a sound speed for diffuse interstellar clouds about 0.7 \kms\ 
\citep{spitzer} the observed components are either significantly hotter than a standard
diffuse cloud or are subject to supersonic turbulent motions.

\begin{figure*}
\centering
\includegraphics[width=15.cm]{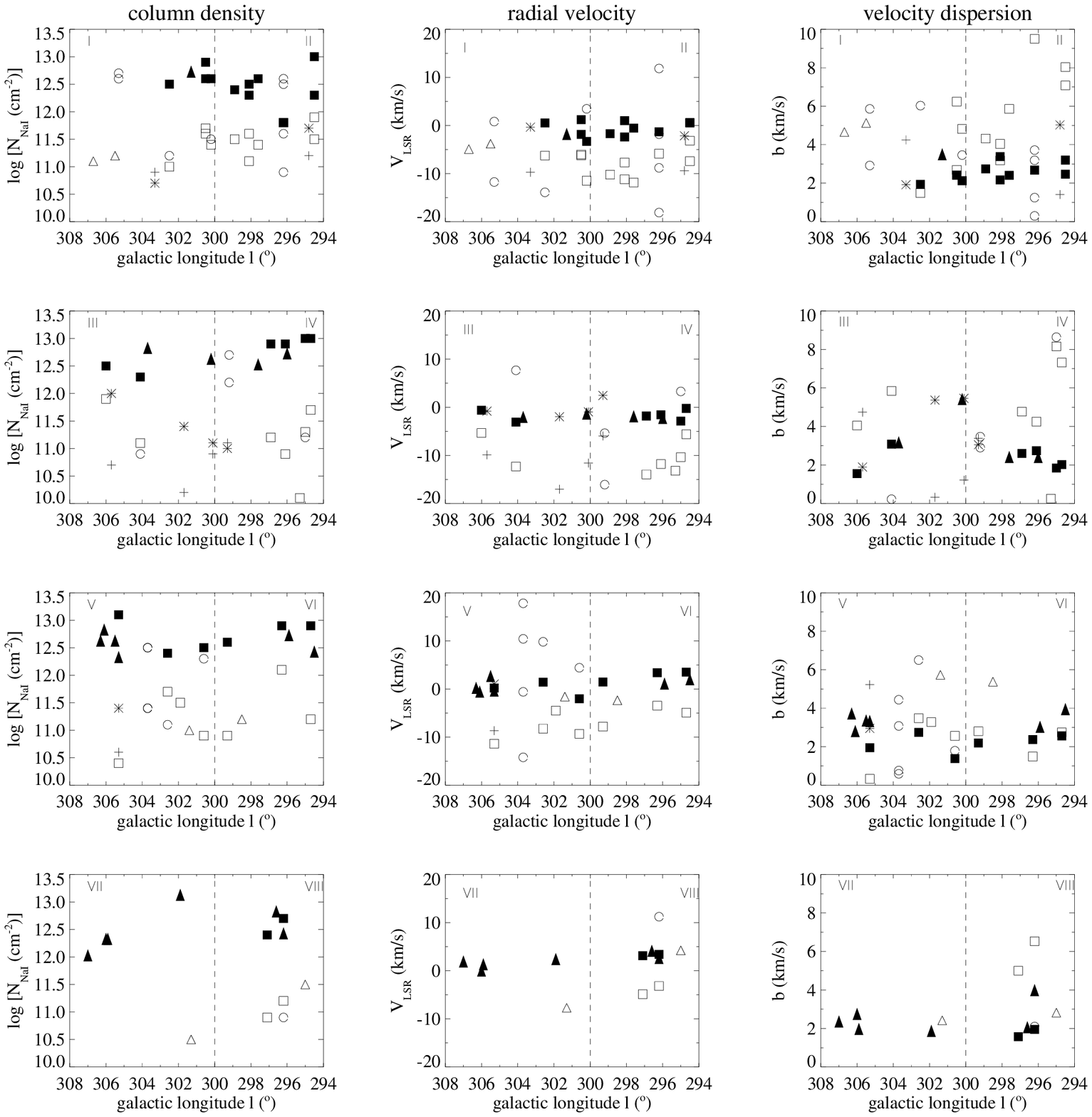}
\caption{{\bf a} Distribution of \logn\ ({\it left}), \vlsr\ ({\it middle}) and
b ({\it right}) with the Galactic Longitude. The symbols have the same meaning
as for Fig.~\ref{dis_nbv} and the dashed lines indicate the dividing point 
between the sub-areas. The {\em weak} and the {\em strong} components form 
two extended interstellar components, that cover the whole connecting area. 
There seems to have no dependence with the Galactic longitude, but the radial 
velocity may be increasing with the latitude}
\label{lgal_nbv}
\end{figure*}

\addtocounter{figure}{-1}
\begin{figure*}
\centering
\includegraphics[width=16.5cm]{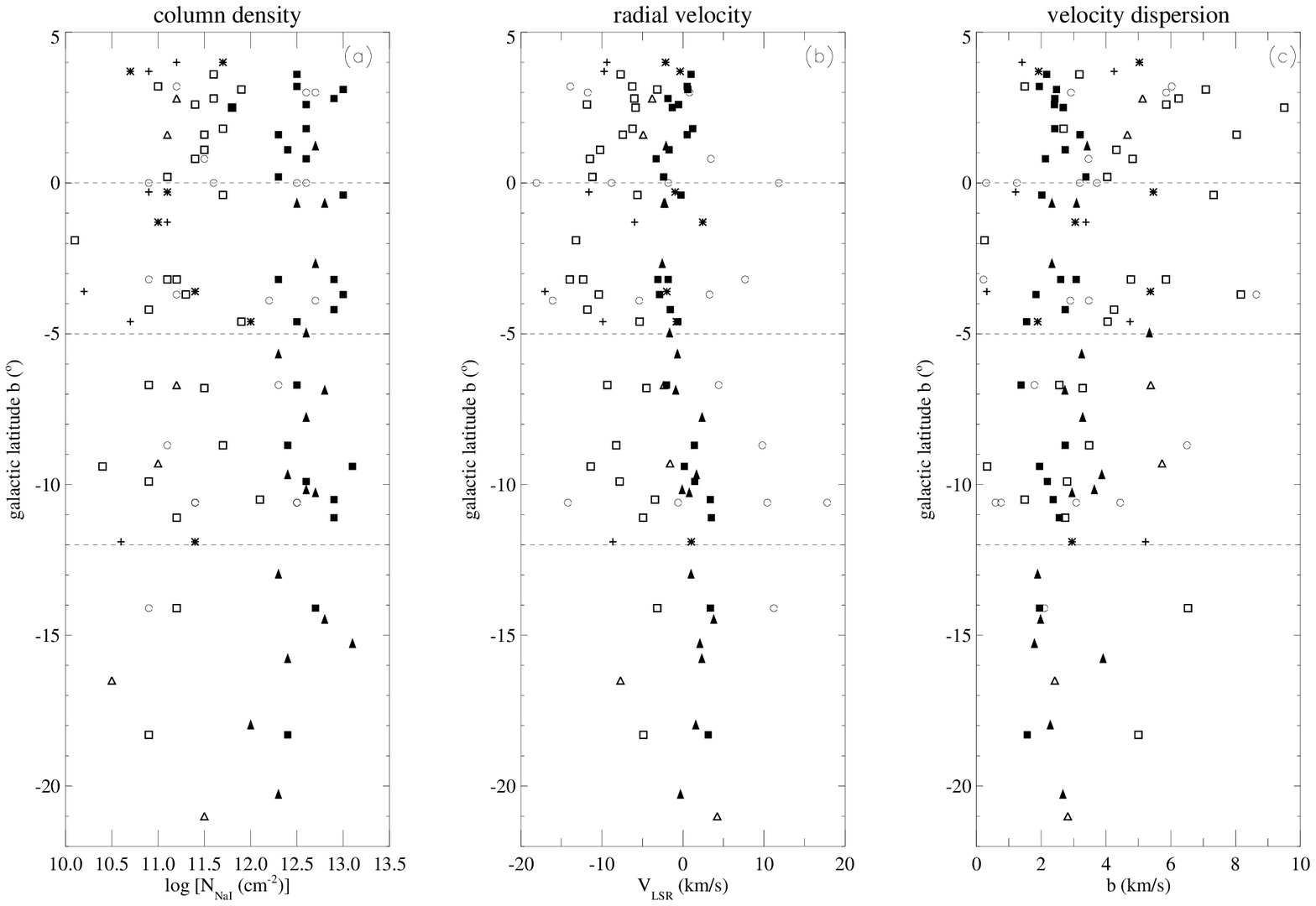}
\caption{{\bf b} Distribution of \logn\ ({\it left}), \vlsr\ ({\it middle}) and
b ({\it right}) with the Galactic Latitude. The symbols have the same meaning
as for Fig.~\ref{dis_nbv} and the horizontal dashed lines indicate the 
dividing point between the sub-areas. The {\em weak} and the {\em strong} 
components are present in the whole latitude range. Note that radial velocity 
of the {\em strong} component seems to increase from $-$3 \kms\ near the 
galactic plane to +3 \kms\ near the southern parts}
\end{figure*}

\subsection{The gas sheet-like features}

To look for a possible association of the interstellar gas components with the
observed dust features in this direction, \logn, \vlsr\ and b as a function of
the $E(b-y)$ colour excess have been plotted in Fig.~\ref{eby_nbv}. The 
columns and symbols have the same meaning as for Fig.~\ref{dis_nbv}. It can be 
seen that the {\em strong} component is only picked up by the reddened stars, 
i.e., those with $E(b-y) \geq 0\fm050$, whereas the {\em weak} component is 
picked up by both reddened and unreddened stars, suggestting that dust and 
gas apparently have the same distribution along the line of sight.

The average column density of the {\em weak} component (\logn\ \ $\sim$ 11.2
\cm2), would correspond to $N_{\rm H}$ $\sim 3 \times 10^{19}$ \cm2. It is
worth noticing at this point that, in Fig.~6a of Paper I, around 
60-70 pc the dispersion of the $E(b-y)$ increases, with the lower limit 
becoming slightly shifted to positive values. This indicates that the 
spectroscopy has provided a finer tuning of the existing low column density 
components, more difficult to disentagle based only on the colour excess 
values. On the other hand, it shows that the observed increase in the 
dispersion of $E(b-y)$ is real, and not due to observational errors, attesting
the high quality of the photometric data and method.

There appears to be a common envelope of minimum and maximum column density,
centered in a narrow distance slot around 120-150 pc, that remain unchanged up
to 350-400 pc, suggestting that a second volume of low density may have been
reached. The complete range of column densities from \logn $\sim$ 10.7 to 11.7,
is also shown by the stars centered around 60 pc, suggesting a sheet-like
structure.

Interestingly, the minimum and maximum column densities of the {\em strong}
component (\logn\ = 12.3 and 13.2 \cm2) correspond to $E(b-y) \sim 0\fm050$ and
0\fm300, respectively, if it is assumed that the standard relationship between
\logn\ and log $N_{\rm H}$ \citep{ferlet} and $E(b-y)$  and log $N_{\rm H}$
\citep{knude78} are valid. This may suggest that the high column densities 
may be the gas related to the dust sheet-like structure, reported in Paper I.
Since the Coalsack, Musca and Chamaeleon are also located around 150 $\pm$ 
30 pc they seem also to be part of the gas composing the sheet.

Assuming that the observed stars are within different dark clouds the column 
densities, radial velocities and velocity dispersions, shown in 
Figs.~\ref{spectra} to \ref{dis_nbv}, could be either the effect of a common 
structure containing the embedded molecular clouds or just an effect of 
different superposing structures. In order to investigate how the parameters 
(\logn, \vlsr, b) are distributed over the studied area we have used the eight
sub-areas delineated by the horizontal and vertical dashed lines and 
identified by the roman numbers in Fig.~\ref{obpos}. 

Figures \ref{lgal_nbv}a and b show the distribution of \logn\ ({\it left}), 
\vlsr\ ({\it middle}) and b ({\it right}) against the Galactic longitude and 
latitude, respectively, for the eight sub-areas. The symbols have the same 
meaning as for Fig.~\ref{dis_nbv}, and the vertical and horizontal dashed 
lines indicate the division between the sub-areas. The {\em strong} and 
{\em weak} components are present -- and well separated -- everywhere. The 
gas components form two extended interstellar features, that cover the whole 
connecting area.

Note, however, that the radial velocity of the {\em strong} component seems to
increase from $-3$ \kms\ near the galactic plane to +3 \kms\ near the southern
parts. This trend would be expected from an expanding bubble with the center
around $l \approx 309\degr$, $b = -9\degr$. This fact is quite interesting 
given the fact that high thermal pressure has been measured for the hot gas 
phase for a nearby direction ($l = 309\degr$, $b=-15\degr$) from a combination
of ROSAT 0.25 keV and optical data (Knude et al. -- in preparation).

\section{Discussion}

The analysis of the identified components indicates that the interstellar gas is
distributed in two extended sheet-like structures permeatting the whole area,
one at $d \leq$ 60 pc and another around 150 pc from the Sun. The nearby feature
is approaching the Sun with average radial velocity of $-$7 \kms, has low
average column density (\logn $\approx$ 11.2 \cm2) and is either hotter or more
turbulent than the {\em strong} component, with velocity dispersion b $\approx$
5 \kms. The more distant feature has column densities between 12.3 $\leq$ \logn\
$\leq$ 13.2, average velocity dispersion b $\approx$ 2.5 \kms\ and seems
associated to the dust sheet observed towards the Coalsack, Musca and
Chamaeleon direction. Its velocity is centered around 0 \kms, but there is a
trend for increasing from $-$3 \kms\ near $b = 1\degr$ to 3 \kms\ near
$b = -18\degr$.

There seems to be a common envelope of minimum and maximum column density,
centered in a narrow distance slot around 120 -- 150 pc, that remains unchanged
up to 350 pc, suggestting that a volume of low density may have been reached.
After that, a third group of high negative and positive velocity components
seemingly arise around 300-350 pc from the Sun, however, without indications
of an extended structure.

\subsection{The nearby, low column density feature}

Details of the overall distribution of the ISM ($d \leq$ 500 pc) along the
studied direction have been extensively discussed in Paper I. The kinematics of
the nearby gas has been reviewed by \citet{frisch86}. Details of the
velocity structure in the region (360\degr $\leq l \leq$ 295\degr\ and 0\degr
$\leq b \leq$ 30\degr), encompassing most of the Sco-Cen association, have been
discussed by \citet{crawford}.

In the immediate solar neighbourhood the observational data suggest that
the Sun is located within but close to the edge of the Local Interstellar
Cloud (LIC). First identified from its kinematics by \citet{LB92}, the
LIC was modelled by \citet{RL00} as roughly spherical with dimensions of
5--8\,pc, warm (T $\approx 7000$\,K), low density ($n_{\rm HI}<$ 0.1 cm$^{-3}$) 
and mostly neutral ($n_{\rm e}/n_{\rm HI} < 0.5$). Its largest column density
is about $2 \times 10^{18}$\,cm$^{-2}$ towards $l=157\degr$, $b=-25\degr$.

Moving at a heliocentric velocity of $\approx  26$ \kms\ towards 
$l \approx 186\degr$, $b \approx -16\degr$ (cf. \citealt{lallement95,
lallement98}), the LIC flows through the solar system and resonantly 
scatters solar H{\sc i} (Lyman $\alpha$) and He{\sc i} ($\lambda$584) 
radiation. This flow, is referred to as the local interstellar wind (LISW).

The LIC and other warm clouds with similar temperatures but a wide range
of metal depletions, are embedded in an irregularly shaped region, whose
radius ranges from 30 to 300 pc, and is deficient in dense neutral hydrogen
compared to the galactic average (e.g. \citealt{frisch83, paresce,
snowden90, tinbergen, warwick, welsh94}). Usually called Local Bubble, this 
cavity is thought to contain hot (T $\approx$ 10$^{6}$ K), low-density 
($n_{\rm HI}$ $\leq$ 0.025 cm$^{-3}$) gas and as recently shown by the ROSAT 
and EUVE shadowing experiments, coexists with the neutral atomic and 
molecular gas within its interior (e.g. \citealt{bowyer,kerp,snowden91,
snowden95,snowden98,wang}). It is, however, worth noticing that 
\citet{cravens} has recently shown that heliospheric X-ray emission can 
account for about 25\%-50\% of the observed soft X-ray background intensities,
which may put in doubt if the Local Bubble really contains the million
degree hot gas.

Optical interstellar absorption line studies within 100\,pc of the Sun show
evidence of a flow of material analogous to the LISW. Assuming that the Sun 
is embedded within a single coherently moving interstellar wind, from a 
direction ($l_{\rm w}$,$b_{\rm w}$) with velocity $v_{\rm w}$,
\citet{crutcher82} obtained the following vector ($l_{\rm w}$, $b_{\rm w}$,
$v_{\rm w}$) = (345\degr, $-$10\degr, $-$15 \kms). In a similar analysis 
\citet{frisch86} obtained (354\degr, +3\degr, $-$12 \kms) and \citet{crawford} 
obtained (301\degr, +59\degr, $-$9 \kms). A possible description for this 
flow assumes that the LISM is material which belongs to the surface of an 
expanding shell centered on the Sco-Cen association. \citet{crawford} has 
obtained an expansion velocity of 7 -- 9 \kms\, assuming the center at 
($l=320\degr$, $b= +10\degr$), the distance to the Sun of 140 pc and the 
radius of 110 pc, as suggested by \citet{degeus}. 

Another possible explanation for the existence of the neutral clouds and
their predominant flow away from the Sco-Cen association is a consequence of 
the interaction between the Local and Loop\,I Bubbles (see Sect.\,\ref{LBLI})
and subsequent local fragmentation of the interaction zone \citep{BFE00}.

Since the closest star in our sample is at 65 pc the LIC cannot be addressed
properly, neither the distance of the lower column density component. However,
according to \citet[and references therein]{welsh94} the column densities
of the nearby feature observed in this work should not be much closer than 50
pc from the Sun.

For the latitude/longitude region observed in our work the velocities 
predicted by the interstellar wind flow vector and the expanding shell 
models give essentially the same results \citep[cf. his Fig. 3]{crawford}. 
In any case the velocity of the low column density component
is consistent with the previous results, and indicate an outflow from the
Sco-Cen association.

A more recent discussion by \citet{cha} also using \nad\ line data and
{\it Hipparcos} distances proposes components at three distinct heliocentric
velocities in the Puppis-Vela region. Their component A with the velocity
in the range from +6 to +9\,\kms\ covers part of the region presently 
discussed.

\begin{figure}
\centering
\includegraphics[width=7.cm]{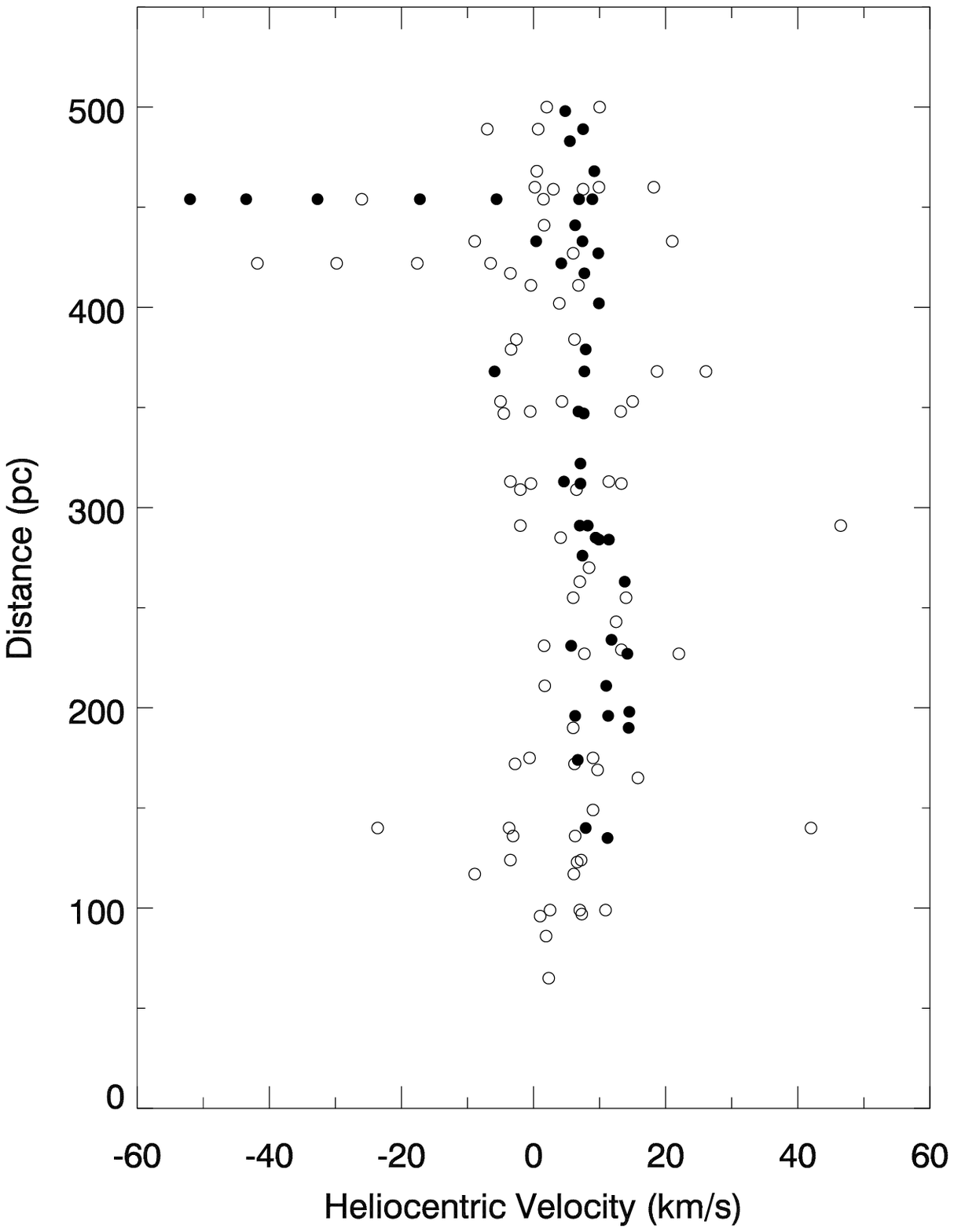}
\caption{Heliocentric velocities of the interstellar clouds components as a
function of stellar distance. Filled circles represent components for which
\logn$ \ge 12.5$\,cm$^{-2}$, while open circles represent components for
which \logn$ < 12.5$\,cm$^{-2}$. Note the well defined velocity distribution
for the {\it strong} component and that it appears for stars beyond $\sim
140$\,pc.}
\label{vel_dist}
\end{figure}

From the distances of Table 1 and heliocentric velocities from Table 2
we construct Fig.~\ref{vel_dist} of V$_{\sun}$ vs. distance and notice that 
\citeauthor{cha}'s component A is seen in two stars closer than 100\,pc and in 
fact seems present in all stars all the way to 450\,pc, this implies that 
component A covers all of our surveyed region, indeed a sheet nearer than 
about 100\,pc. Two stars marginally have absorption features at component B's 
lower velocity limit (12\,\kms), these stars are at 100\,pc and 130\,pc 
respectively. Otherwise component B is seen in most stars beyond 180\,pc; 
the upper distance limit is very sharply defined by component B's upper 
limit at 15\,\kms. No stars beyond $\approx$350\,pc shows any indication 
of component B. Component C (21 -- 23\,\kms) is virtually absent from our 
region and is only observed for galactic longitude less than 275\degr\ in 
\citeauthor{cha}'s survey, far from our region. Four stars closer than 100\,pc
show absorption in a narrow interval from +1 to +4\,\kms, this interval
is hardly represented between 100 and 400\,pc from where it is again present
in almost any star. A fifth group of absorption has velocities between 
$-$2 and $-$4\,\kms\ for three stars between 120 and 140\,pc, lines in 
this interval disappear for larger distances but reappear between 300 and 
400\,pc. Locally, that is for distances less than 150\,pc, we confirm 
three interstellar kinematic groups with absorption lines in the intervals: 
[$-$5,0], [0,5] and [5,10] the second of which represents the volume within 
100\,pc whereas the other two both are present between 120 and 150\,pc. 
The local ISM thus seems more complicated than proposed in \citet{cha}
investigation. One could speculate whether the interval [0,5] is 
representing the local undisturbed material and [5,10] and [$-$5,0] the 
expansion of the Local and the Sco--Cen Bubbles respectively?

Fig.~\ref{vel_dist} also indicates that the distribution of heliocentric
velocities for the diffuse components changes character between 300 and 
350\,pc. Closer than 300\,pc there is a set of velocities exceeding 10\,\kms,
this set disappears beyond 350\,pc. On the other hand a new set with
heliocentric velocities smaller than 0\,\kms\ appears beyond 300\,pc. Could 
this kinematic transition at 300--350\,pc be connected to the expansive 
movement of the backside of the Lower Centaurus--Crux feature being countered 
by an opposite motion? The expansion has ceased at about 350\,pc and the 
counter motion halting the expansion is first noticed at about 300\,pc.

\subsection{The interface between the Local and Loop I bubbles} \label{LBLI}

A soft X-ray shadow has recently been discovered on the edges of the Loop I
Bubble, cast by a warped annular volume of dense neutral matter that supposedly
formed during its collision with the Local Bubble \citep{egger}. 
The shadow counterpart, a huge H{\sc I} ring was also identified on the 
\citet{dickey} data.

\citet{egger} suggested that the steep increase of the column
density caused by the annular feature, from less than 10$^{20}$ to $\geq$ 7
$\times 10^{20}$, occurs at a distance of $\approx$ 70 pc from the Sun. This
distance is thought to be supported by results from optical and UV spectral
analysis of stars near the center of Loop I ($310\degr \leq l \leq 330\degr$
and $15\degr \leq b \leq 25\degr$) by \citet{centu}, who found the
presence of a neutral gas wall of $N_{\rm H} \sim 10^{20}$ \cm2\ at a distance
of 40 $\pm$ 25 pc. See also the determination of the radial extent of the 
Local Hot Bubble recently estimated by \citet{snowden98} based on the 
{\it ROSAT} and {\it IRAS} all-sky survey data.

However, for the hydrogen column density of the annular feature, the colour
excess and the \nai\ column densities obtained in this work suggest that the
interaction zone between the two bubbles is located around 120-150 pc from the
Sun. This result is corroborated by \citet{sfeir} -- an improved 
version of the 3D mapping of the Local Bubble cavity has been recently 
obtained by \citet{lallement03} -- where it is found that for the region 
investigated here the 20--50\,m\AA\ isocontours at $b \sim -20\degr$ occurs 
at $d\approx 100-150$\,pc, but at $b \sim -45\degr$ the distance is close to
70\,pc.  The hydrogen column densities and velocities observed by \citet{centu} 
are more consistent with our lower column density component at $d \leq$ 60 pc.
In the same strip of latitude, the wall of neutral gas has a column density 
$N_{\rm H} \sim 10^{21}$ \cm2 , as suggested by \citet{iwan}. Moreover,
the hydrogen column densities towards the interaction ring, used by 
\citet[cf. their Fig. 4]{egger} suggest a value around 3 $\times$
10$^{19}$ \cm2\ at 70 pc, also more consistent with our nearby feature. Since
\citet{centu} have observed only unreddened directions our results
cast some doubt on the distance to the interaction zone being located around 70
pc. It seems that the annular region is twisted and folded, with different
directions having different distances.

In addition to the velocities indicative of the flow of material in the last
section, some of the more distant stars also have components at nearly 0 \kms\
in the LSR \citep{frisch86}. In the longitude range $360 \leq l \leq 345$
there are a number of low velocity ($-4 \leq$ \vlsr $\leq +4$ \kms) components
with generally strong \nai\ lines, consistent with the proximity of the $\rho$
Oph and Lupus clouds \citep{crawford}. In the longitude range $325 \leq l \leq
295$ there are also several components, in the positive latitudes, with LSR
velocities which are small but clearly negative ($-4 \leq$ \vlsr $\leq 0$ \kms).

As mentioned in Sect.\,\ref{kinem}, the feature around 120 -- 150 pc observed 
in our work, has column densities suggestting that the gas is associated to 
the dust sheet observed towards the Coalsack, Musca and Chamaeleon direction. 
It is certainly not appropriate to make generalizations exclusively from our 
data, but the fact that the velocities observed here are also in the same 
range ($-3 \leq$ \vlsr $\leq +3$ \kms), suggests that the dust and gas feature
around 120 -- 150 pc, seems to be part of an extended large scale feature of 
similar kinematic properties, supposedly identified with the interface of the 
Local and Loop I bubbles.

\section{Conclusions}

The investigation of the interstellar gas components towards the Southern
Coalsack, Chamaeleon and Musca dark have produced the following results:

\begin{itemize}

\item The interstellar gas is distributed in two extended sheet-like structures
permeatting the whole searched area, one at $d \leq$ 60 pc and another around
120 -- 150 pc from the Sun.

\item The nearby feature is approaching to the Sun with average radial velocity
of $-7$ \kms, has low average column density \logn $\approx$ 11.2 \cm2 and
has velocity dispersion b $\approx$ 5 \kms. The more distant feature has column
densities between 12.3 $\leq$ \logn\ $\leq$ 13.2, average velocity dispersion
b $\approx$ 2.5 \kms\ and seems associated to the dust sheet observed towards
the Coalsack, Musca and Chamaeleon direction. Its velocity is centered around
0 \kms, but there is a trend for increasing from $-$3 \kms\ near $b = 1\degr$ to
3 \kms\ near $b = -18\degr$.

\item In aggreement with several independent investigations the nearby low
column density feature indicates a general outflow from the Sco-Cen association,
while the dust and gas feature around 120 -- 150 pc seem to be part of an
extended large scale feature of similar kinematic properties, supposedly
identified with the interaction zone of the Local and Loop I bubbles.  
Assuming that the interface and the ring-like volume of dense neutral matter 
have similar properties, our results suggest that the interaction zone between
the bubbles is twisted and folded.
\end{itemize}

\section*{Acknowledgments}
This paper is based on observations collected at the European Southern
Observatory (ESO, La Silla, Chile). The staff at ESO, both in La Silla and 
Garching are thanked for the assistence during the observing runs. Dr. Welty 
is thanked for the line-fitting program and Dr. Egger is thanked for the 
ring-like feature contours. The Brazilian Agencies CNPq and FAPEMIG, and the 
Danish NBIfAFG are acknowledged for supporting this research. W. Corradi 
wishes to express his gratitude to the NBIfAFG members for the invaluable 
help during the development of this work in Denmark.


\bibliographystyle{mn2e}

{}

\end{document}